\newcommand{\nmu}{\nabla_{\mu}}
\newcommand{\nmo}{\nabla^{\mu}}
\newcommand{\nnu}{\nabla_{\nu}}

\newcommand{\vpas}{4\pi\alpha_{{\rm S}}}

\newcommand{\MA}{{\mathcal A}}
\newcommand{\MAmu}{{\mathcal A}_{\mu}}

\newcommand{\MAnu}{{\mathcal A}_{\nu}}

\newcommand{\Aaomu}{A^{\alpha}_{\;\;\mu}}

\newcommand{\Aamu}{A^{a}_{\;\;\mu}}
\newcommand{\Aemu}{A^{1}_{\;\;\mu}}
\newcommand{\Aenu}{A^{1}_{\;\;\nu}}
\newcommand{\Azmu}{A^{2}_{\;\;\mu}}
\newcommand{\Aznu}{A^{2}_{\;\;\nu}}
\newcommand{\Bmu}{B_{\mu}}
\newcommand{\Bsmu}{B^{*}_{\mu}}
\newcommand{\Bmo}{B^{\mu}}
\newcommand{\Bsmo}{B^{*\mu}}
\newcommand{\Bnu}{B_{\nu}}

\newcommand{\Bsnu}{B_{\nu}^{*}}

\newcommand{\MDmu}{{\mathcal D}_{\mu}}
\newcommand{\MDnu}{{\mathcal D}_{\nu}}
\newcommand{\MDmo}{{\mathcal D}^{\mu}}
\newcommand{\MDno}{{\mathcal D}^{\nu}}

\newcommand{\MFmunu}{{\mathcal F}_{\mu\nu}}

\newcommand{\exFmunu}{{}^{(ex)}\!F_{\mu \nu}}

\newcommand{\Fmunu}{F_{\mu \nu}}

\newcommand{\Famunu}{F^{a}_{\;\;\mu\nu}}

\newcommand{\Femunu}{F^{1}_{\;\;\mu\nu}}
\newcommand{\Fzmunu}{F^{2}_{\;\;\mu\nu}}

\newcommand{\xefnu}{{}^{(xe)}\!f_{\nu}}

\newcommand{\Gmunu}{G_{\mu\nu}}
\newcommand{\Gmonu}{G^{\mu}_{\;\;\nu}}

\newcommand{\Gsmunu}{G^{*}_{\mu\nu}}
\newcommand{\Gsmonu}{G^{*\mu}_{\;\;\nu}}

\newcommand{\MHmu}{{\mathcal H}_{\mu}}

\newcommand{\MHmo}{{\mathcal H}^{\mu}}
\newcommand{\MHnu}{{\mathcal H}_{\nu}}

\newcommand{\MJmu}{{\mathcal J}_{\mu}}

\newcommand{\jmu}{j_{\mu}}

\newcommand{\jalomu}{j^{\alpha}_{\;\;\mu}}
\newcommand{\jaomu}{j^{a}_{\;\;\mu}}
\newcommand{\jaumu}{j_{a\mu}}

\newcommand{\jeomu}{j^{1}_{\;\;\mu}}

\newcommand{\jzomu}{j^{2}_{\;\;\mu}}

\newcommand{\jdomu}{j^{3}_{\;\;\mu}}
\newcommand{\jvomu}{j^{4}_{\;\;\mu}}

\newcommand{\jeonu}{j^{1}_{\;\;\nu}}
\newcommand{\jzonu}{j^{2}_{\;\;\nu}}

\newcommand{\jalumu}{j_{\alpha\mu}}

\newcommand{\jeumu}{j_{1\mu}}
\newcommand{\jzumu}{j_{2\mu}}

\newcommand{\jbumu}{j_{\beta\mu}}
\newcommand{\jalonu}{j^{\alpha}_{\;\;\nu}}
\newcommand{\jbomu}{j^{\beta}_{\;\;\mu}}

\newcommand{\exjmu}{{}^{(ex)}\!j_{\mu}}

\newcommand{\exjmo}{{}^{(ex)}\!j^{\mu}}

\newcommand{\Kaubu}{K_{\alpha\beta}}

\newcommand{\Kaobo}{K^{\alpha\beta}}

\newcommand{\keumu}{k_{1\mu}}
\newcommand{\kzumu}{k_{2\mu}}

\newcommand{\LRST}{L_{{\rm RST}}}

\newcommand{\LD}{L_{{\rm D}}}

\newcommand{\LG}{L_{{\rm G}}}

\newcommand{\Tmunu}{T_{\mu\nu}}
\newcommand{\MTmunu}{\mathcal T_{\mu\nu}}

\newcommand{\TTmunu}{{}^{(T)}\!T_{\mu\nu}}
\newcommand{\DTmunu}{{}^{(D)}\!T_{\mu\nu}}
\newcommand{\GTmunu}{{}^{(G)}\!T_{\mu\nu}}

\newcommand{\Gmu}{\Gamma_{\mu}}

\newcommand{\gmu}{\gamma_{\mu}}

\newcommand{\Gnu}{\Gamma_{\nu}}

\newcommand{\M}{{\mathcal{M}}}

\newcommand{\Ai}{{}^\textrm{[1]}A_0(r)}
\newcommand{\Er}{ {}^\textrm{[1]}E_r(r)}
\newcommand{\Ac}{{}^\textrm{[c]}A_0(r)}
\newcommand{\Accc}{{}^\textrm{[cc]}A_0(r)}
\newcommand{\AS}{{}^\textrm{[S]}A_0(r)}

\newcommand{\eAo}{{}^{(1)}\!A_0}
\newcommand{\zAo}{{}^{(2)}\!A_0}

\newcommand{\Fbomono}{F^{\beta\mu\nu}}

\newcommand{\kaumu}{k_{a\mu}}

\newcommand{\ako}{{}^{(a)}\! k_0}

\newcommand{\TToo}{{}^{(T)}\!T_{00}}

\newcommand{\DToo}{{}^{(D)}\! T_{00}(\vec{r})}

\newcommand{\GToo}{{}^{(G)}\! T_{00}(\vec{r})}

\newcommand{\zetajml}{\zeta^{j,m}_{\;\; l}}
\newcommand{\zetappn}{\zeta^{\frac{1}{2},\frac{1}{2}}_{\;\;\, 0}}
\newcommand{\zetappe}{\zeta^{\frac{1}{2},\frac{1}{2}}_{\;\;\, 1}}
\newcommand{\zetapmn}{\zeta^{\frac{1}{2},-\frac{1}{2}}_{\;\;\, 0}}
\newcommand{\zetapme}{\zeta^{\frac{1}{2},-\frac{1}{2}}_{\;\;\, 1}}

\newcommand{\sdot}{\,{\scriptscriptstyle{}^{\bullet}}\,}

\newcommand{\Falomunu}{F^{\alpha}_{\;\;\mu\nu}}

\newcommand{\DGmu}{\textnormal I \! \Gamma_\mu}
\newcommand{\DGmo}{\textnormal I \! \Gamma^\mu}
\newcommand{\DGnu}{\textnormal I \! \Gamma_\nu}

\documentclass[preprint,eqsecnum,aps,showpacs]{revtex4}
\usepackage{epsfig,amssymb,amsmath}
\usepackage[a4paper]{geometry}

\begin{document}

\title{\LARGE{Positronium Groundstate \\in\\Relativistic Schr\"odinger Theory}\\[6ex]}
\author{T. Beck, M. Mattes, and M. Sorg} \affiliation{{II}.\ Institut f\"ur Theoretische
  Physik der
  Universit\"at Stuttgart\\Pfaffenwaldring 57\\ D-70550 Stuttgart\\ Germany\\
  {\rm e-mail:} {\tt sorg@theo2.physik.uni-stuttgart.de}}

\maketitle

\pagebreak

\begin{center}\Large Abstract\end{center}

The usefulness of the Relativistic Schr\"odinger Theory (RST) is studied in the field of
atomic physics. As a concrete demonstration, the positronium groundstate is considered in
great detail; especially the groundstate energy~$E_0$ is worked out in the
non-relativistic approximation and under neglection of the magnetic interactions between
the positron and the electron. The corresponding RST prediction~$(E_0\simeq -6,48\,[eV])$
misses the analogous conventional Schr\"odinger result~$(E_0\simeq -6,80\,[eV])$ but is
closer to the latter than the corresponding Hartree approximation~$(-2,65\,[eV])$. The
missing binding energy of $6,80-6,48=0,32 [eV]$ can be attributed to the approximative use
of an SO(3) symmetric interaction potential which in RST, however, is actually only SO(2)
invariant against rotations around the z-axis. It is expected that, with the correct use
of an anisotropic interaction potential due to the SO(2) symmetry, the RST predictions
will come even closer to the conventional Schr\"odinger result, where however the
mathematical structure of RST relies on exotic (i.e.\ double-valued) wave functions and on
the corresponding unconventional interaction potentials (e.g.\ Struve-Neumann potential).

\pagebreak

\section{Introduction and Survey of Results}
The general philosophy, being well accepted by most of the physicists, says that any new
theory must be in agreement with the successes of the existing theories of long standing.
There are many instances in the development of physics which can serve as a demonstration
of this dogma: general relativity vs.\ Newtonian gravitation theory~\cite{We,HaIs}, special
relativity vs.\ Newtonian mechanics~\cite{Sa,Ar}, Maxwellian electrodynamics vs.\
Faraday's theory of magnetism and Gau\ss' law of electrostatics~\cite{Ja}, or Heisenberg's
matrix mechanics and Schr\"odinger's wave mechanics vs.\ Bohr's and Sommerfeld's old
quantum theory~\cite{Jam} etc. The present paper is intended to present a further example
of this hierarchical ``inclusion'' of physical theories; namely the recently established
\textbf{R}elativistic \textbf{S}chr\"odinger \textbf{T}heory
(RST)~\cite{SSMS}-\cite{BeSo2}, as a non-standard \emph{relativistic} generalization of
the ordinary \emph{non-relativistic} Schr\"odinger quantum mechanics for many-particles
systems, must be required to reproduce the well-known and generally accepted results of
the conventional quantum mechanics! Clearly, such a great goal can hardly be achieved in
full generality by one single paper; but for a first demonstration of this one would be
satisfied with a \emph{special} but nevertheless typical situation so that there may arise
some confidence into the validity of the inclusion dogma for the \emph{general} case of
RST.

Our choice of such a demonstration refers to the binding energy~$(E_0)$ of the positronium
groundstate in the non-relativistic approximation and with neglection of the magnetic
(i.e.\ spin-spin) interactions between the electron and the positron. In this lowest order
of approximation, the conventional Schr\"odinger theory predicts the binding energy~$E_0$
(= -6,80 eV) by means of a very simple argument~\cite{Me}: The two-particle wave
function~$\Psi(\vec{r}_1,\vec{r}_2)$ is factorized into the
product~$\psi(\vec{R})\cdot\psi(\vec{r})$ when written down in terms of the center-of-mass
coordinate~$\vec{R}\ \left(\doteqdot\frac{1}{2}\cdot(\vec{r}_1+\vec{r}_2)\right)$ and
relative coordinate~$\vec{r}\ \left(\doteqdot (\vec{r}_1-\vec{r}_2)\right)$;
and then the conventional two-particle Schr\"odinger equation (see equation~(\ref{6.14'}) below)
yields the ordinary Coulomb force problem for the internal motion written in terms of the
relative coordinate~$\vec{r}$. However the solution of this standard problem in atomic
physics is well-known and the conventional groundstate energy for positronium is easily
found as
\begin{equation}
  \label{eq:I1}
  E_0\left|\right._{\rm conv.} = -\frac{1}{4}\frac{e^2}{a_{\rm B}} = -\frac{1}{4}\alpha_{\rm S}^2Mc^2 
 \backsimeq -6,80\;[\mathrm{eV}]\ ,
\end{equation}
where~$a_{\mathrm{B}}$ is the Bohr radius~$(= \frac{\hbar^2}{Me^2})$
and~$\alpha_{\mathrm{S}}(=\frac{e^2}{\hbar c})$ is the fine structure constant. Indeed
this result is nothing else than the hydrogen groundstate energy due to a fixed nucleus,
with merely the electron mass~$M$ being replaced by the reduced mass~$\frac{M}{2}$ on
account of the comoving positron. Clearly there exists an extended literature for
improving these semiclassical predictions for both the hydrogen~\cite{JoSo} and
positronium~\cite{Kar} semiclassical groundstate energy~$E_0$, namely by taking into
account various kinds of QED corrections. However the point here is that one would like to
see first the coincidence of the results in the lowest-order of approximation before one
compares the higher-order predictions of the various theoretical approaches. This is the
reason why the intention of the present paper must consist in elaborating (within the RST
formalism) the positronium groundstate energy~$E_0$ in the \emph{non-relativistic approximation} 
and with \emph{neglection of the spin!} According to the inclusion dogma, this RST result
should then agree with the conventional prediction~(\ref{eq:I1})

However the calculation of the non-relativistic~$E_0$ is within the RST formalism not so
simple as in the conventional approach since both kinds of approximations do require some
subtle considerations: First, it has been asserted in some preceding
papers~\cite{PrSo}-\cite{SchMaSo} that the non-relativistic limit of RST coincides with
the well-known Hartree-Fock approach.  However the problem is here that the latter
approach is an approximation method whose predictions do not always come close enough to
the corresponding predictions of the standard Schr\"odinger quantum mechanics. A drastic
example for this deficiency refers just to the positronium groundstate~$E_0$ which is
missed by the Hartree approximation by roughly 50\%! (see Sect.~VI.~B). Thus if, for any
physical situation, the non-relativistic limit of RST would agree with the Hartree-Fock
approach, RST would be afflicted with the same deficiencies as HF; and it would therefore
probably not be worth while to further elaborate the new formalism. However the subsequent
RST treatment of positronium demonstrates the occurrence of \emph{``exotic'' states} and
the corresponding interaction potentials which shift the corresponding RST predictions
closer towards the conventional Schr\"odinger results (and thus also closer to the
observational data) than it is possible for the original Hartree-Fock method (Sect.~VI). 

A second point of concern refers to the spherical symmetry of the (internal) groundstate
of the conventional approach. This spherical symmetry is of course due to the corresponding
symmetry of the (internal) Coulomb potential. However, if one approaches the
non-relativistic limit of RST from the side of the fully relativistic formalism, it is not
a matter of course that one will encounter a spherically symmetric interaction potential
for the two particles. The reason for this is that the relativistic approach takes account
of the particle spin ($\leadsto$ Dirac equation); and usually there emerges some preferred
direction in three-space when angular momentum (spin or orbital) comes into play in
connection with the bound states ($\leadsto$ angular momentum conservation).  Especially
for the present situation with RST, the preferred space direction leaves its imprint on the
electric interaction potential in such a way that its spin-induced anisotropy survives the
non-relativistic limit. Therefore, even in this limit, we have to deal with an anisotropic
potential (see the Struve-Neumann potential~(\ref{4.17})-(\ref{4.18})); but we will neglect
this anisotropy and will take into account only the spherically symmetric part in order that
our non-relativistic treatment displays the same symmetry as the conventional
theory (i.e.\ the spherical symmetry of the internal Coulomb potential). Of course it
should be clear that one has to pay some prize for such a forceful approximation
procedure, and this means concretely that our non-relativistic RST prediction for the
positronium groundstate energy~$E_0$ will adopt a certain inaccuracy (see the discussion
of this point in the Appendix).  Actually, in place of the expected 6,80~[eV] of
equation~(\ref{eq:I1}), one ultimately arrives at 6,48~[eV] in the spherically symmetric
approximation, so that the elaboration of the proper RST prediction will require a more
efficient approximation technique for anisotropic potentials (not to be treated in
the present paper).  

The results, being described so far, are worked out by the following arrangement:

\begin{center}
\emph{\textbf{A.\ Relativistic Schr\"odinger Theory}}
\end{center}

In \textbf{Sect.~II}, the general RST dynamics for two-particle systems is briefly
sketched in order that the paper be sufficiently self-contained (for more details see the
preceding papers~\cite{SSMS}-\cite{BeSo2},~\cite{PrSo}-\cite{GrBeMa}). Here the key points
refer to the coupled matter and gauge field equations (i.e.\ \emph{Relativistic
Schr\"odinger Equation}~(\ref{2.14}) and non-Abelian \emph{Maxwell
Equation}~(\ref{2.35})), and also to the specific gauge structure of the theory which
admits to treat the electromagnetic and exchange forces on the same footing. Furthermore
the general RST formalism provides one with the energy-momentum densities~$T_{\mu\nu}$
(e.g.\ for matter, see~(\ref{2.19'}) below), the time components~($T_{00}$) of which serve
later on to define the energy~$E_{\mathrm{T}}$~(\ref{5.2a}) of the bound stationary field
configurations.

\begin{center}
  \emph{\textbf{B.\ Non-Standard Eigenvalue Problem}}
\end{center}

\textbf{Sect.~III} presents the \emph{mass eigenvalue equations} for the stationary
two-fermion systems. The crucial point here is that, at least for the ground state, both
particles must occupy physically equivalent states which can differ at most with respect
to the relative orientation of their magnetic fields~$H_a(\vec{r}),a=1,2$. This simple
logical requirement leads to two classes of possible field configurations:
\emph{ortho-positronium}~(\ref{3.7a}) and \emph{para-positronium}~(\ref{3.7b}). Naturally
the two wave functions~$\psi_a(\vec{r})\,(a=1,2)$ due to both classes turn out to be very
similar but are not identical as long as the magnetic (i.e.\ spin-spin) interactions of
both particles are taken into account, see the eigenvalue equations for
ortho-positronium~(\ref{3.29a})-(\ref{3.29b}) and for
para-positronium~(\ref{3.32a})-(\ref{3.32b}). If, however, the magnetic interactions are
neglected, the spin-up and spin-down components of the wave functions~$\psi_a(\vec{r})$
become decoupled and this leads to a simpler eigenvalue system,~(\ref{3.51a})-(\ref{3.51d}),
which has the advantage to admit \emph{spherically symmetric} solutions. Their
non-relativistic limit can easily be obtained, cf.~(\ref{3.60a}) or~(\ref{4.7}), and does
serve later on as a trial function for minimalizing the groundstate energy
functional~(Sect.~VI).

The most important point, however, refers here to the fact that the
solutions~$\varphi_{\pm}(r,\theta,\phi)$ of the mass eigenvalue system are
double-valued~\cite{BeSo2}, cf.~(\ref{3.49'}); and their non-relativistic
limit~$\widetilde{R}$~(\ref{3.60a}) does not obey an ordinary one-particle Schr\"odinger
equation (as is usually the case with the Hartree-Fock approximations) but it obeys a
modified form thereof, see equation~(\ref{3.62}). This is \emph{one} of the origins of the
potentiality of RST to predict the energy of the bound systems in better agreement with
the experimental data (Sect.~VI) than is possible for the Hartree-Fock approach!

\begin{center}
  \emph{\textbf{C.\ Unconventional Gauge Potentials}}
\end{center}

\textbf{Sect.~IV} reveals the \emph{second} origin of the superiority of RST over the Hartree-Fock
approach: this refers to the gauge potential~$A_\mu$ which mediates the interaction
between the two particles. Since the magnetic interactions are neglected, one is concerned
here exclusively with the time component~$A_0(\vec{r})$ (of~$A_\mu$) and this is linked to
the charge density~$j_0(\vec{r})$ of the particles by a \emph{Poisson equation},
cf.~(\ref{3.17}) below, which itself is to be deduced from the Maxwell
equation~(\ref{2.35}). However the point here is that the RST ground state of positronium
is of a rather \emph{exotic} nature, namely in the sense that the bosonic character of the
bound two-fermion system is transferred to anyone of its constituents: they individually own
an integral quantum number of total angular momentum~$(J_z=0)$, see equation~(\ref{3.41}).

A consequence of this exotic behavior of the fermions refers to the electrostatic
potential~$A_0(\vec{r})$~(\ref{4.8}) being generated by the corresponding electric charge
distribution~$j_0(\vec{r})$. On principle, such an exotic potential~$A_0(\vec{r})$ must
always be anisotropic (even for the groundstate), but in the Appendix it is demonstrated
that for a first rough estimate of the positronium groundstate energy~$E_0$ it is
sufficient to use only the isotropic part~${}^{(I)}A_0(\vec{r})$ of the electrostatic
interaction potential. This spherically symmetric part~${}^{(I)}A_0(\vec{r})$ is defined
by equation~(\ref{4.11}); and when the simple exponential
function~$\widetilde{R}(r)$~(\ref{4.7}) is taken as the (non-relativistic) trial wave
function, the associated potential is the \emph{Struve-Neumann
  potential}~$\Ai$~(\ref{4.17})-(\ref{4.18}) being exhibited in Fig.~1. In contrast to
their Hartree-Fock counterparts, these exotic RST potentials are still singular at the
origin~$(r=0)$, but less irregular as the Coulomb potential~$(\sim r^{-1})$, so that their
energy content is finite.

\begin{center}
\emph{\textbf{D.\ Groundstate Energy}}
\end{center}

Concerning the physical relevance of any new theory of quantum matter, perhaps the most
important object is here the \emph{energy functional}~$(E_{\rm T})$ whose value upon the
solutions of the eigenvalue problem yields the energy being carried by the bound
stationary field configurations. Clearly, the reason is that the set of these energy values
constitutes the theoretical level system which may then be opposed to the observational
data in order to test the usefulness of the theory. Therefore setting up the right energy
functional is one of the crucial points of any particle theory. In this sense,
\textbf{Sect.~V} presents a thorough inspection of the RST energy
functional~$E_{\mathrm{T}}$, especially concerning its non-relativistic approximation,
because the \emph{non-relativistic} groundstate energy of positronium is the main concern of the
present paper. According to the general expectation, the desired energy
functional~$(E_\mathrm{T})$ should be equipped with the property that its value upon the
groundstate solution of the eigenvalue problem yields the \emph{minimally possible} energy
value~($E_0$, say) in comparison to all other field configurations. In ordinary,
non-relativistic quantum mechanics such a \emph{principle of minimal energy} is
well-known, i.e. the \emph{Ritz variational principle}~\cite{Fl,Ba}, whose associated
Euler-Lagrange equation is just the conventional Schr\"odinger equation. 

Thus the task is now to find first the right relativistic generalization of the Ritz
principle within the RST framework; and afterwards one may look for its non-relativistic
approximation in order to calculate with its help the positronium groundstate energy. But
here the interesting point is now that the non-relativistic approximation of the \emph{RST
  principle of minimal energy} does not lead us back to its philosophical origin (i.e.\
the Ritz variational principle); rather, a new form of variational principle is found
which yields its associated groundstate energy closer to the experimental data than is
obtained by the well-known Hartree (or Hartree-Fock) approach. This RST result is obtained
by being satisfied with the \emph{spherically symmetric approximation} of the groundstate
wave functions (Appendix). It remains to be tested whether or not the present RST
principle of minimal energy would yield even exact coincidence with the conventional
Schr\"odinger energy eigenvalue if the anisotropy of the RST groundstate is taken into account.

The RST principle of minimal energy is constructed by the following steps: First, the
coupled RST system of eigenvalue and gauge field equations is identified as the set of
Euler-Lagrange equations being due to the well-known Hamilton-Lagrange action principle,
see equations~(\ref{5.8})-(\ref{5.11}). For this variational procedure, the imposition of
some constraint is not necessary, neither for the wave function~$\Psi$ nor for the gauge
field~$\MA_\mu$. However the stationarity of the action~$W_{\mathrm{RST}}$ with respect to
the scaling transformations of the electric gauge potential~$A_0$ yields a certain
relationship between the latter potential and the matter field (i.e.\ the \emph{Poisson
  identity}~(\ref{5.20})) which plays an important part for the subsequent construction of
the RST principle of minimal energy! Next, the \emph{mass
  functional}~$M_{\mathrm{T}}c^2$~(\ref{5.21}) is constructed as an intermediate step,
where this functional is to be considered as the immediate relativistic generalization of
the conventional Ritz variational principle. Here, the first constraint comes into play
(i.e.\ the normalization of the wave functions, see equation~(\ref{5.22})).

Although the mass functional correctly reproduces the mass eigenvalue system in form of
the corresponding Euler-Lagrange equations, it nevertheless cannot be identified with the
wanted energy functional~$(E_{\mathrm{T}})$ because the gauge field equations cannot be
deduced from it by means of the usual variational procedure. The solution of the problem
is finally attained by reconsidering the total field energy~$E_\mathrm{T}$ as the spatial
integral of the total energy-momentum density~$\TToo(\vec{r})$, see
equation~(\ref{5.2a}). Indeed if the stationary field configurations of this
functional~$E_\mathrm{T}$ are determined, under \emph{both} constraints of wave function
normalization \emph{and} Poisson identity, then the coupled system of mass eigenvalue
\emph{plus} gauge field equations is properly recovered for the stationary bound systems.
Thus, incorporating the two constraints by the method of Lagrangian multipliers, the final
form of the desired RST energy functional~$\widetilde{E}_\mathrm{T}$ is given by
equation~(\ref{5.43}).

\begin{center}
  \emph{\textbf{E.\ Numerical Test}}
\end{center}

Clearly, the usefulness of any theoretical construction can be evaluated only by opposing
its predictions to the observational data. Therefore the value of the obtained energy
functional~$\widetilde{E}_\mathrm{T}$~(\ref{5.43}) upon the chosen trial
function~$\widetilde{R}(r)$~(\ref{4.7}) for the positronium groundstate is calculated in
\textbf{Sect.VI} and the result is compared to the analogous procedure for the
Hartree-Schr\"odinger approximation method (since the spin is neglected here, the
Hartree-Fock approach reduces to the much more simpler Hartree approximation method). The
difference between RST and the Hartree-Schr\"odinger approach lies mainly in the
interaction potential of the electron and positron (see
equations~(\ref{4.17})-(\ref{4.18}) and~(\ref{6.25}) together with fig.~2) where the
competition ends in favour of RST: the RST predictions for the positronium groundstate
binding energy~$(-E_0)$ amount to 6,48~[eV] and therefore are closer to the experimental
value (6,80~[eV]) than the predictions of the simple Hartree-Schr\"odinger approach
(2,65~[eV]), see fig.~2. Moreover it is demonstrated in the Appendix that the
residual RST inaccuracy of (6,80 - 6,48) = 0,32~[eV] is due to the neglected
anisotropy of the Struve-Neumann potential~$\Ai$, so that one may legitimately expect a
higher degree of coincidence between the RST predictions and the experimental data as soon
as the RST approximation techniques become improved.

\section{Two-Fermion Systems in RST}
\indent RST, in general, provides a relativistic description for an arbitrary assemblage
of positive and negative charges \cite{BeSo}; but for the present treatment of positronium
one is concerned with one positively and one negatively charged Dirac particle.
\subsection{Conservation Laws}
Therefore the {\it total velocity operator} $\DGmu$ for such a two-fermion system must
appear as the direct sum of two ordinary Dirac matrices $\gmu$:
\begin{equation}\label{2.1}
\DGmu = (-\gmu) \oplus \gmu \ ,
\end{equation}
where the minus sign for the first particle (i.e. the positron) is merely due to
convention (the preceding papers were concerned with N-electron systems for which the
total velocity operator has been adopted as $\Gmu = \gmu \oplus \gmu \oplus \dots$). With
the existence of such a velocity operator $\DGmu$ (\ref{2.1}) it becomes possible to
associate a conserved current $\jmu$ with any two-particle wave function $\Psi(x)$ which
itself is the direct sum (Whitney sum) of the single-particle wave functions $\psi_a$
($a=1,2$)
\begin{equation}\label{2.2}
\Psi= \psi_1 \oplus \psi_2 \ ,
\end{equation}
namely by means of the following prescription
\begin{equation}\label{2.3}
\jmu = \bar \Psi \, \DGmu \Psi \ .
\end{equation}
Thus the desired conservation law may be written down in local form as
\begin{equation}\label{2.4}
\nabla^\mu j_\mu = 0 \ ,
\end{equation}
and the associated {\it total charge} can be defined through
\begin{equation}\label{2.5}
z=\int\limits_{(S)} j_\mu \, dS^\mu
\end{equation}
where the space-like three-dimensional hypersurface ($S$) is arbitrary just on account of
the local law (\ref{2.4}). Clearly, for the considered positronium, the total charge $z$
(\ref{2.5}) is zero because both particles (i.e. electron and positron) carry opposite
charges.

\indent Besides the charge conservation law (\ref{2.4}), there are further conservation
laws which have to be obeyed by a closed two-particle system, e.g. the energy-momentum
conservation law
\begin{equation}\label{2.6}
\nabla^\mu \, \TTmunu = 0 
\end{equation}
where $\TTmunu$ is the {\it total energy-momentum density} of the two-particle system.
Clearly when this system is non-closed in the sense that it is acted upon by an external
field $\exFmunu$ generated by the external four-current $\exjmu$, then the source of
$\TTmunu$(\ref{2.6}) becomes non-zero and actually is determined by the well-known Lorentz
force $\xefnu$, i.e.
\begin{subequations}\label{2.7}
\begin{align}
&\nabla^\mu \, \TTmunu = - \xefnu \label{2.7a} \\
&\xefnu = - \hbar c \, F_{\mu\nu} \cdot \exjmo \label{2.7b} \ .
\end{align}
\end{subequations}
According to Newton's {\it actio = reactio}, this is the reactive force to that force by
which the {\it total field} $\Fmunu$ of the two-particle system pulls at the external
current $\exjmu$ (see the discussion of this point in the preceding
papers~\cite{SSMS}-\cite{BeSo2} ). However, the presently considered positronium is
assumed to be closed with no external field being present (i.e. $\exFmunu=0$; $\exjmu=0$),
so that the energy-momentum conservation (\ref{2.6}) is strictly valid.

\indent The problem is now to define the total energy-momentum density $\TTmunu$ of the
two-particle system in such a way that the required conservation law (\ref{2.6}), and also
the charge conservation (\ref{2.4}), does automatically hold as an implication of the RST
dynamics to be readily set up. The solution to this problem is obtained by first building
up the total density $\TTmunu$ from a matter part $\DTmunu$ and a gauge field part
$\GTmunu$, i.e.
\begin{equation}\label{2.7'}
\TTmunu=\DTmunu+\GTmunu \ ,
\end{equation}
and then one introduces an {\it energy-momentum operator} $\MTmunu$ for the matter part
which serves to define the matter density $\DTmunu$ through
\begin{equation}\label{2.8}
\DTmunu= \bar \Psi \, \MTmunu \Psi \ ,
\end{equation}
quite analogously to the procedure with the total current $j_\mu$ (\ref{2.3}). The problem
has thus been deferred from the \emph{density} $\TTmunu$ to the \emph{operator} $\MTmunu$
and the gauge field density $\GTmunu$ (see below); but it can be demonstrated that the
right choice for the matter operator is the following \cite{BeSo,GrBeMa}:
\begin{equation}\label{2.9}
\MTmunu =\frac{1}{4} \Big\{\DGmu \MHnu + \bar \MHnu \DGmu + \DGnu \MHmu + \bar \MHmu \DGnu \Big \} \ .
\end{equation}
Here a new object is introduced, i.e. the {\it Hamiltonian} $\MHmu$, which will require
its own field equations.

\subsection{Hamiltonian Dynamics}
\indent The RST dynamics must be set up now in such a way that the above mentioned
conservation laws are automatically implied. Evidently, the RST kinematics refers to four
fields: total velocity operator $\DGmu$ (\ref{2.1}), wave function $\Psi$ (\ref{2.2}),
gauge field $\MFmunu$ (see, e.g., (\ref{2.7b})), and Hamiltonian $\MHmu$ (\ref{2.9}).
Thus, anyone of these objects requires the specification of a field equation, and the
totality of these field equations will then constitute the {\it RST dynamics}.

\indent Concerning the Hamiltonian $\MHmu$, its dynamical equations consist of the {\it
  integrability condition}
\begin{equation}\label{2.10}
\MDmu \MHnu - \MDnu \MHmu + \frac{i}{\hbar c} \big [ \MHmu, \MHnu \big] = i\, \hbar c \, \MFmunu
\end{equation}
and of the {\it conservation equation}
\begin{equation}\label{2.11}
  \MDmo \MHmu - \frac{i}{\hbar c} \MHmo \MHmu = - i \, \hbar c \, \Big \{ \Big(\frac{\M
    c}{\hbar} \Big)^2   + \Sigma^{\mu\nu} \MFmunu \Big\} \ ,
\end{equation}
where $\mathcal M$ is the {\it mass operator} and $\Sigma_{\mu\nu}$ are the $Spin(1,3)$
generators constructed by means of the velocity operator $\DGmu$ as
\begin{equation}\label{2.12}
\Sigma_{\mu\nu} \doteqdot \frac{1}{4} \big[ \DGmu, \DGnu \big] \ .
\end{equation}
It can be shown that the Hamiltonian dynamics (\ref{2.10})-(\ref{2.11}) guarantees the
validity of both the {\it bundle identities} such as, e.g., for the wave function $\Psi$
\begin{equation}\label{2.13}
\Big[\MDmu \MDnu - \MDnu \MDmu \Big ] \Psi = \MFmunu \cdot \Psi
\end{equation}
and also the validity of the desired conservation laws (\ref{2.4}) and (\ref{2.6}). But
clearly, for the latter purpose one has to specify also an appropriate field equation for the
matter field $\Psi$.

\subsection{Matter Dynamics}
The field equation for the wave function $\Psi$ is adopted as the
\emph{\textbf{R}elativistic \textbf{S}chr\"odinger \textbf{E}quation} (RSE)
\begin{equation}\label{2.14}
i \, \hbar c \, \MDmu \Psi = \MHmu \Psi \ .
\end{equation}
Differentiating this equation once more and applying also the conservation equation
(\ref{2.11}) recasts the RSE (\ref{2.14}) to the \emph{\textbf{K}lein \textbf{G}ordon
  \textbf{E}quation} (KGE)
\begin{equation}\label{2.15}
\MDmo \MDmu \Psi + \Big(\frac{\M c}{\hbar}\Big)^2 \Psi = - \Sigma^{\mu\nu} \MFmunu \Psi \ .
\end{equation}
However, there is another nice possibility in order to eliminate the Hamiltonian $\MHmu$
from the RSE (\ref{2.14}). Namely, one can show that the differential form (\ref{2.11}) of
the conservation equation is equivalent to the following algebraic formulation:
\begin{equation}\label{2.16}
\M c^2 = \bar \MHmo \DGmu = \DGmo \MHmu \ .
\end{equation}
From here it may be seen that both the mass operator $\M$ and the velocity operator
$\DGmu$ are assumed to be Hermitian (i.e. $\M = \bar \M$, $\DGmu = \bar \DGmu$), but not
the Hamiltonian $\MHmu$ ($\neq \bar \MHmu$). But now multiply through the RSE (\ref{2.14})
from the left by the velocity operator $\DGmu$ and use the algebraic conservation equation
(\ref{2.16}) in order to find the \emph{\textbf{D}irac \textbf{E}quation} (DE) for the
two-particle wave funcion $\Psi$:
\begin{equation}\label{2.17}
i \hbar \, \DGmo \MDmu \Psi = \M c\, \Psi \ .
\end{equation}

\indent However once the DE is established, it is very instructive to verify the desired
charge conservation (\ref{2.4}) just with its help (the verification of the
energy-momentum conservation (\ref{2.6}) is a little bit more complicated, see e.g.
ref.\cite{GrBeMa}). Indeed, carry through the differentiation process in equation
(\ref{2.4}) and find with the help of just the DE (\ref{2.17}) that the charge
conservation actually holds, provided the velocity operator $\DGmu$ is covariantly
constant:
\begin{equation}\label{2.18}
\MDmu \DGnu = 0 \ .
\end{equation}
In a similar way, one verifies the energy-momentum conservation (\ref{2.6}) under the
additional condition of covariant constancy of the mass operator $\M$
\begin{equation}\label{2.19}
\MDmu \M = 0 \ .
\end{equation}
These constancy conditions do attribute the status of {\it absolute objects} to $\DGmu$
and $\M$ which however is a less dramatic insight because the Dirac matrices $\gamma_\mu$
and the particle rest masses $M$ are always taken as absolute objects in most of the
modern theories of elementary particles. Finally observe also that the DE (\ref{2.17}) can
be used for the elimination of the Hamiltonian $\MHmu$ from the matter density $\DTmunu$
(\ref{2.8})-(\ref{2.9}) which then adopts the following shape
\begin{equation}\label{2.19'}
  \DTmunu = \frac{i\, \hbar c}{4} \big \{ \bar \Psi \DGmu \big( \MDnu \Psi \big ) 
- \big( \MDnu \bar \Psi \big )\DGmu \Psi + \bar \Psi \DGnu \big( \MDmu \Psi\big) 
- \big( \MDmu \bar \Psi \big) \DGnu \Psi \big\} \ .
\end{equation}
This form will subsequently be used for the calculation of the positronium ground state
energy.

\indent However, what is a really interesting point with those constancy conditions is the
fact that they reduce the {\it structure group} of the theory for {\it non-identical}
particles, whereas no reduction occurs for {\it identical} particles! In physical terms,
this reduction implies the elimination of the {\it exchange forces} between the
non-identical particles, whereas the identical particles undergo both the conventional
gauge forces (here: electromagnetic interactions) and the exchange forces! This effect
will readily become elucidated by inspecting now the gauge field equations.

\subsection{Electromagnetic and Exchange Forces}
An important question in connection with the positronium system refers to the phenomenon
of exchange forces between the electron and the positron. In the conventional theory
\cite{Me}, the Schr\"odinger-Hamiltonian $\hat H(\vec r_1, \vec r_2)$ is symmetric under
exchange of the one-particle variables $(\vec r_1, \vec p_1) \leftrightarrow (\vec
r_2,\vec p_2)$
\begin{equation}\label{2.19''}
\hat H = \frac{1}{2M} ( \vec p_1^{\, 2} + \vec p_2^{\, 2} ) - \frac{e^2}{\vert \vec r_1 - \vec r_2 \vert} \ ,
\end{equation}
and therefore the energy eigenfuntions $\Psi(\vec r_1,\vec r_2)$ should be either
symmetric or antisymmetric under particle exchange ($\vec r_1 \leftrightarrow \vec r_2$).
Properly speaking, such a symmetry of the energy eigenfunctions signals the presence of
exchange forces (in addition to the usual electromagnetic interactions) with the
corresponding shift of the energy levels due to the {\it exchange energy}.

\indent On the other hand, the conventional non-relativistic treatment of positronium
makes use of the transformation from the single-particle coordinates $\vec r_1, \vec r_2$
to the center of mass coordinate $\vec R$ and relative coordinate $\vec r$, and then there
arises a one-particle eigenvalue problem for the bound states being written in terms of
the relative coordinate $\vec r$ alone. The corresponding positronium spectrum emerges
thus as the usual hydrogen level system (Coulomb force problem) with merely the electron
mass $M$ being substituted by the reduced mass $M/2$. It should be immediately evident
that for such an effective {\it one-particle} problem there cannot emerge exchange effects
because the latter do always refer to {\it many-particle} systems! For the subsequent RST
treatment of the positronium spectrum, the exchange forces between the electron and
positron are also suppressed but on account of quite different reasons. Therefore it is
very instructive to first inspect this RST mechanism of suppression of the exchange
interactions.

\indent Observe also that the conventional Hamiltonian $\hat H$ (\ref{2.19''}) works with
the exact Coulomb force between both particles, with no shielding effects included. In
contrast to this, the subsequent RST treatment will lead to some weakening (but not complete
regularization) of the Coulomb singularity.

\indent The coupling of the RST matter field $\Psi$ to the gauge field $\MAmu$ ({\it
  bundle connection}), with associated {\it curvature} $\MFmunu$
\begin{equation}\label{2.20}
\MFmunu =\nabla_\mu \MAnu - \nabla_\nu \MAmu + [ \MAmu, \MAnu ] \ ,
\end{equation}
occurs as usual via the principle of {\it minimal coupling}; i.e. the gauge-covariant
derivative of $\Psi$ in the RSE (\ref{2.14}) or DE (\ref{2.17}) is written as
\begin{equation}\label{2.21}
\MDmu \Psi = \partial_\mu \Psi + \MAmu \Psi \ .
\end{equation}
Since the two-particle wave function $\Psi$ is here the (Whitney) sum of the two
single-particle wave functions $\psi_a$ ($a=1,2$), the abstract derivative (\ref{2.21})
may be recast into component form and then looks as follows \cite{BeSo}:
\begin{subequations}\label{2.22}
\begin{align}
D_\mu \psi_1 &= \partial_\mu \psi_1 - i \, \Azmu \psi_1 - i \, B_\mu \psi_2 \label{2.22a} \\
D_\mu \psi_2 &= \partial_\mu \psi_2 - i \, \Aemu \psi_2 - i \, B^*_\mu \psi_1 \label{2.22b} \ .
\end{align}
\end{subequations}
Here the anti-Hermitian bundle connection $\MAmu$ ($=-\bar\MAmu$) takes its values in the
four-dimensional Lie algebra $\mathfrak u(2)$ of the two-particle {\it structure group}
$U(2)$ and may therefore be decomposed with respect to some appropriate basis $\{\tau_1,
\tau_2; \chi, \bar \chi \}$ of $\mathfrak u(2)$ as follows
\begin{equation}\label{2.23}
\MAmu = \sum\limits^2_{a=1} \Aamu \tau_a + B_\mu \chi - B_\mu^* \bar \chi \ .
\end{equation}

\indent Clearly a similar decomposition does apply also to the curvature $\MFmunu$
(\ref{2.20}) of $\MAmu$, i.e.
\begin{equation}\label{2.24}
\MFmunu = \sum\limits^2_{a=1} \Famunu \tau_a + \Gmunu \chi - \Gsmunu \bar \chi \ ,
\end{equation}
where the curvature components ({\it field strengths}) are given in terms of the
connection components ({\it gauge potentials}) through
\begin{subequations}\label{2.25}
\begin{align}
\Femunu &= \nmu\Aenu - \nnu\Aemu + i\left[\Bmu\Bsnu - \Bnu\Bsmu\right] \label{2.25a}\\
\Fzmunu &= \nmu\Aznu - \nnu\Azmu - i\left[\Bmu\Bsnu - \Bnu\Bsmu\right] \label{2.25b}\\
\Gmunu  &= \nmu\Bnu - \nnu\Bmu + i\left[\Aemu - \Azmu\right]\Bnu - i\left[\Aenu - \Aznu\right]\Bmu \label{2.25c} \\
\Gsmunu  &= \nmu\Bsnu - \nnu\Bsmu - i\left[\Aemu - \Azmu\right]\Bsnu + i\left[\Aenu - \Aznu\right]\Bsmu \label{2.25d} \ .
\end{align}
\end{subequations}
The physical meaning of such a decomposition refers to the subdivision of the totality of
gauge interactions into the {\it electromagnetic forces} (mediated by the {\it
  electromagnetic potentials} $\Aamu$ ) and into the {\it exchange forces} (mediated by
the {\it exchange potential} $B_\mu$). Observe here that both wave functions $\psi_1$ and
$\psi_2$ become directly coupled to each other by the exchange potential $B_\mu$ via the
minimal coupling (\ref{2.22a})-(\ref{2.22b}).

\indent The interesting point with the exchange forces refers now to the fact, that they
can act exclusively among {\it identical} particles (sharing identical masses and
charges). Or conversely, for {\it non-identical} particles (such as the constituents of
our positronium system) the exchange forces must vanish (i.e. $B_\mu\equiv 0$) so that the
non-identical particles can feel only the usual electromagnetic forces. The reason for
this is due to the former requirement of covariant constancy of the total velocity
operator $\DGmu$ (\ref{2.18}) and the mass operator $\M$ (\ref{2.19}). Indeed since these
objects are of absolute (i.e. non-dynamical) character, they can be thought to be constant
over space-time and consequently those gauge covariant constancy conditions imply the
commutativity of those objects with the structure algebra being spanned by the above
mentioned generators $\{ \tau_\alpha \} = \{ \tau_a; \chi, \bar \chi \}$:
\begin{subequations}\label{2.26}
\begin{align}
[\M, \tau_\alpha ] &= 0 \label{2.26a} \\
[\DGmu, \tau_\alpha ] &= 0 \label{2.26b} \\
(\alpha =1,&\dots, 4) \ . \nonumber
\end{align}
\end{subequations}
This property of commutativity is safely ensured when both particles have identical masses
and charges so that the absolute objects become proportional to unity $\mathbf 1_{(8)}$ of
the eight-dimensional representation of the Clifford algebra $\mathbb C(1,3)$
\begin{equation}\label{2.27}
\Gmu\Gnu+\Gnu\Gmu = \DGmu\DGnu+\DGnu\DGmu =2g_{\mu\nu}\cdot \mathbf 1_{(8)} \ ,
\end{equation}
i.e. when the mass and velocity operators adopt the following shape:
\begin{subequations}\label{2.28}
\begin{align}
\M &\Rightarrow M \cdot \mathbf 1_{(8)} \label{2.28a} \\
\DGmu &\Rightarrow \Gmu = \gamma_\mu \cdot \mathbf 1_{(8)} \ . \label{2.28b}
\end{align}
\end{subequations}
Thus for identical particles, the commutation conditions (\ref{2.26a})-(\ref{2.26b}) are
trivially satisfied and the gauge potential $\MAmu$ (\ref{2.23}) can sweep out the whole
structure algebra $\mathfrak u(2)$, with non-trivial exchange fields $B_\mu$.

\indent However when the particles carry opposite charges, the velocity operator $\DGmu$
(\ref{2.1}) is to be used in place of $\Gmu$ ($=\gamma_\mu \oplus \gamma_\mu$); and
furthermore, when both particles have different masses $M_a$ ($a=1,2$), the mass operator
is no longer proportional to unity but adopts the more general shape
\begin{equation}\label{2.29}
\M = i \sum\limits_{a=1}^{2} M_a \tau_a \ .
\end{equation}
Naturally for this general situation, the absolute objects $\M,\DGmu$ do not commute with
all four generators $\tau_\alpha$ of the original structure algebra $\mathfrak u(2)$ which
therefore must be reduced to a certain subalgebra $\mathfrak u'(2) \in \mathfrak u(2)$ so
that the commutativity requirement does then hold with respect to that subalgebra
$\mathfrak u'$. For the present positronium case, this reduction process cuts the original
structure group $U(2)$ down to its maximal Abelian subgroup $U'(2)=U(1)\times U(1)$ which
is due to the two {\it electromagnetic generators} $\tau_a$ ($a=1,2$). The latter are
obeying the following commutation relations:
\begin{align}\label{2.30}
[\tau_1, \tau_2 ] &= 0 & [\chi, \bar\chi]&=-i\,(\tau_1-\tau_2) \nonumber \\
[\tau_1, \chi ] &= i \, \chi & [\tau_2, \chi]&=-i \, \chi\\
[\tau_1, \bar \chi ] &= -i\, \bar \chi & [\tau_2, \bar\chi]&=i \, \bar \chi \nonumber \ .
\end{align}
According to this reduction process, the bundle connection $\MAmu$ (\ref{2.23}) and its
curvature $\MFmunu$ (\ref{2.24}) become projected to the abelian subalgebra $\mathfrak
u(1) \oplus \mathfrak u(1)$, i.e. the exchange potential $B_\mu$ is to be omitted:
\begin{subequations}\label{2.31}
\begin{align}
\MAmu &\Rightarrow \sum\limits^2_{a=1} \Aamu \tau_a = \Aemu \tau_1 + \Azmu \tau_2 \label{2.31a} \\
\MFmunu &\Rightarrow \sum\limits^2_{a=1} \Famunu \tau_a = \Femunu \tau_1 + \Fzmunu \tau_2 \label{2.31b} \\
\Famunu &\Rightarrow \nabla_\mu \Aamu - \nabla_\nu \Aamu \label{2.31c} \ .
\end{align}
\end{subequations}
This situation of missing exchange interaction does apply to the subsequent treatment of
the positronium system consisting of a positron and an electron which, it is true, have
the same mass $M$ (\ref{2.28a}) but are oppositely charged so that the velocity operator
$\DGmu$ (\ref{2.1}) is to be applied in place of its identical-particle form $\Gmu$!

\indent A similar argument for the reduction of the structure group comes from the
inspection of the RST currents $j_{\alpha\mu}$ ($\alpha=1\dots 4$) which are quite
generally defined in terms of the {\it gauge velocity operators} $\upsilon_{\alpha\mu}$
through
\begin{align}\label{2.32}
\jalumu &\doteqdot \bar \Psi \, \upsilon_{\alpha\mu} \Psi \\
\big(\upsilon_{\alpha\mu} &\doteqdot \frac{i}{2} \{ \tau_\alpha, \DGmu \} \big) \nonumber \ .
\end{align}
The point with these currents is here that (by explicit differentiation with reference to
the RST dynamics) their source equations are found to be of the following form \cite{BeSo}
\begin{equation}\label{2.33}
D^\mu \jalumu = - \frac{i}{\hbar c} \bar \Psi \, [\M c^2, \tau_\alpha ] \Psi \ .
\end{equation}
But since these RST currents act as the sources of the gauge fields $\Aaomu$ (see the
non-Abelian Maxwell equations below), the currents $\jalumu$ must themselves have
vanishing source from consistency reasons:
\begin{equation}\label{2.34}
D^\mu \jalumu \equiv 0 \ .
\end{equation}
However, this requirement can be satisfied again by reducing the structure group in such a
way that the commutator (\ref{2.26a}) vanishes and the desired gauge continuity equation
(\ref{2.34}) becomes then implied by the general relation (\ref{2.33}).

\indent Thus, the suppression of exchange forces among non-identical particles is seen to
be supported by the intrinsic logic of RST, whereas in the conventional theory the
emergence/suppression of exchange effects for (non-)identical particles is a purely
kinematical effect being induced by postulating the (anti)symmetrization of wave functions
or (anti)commutation of field operators, resp.

\subsection{Gauge Field Equations}
\indent The RST dynamics presented up to now is not yet a closed system because one
finally has to specify a field equation for the gauge potential $\MAmu$. The choice of
such a field equation must necessarily undergo certain constraints because it must be
compatible with the already specified RST dynamics, especially with the conservation laws.
Our choice is the non-Abelian Maxwell equation for the curvature $\MFmunu$~(\ref{2.20})
\begin{equation}\label{2.35}
\MDmo \MFmunu = -4\pi \, i \, \alpha_{\rm S} \, \mathcal J_\nu \ .
\end{equation}
Here $\alpha_{\rm S}$ is the fine structure constant ($e^2/\hbar c$); and the current
operator $\mathcal J_\mu$ may be decomposed with respect to the chosen Lie algebra basis
$\{\tau_\alpha\}$ as
\begin{equation}\label{2.36}
\mathcal J_\mu = i \, \jalomu \tau_\alpha
\end{equation}
so that the abstract equation (\ref{2.35}) reads in component form
\begin{equation}\label{2.37}
D^\mu \Falomunu = \vpas \, \jalonu \ .
\end{equation}
As a brief consistency check of the chosen gauge field dynamics, convince yourself of the
fact that those continuity equations (\ref{2.34}) are an implication of the Maxwellian
equations (\ref{2.35}), or (\ref{2.37}), resp. To this end, simply apply the bundle
identity for the curvature $\MFmunu$
\begin{equation}\label{2.37'}
\MDmo \MDno \MFmunu \equiv 0
\end{equation}
to the Maxwell equations (\ref{2.35}) and thus find the desired source equations
\begin{equation}\label{2.38}
\MDmo \MJmu \equiv 0 \Leftrightarrow D^\mu \jalomu \equiv 0 \ .
\end{equation}

\indent It is true, these source equations (\ref{2.38}) for the {\it Maxwell currents}
$\jalomu$ are not strictly identical to the corresponding source equations (\ref{2.34})
for the RST currents $\jalumu$ (\ref{2.32}). But it is possible to introduce a fibre
metric $\Kaubu$ in the Lie algebra bundle, which is invariant under the structure group
and is also covariantly constant \cite{SSMS}
\begin{equation}\label{2.39}
D_\mu \Kaubu \equiv 0 \ .
\end{equation}
This, namely, provides one with the possibility to identify the Maxwell and RST currents
as contra- and covariant versions of each other:
\begin{subequations}\label{2.40}
\begin{align}
\jalomu &= \Kaobo \jbumu \label{2.40a} \\
\jalumu &= \Kaubu \jbomu \label{2.40b} \ .
\end{align}
\end{subequations}
Clearly through this identification, both forms of continuity equations (\ref{2.34}) and
(\ref{2.38}) become now equivalent and this does support the desired compatibility of the
matter and gauge field dynamics.

\indent For the sake of generality, one may first wish to include the exchange interactions (i.e.
putting $B_\mu \neq 0$) so that the Maxwell equations (\ref{2.37}) read explicitly
\begin{subequations}\label{2.41}
\begin{align}
  &\nmo\Femunu + i\left[\Bmo\Gsmunu - \Bsmo\Gmunu\right] = \vpas \, \jeonu \label{2.41a} \\
  &\nmo\Fzmunu - i\left[\Bmo\Gsmunu - \Bsmo\Gmunu\right] = \vpas \, \jzonu \label{2.41b} \\
  &\nmo\Gmunu + i\left[\Aemu - \Azmu\right]\Gmonu - i \left[\Femunu - \Fzmunu \right] \Bmo = \vpas \, g_\nu \label{2.41c} \\
  &\nmo\Gsmunu - i\left[\Aemu - \Azmu\right]\Gsmonu + i \left[\Femunu - \Fzmunu \right]
  \Bsmo = \vpas \, g^*_\nu \label{2.41d} \ ,
\end{align}
\end{subequations}
where the {\it exchange currents} $\jdomu, \jvomu$, (\ref{2.36}) are denoted by
\begin{subequations}\label{2.42}
\begin{align}
\jdomu &\doteqdot g_\mu \label{2.42a} \\
\jvomu &\doteqdot -g^*_\mu \label{2.42b} \ .
\end{align}
\end{subequations}
However, since the positronium system consists of two non-identical particles which are
not able to feel the exchange force, one has to drop the exchange potential $\Bmu$ and
thus the gauge field system (\ref{2.41a})-(\ref{2.41d}) simplifies to
\begin{subequations}\label{2.43}
\begin{align}
&\nmo \Femunu = \vpas \, \jeonu \label{2.43a} \\
&\nmo \Fzmunu = \vpas \, \jzonu \label{2.43b} \ .
\end{align}
\end{subequations}
From this reduction process it is clearly seen that the omission of the exchange force
implies the simplification to an Abelian and linear theory referring to the
electromagnetic interactions alone.

\indent Finally, in order to close the whole dynamical system, one has to specify the link
of the Maxwell currents $\jaomu$ ($a=1,2$) to the one-particle wave functions $\psi_a$.
This may be done by first defining the {\it Dirac currents} $\kaumu$ in the usual way
through ($a=1,2$)
\begin{equation}\label{2.44}
\kaumu = \bar \psi_a \gamma_\mu \psi_a \ ,
\end{equation}
and then the RST currents $\jaumu$ (\ref{2.32}) are found to be related to these Dirac currents through
\begin{subequations}\label{2.45}
\begin{align}
j_{1\mu} &= k_{2\mu} \label{2.45a} \\
j_{2\mu} &= -k_{1\mu} \label{2.45b} \ .
\end{align}
\end{subequations}

\indent Thus, the last step for constructing the Maxwell currents $\jaomu$ (\ref{2.40a})
must consist in specifying the fibre {\it sub}metric $K_{ab}$ or $K^{ab}$, resp. In
general, this object depends upon one parameter $u$ (i.e. the {\it self-interaction}
parameter): $K_{ab} = K_{ab}(u)$. However for the present purposes, we want to neglect the
self-energy effects which implies $u=0$; and for this special value of $u$ the
two-dimensional fibre submetric becomes \cite{SSMS}
\begin{equation}\label{2.46}
\big\{K^{ab}(0)\big\} = \big \{ K_{ab}(0) \big \} = \left (\begin{array}{cc} 0 & -1 \\ -1 & 0 \end{array}\right) 
\end{equation}
so that the desired two Maxwell currents $\jaomu$ read in terms of the wave functions $\psi_a$:
\begin{subequations}\label{2.47}
\begin{align}
\jeomu &= -\jzumu =  k_{1\mu} \label{2.47a} \equiv \bar \psi_1 \gamma_\mu \psi_1 \\
\jzomu &= -\jeumu = -k_{2\mu} \label{2.47b} \equiv - \bar \psi_2 \gamma_\mu \psi_2 \ .
\end{align}
\end{subequations}
Thus the first Maxwell current $\jeomu$ correctly refers to a positively charged particle (positron) and the second Maxwell current $\jzomu$ refers to the negatively charged electron.

\indent In this way, the RST dynamics for a two-particle system becomes closed in a
consistent way: the wave function of the first/second particle generates the first/second
Dirac current $\keumu / \kzumu$ according to (\ref{2.44}); then these Dirac currents
$\kaumu$ give rise to the Maxwell currents $\jaomu$ according to
(\ref{2.47a})-(\ref{2.47b}). Next, the Maxwell currents $\jaomu$ act as the sources for
the generation of the gauge potentials $\Aamu$ as the solutions of the (Abelian) Maxwell
equations (\ref{2.43a})-(\ref{2.43b}); and finally the first/second gauge potential $\Aemu
/ \Azmu$ enters the Dirac equation (\ref{2.17}) for the second/first particle via the
gauge-covariant derivatives (\ref{2.22a})-(\ref{2.22b}). Clearly such a simple cross
relation is to be expected from logical reasons, but it must be somewhat modified when the
self-interactions and exchange effects are included \cite{BeSo}-\cite{BeSo2}.

\section{Mass Eigenvalue Equations}
\indent In a stationary bound state, both positronium constituents (i.e. positron and
electron) will be described by wave functions $\psi_a(\vec r, t)$ satisfying the usual
space-time factorization
\begin{subequations}\label{3.1}
\begin{align}
\psi_1(\vec r,t)& = \exp\Big[i\,\frac{M_*c^2}{\hbar} t \Big] \cdot \psi_1(\vec r) \label{3.1a} \\
\psi_2(\vec r,t)& = \exp\Big[-i\,\frac{M_*c^2}{\hbar} t \Big] \cdot \psi_2(\vec r) \label{3.1b} \ .
\end{align}
\end{subequations}
Observe here that the time factors of both constituents are inverse to each other which is
due to the fact that the first particle ($a=1$, positron) obeys the Dirac equation
(\ref{2.17}) with a negative mass term
\begin{equation}\label{3.2}
i\hbar \, \gamma^\mu D_\mu \psi_1 = - M c \, \psi_1 \ ,
\end{equation}
whereas the electron obeys the conventional Dirac equation
\begin{equation}\label{3.3}
i\hbar \, \gamma^\mu D_\mu \psi_2 = Mc \, \psi_2 \ .
\end{equation}
Clearly, this difference is due to the former choice of the total velocity operator
$\DGmu$ (\ref{2.1}) for oppositely charged particles.

\subsection{Positronium Configurations}
\indent The {\it mass eigenvalue} $M_*$ in (\ref{3.1a})-(\ref{3.1b}) is common to both
particles and must be determined from the coupled Dirac equations (\ref{3.2})-(\ref{3.3})
by use of the stationary ansatz (\ref{3.1a})-(\ref{3.1b}). Here it is very advantageous to
split up the Dirac four-spinors $\psi_a(\vec{r})$ into the direct sum of two-component
Pauli spinors ${}^{(a)}\!\varphi_\pm(\vec r)$, i.e. one puts
\begin{equation}\label{3.4}
\psi_a(\vec r) = {}^{(a)}\!\varphi_+(\vec r) \oplus {}^{(a)}\!\varphi_-(\vec r) \ ,
\end{equation}
and then the {\it mass eigenvalue equation} for the first particle (positron) reads as follows \cite{BeSo}-\cite{BeSo2}
\begin{equation}\label{3.5}
i \, (\vec \sigma \sdot \vec \nabla ) {}^{(1)}\!\varphi_\pm(\vec r) + \zAo \cdot {}^{(1)}\!\varphi_\mp(\vec r)
-\vec A_2 \sdot \vec\sigma {}^{(1)}\!\varphi_\pm(\vec r)=\frac{\pm M + M_*}{\hbar}c\cdot {}^{(1)}\!\varphi_\mp(\vec r) \ ,
\end{equation}
and similarly for the second particle (electron)
\begin{equation}\label{3.6}
  i \, (\vec \sigma \sdot \vec \nabla ) {}^{(2)}\!\varphi_\pm(\vec r) + \eAo \cdot
  {}^{(2)}\!\varphi_\mp(\vec r) - \vec A_1 \sdot \vec \sigma {}^{(2)}\!\varphi_\pm(\vec r)
  = -\frac{M_* \pm M}{\hbar}c\cdot {}^{(2)}\!\varphi_\mp(\vec r) \ .
\end{equation}

\indent Now the crucial point with the positronium system is that its energy spectrum is
experimentally found to be essentially a one-particle spectrum~\cite{GrRe}, the energy
levels of which are to be subdivided in RST into two subsets: {\it ortho-positronium} and
{\it para-positronium}. These two classes may be characterized by the relative orientation
of the magnetic fields $H_a(\vec r)$ ($a=1,2$) being emitted by both particles:
\begin{subequations}\label{3.7}
\begin{align}
  &ortho\textnormal-positronium: \vec H_1(\vec r) \equiv \vec H_2(\vec r) \doteqdot \vec
  H_b(\vec r),
  \quad \vec A_1(\vec r) \equiv \vec A_2(\vec r) \doteqdot \vec A_b(\vec r) \label{3.7a} \\
  &para\textnormal-positronium: \ \vec H_1(\vec r) \equiv -\vec H_2(\vec r) \doteqdot \vec
  H_p(\vec r), \quad \vec A_1(\vec r) \equiv -\vec A_2(\vec r) \doteqdot \vec A_p(\vec r)
  \label{3.7b} \ .
\end{align}
\end{subequations}
In contrast to this, the conventional classification refers to the spin orientations and
thus works rather in terms of the singlet (${}^1\rm S_0$) and triplet (${}^3\rm S_1$)
states~\cite{GrRe}.

\indent Since, however, the magnetic fields are after all generated by the Dirac
three-currents $ \vec k_a(\vec r) = \{ {}^{(a)}\!k^j(\vec r) \}$, cf. (\ref{2.44})
\begin{equation}\label{3.7'}
\vec k_a(\vec r) = \bar \psi_a(\vec{r}) \vec \gamma \, \psi_a(\vec{r}) \ ,
\end{equation}
and since both particles carry opposite charges one will expect the following arrangement of the three-currents:
\begin{subequations}\label{3.8}
\begin{align}
&ortho\textnormal-positronium: \quad \ \vec k_1(\vec r) = -\vec k_2(\vec r) \doteqdot \vec k_p(\vec r) \label{3.8a} \\
&para\textnormal-positronium: \quad \ \ \vec k_1(\vec r) = \vec k_2(\vec r) \doteqdot \vec k_b(\vec r) \ . \label{3.8b}
\end{align}
\end{subequations}
Actually in view of such an arrangement, the corresponding {\it Maxwell three-currents}
$\vec j_a(\vec r)$ (\ref{2.47a})-(\ref{2.47b}) become
\begin{subequations}\label{3.9}
\begin{align}
  &ortho\textnormal-positronium: \quad \vec j_1(\vec r) = \vec j_2(\vec r) \doteqdot \vec
  j_b(\vec r) \equiv
  \vec k_p(\vec r) \label{3.9a}\\
  &para\textnormal-positronium: \quad \ \vec j_1(\vec r) = - \vec j_2(\vec r) \doteqdot
  \vec j_p(\vec r) \equiv \vec k_b(\vec r) \ . \label{3.9b}
\end{align}
\end{subequations}
Such an arrangement of the Maxwell currents is immediately plausible, because this kind of
currents is responsible for the generation of the magnetic fields $\vec H_a(\vec r)$
according to the (Abelian) Maxwell equations (\ref{2.43a})-(\ref{2.43b}), the space part
of which reads in three-vector notation
\begin{equation}\label{3.10}
\vec \nabla \times \vec H_a = \vpas \, \vec j_a(\vec r) \ .
\end{equation}
Thus the (anti)parallelity of the Maxwell currents $\vec j_a(\vec r)$
(\ref{3.9a})-(\ref{3.9b}) actually implies the (anti)parallelity of the magnetic fields
$\vec H_a(\vec r)$ (\ref{3.7a})-(\ref{3.7b})

\indent On the other hand, the situation with the electric fields $\vec E_a(\vec r)$ is
less complicated: since both particles are oppositely charged, the Dirac densities $\ako
(\vec r)$ ($a=1,2$) must be identical
\begin{equation}\label{3.11}
  ortho \textnormal{ \& } para\textnormal - positronium: \quad {}^{(1)}\!k_0(\vec r)
  \equiv {}^{(2)}\!k_0(\vec r) 
\doteqdot {}^{(b)}\!k_0(\vec r) \ ,
\end{equation}
because in this case the Maxwell densities ${}^{(a)}\!j_0(\vec r)$
(\ref{2.47a})-(\ref{2.47b}) as the true charge densities will differ in sign as required
for oppositely charged particles:
\begin{equation}\label{3.12}
  ortho \textnormal{ \& } para\textnormal - positronium: \quad {}^{(1)}\!j_0(\vec r)
  \equiv - {}^{(2)}\!j_0(\vec r) 
\doteqdot {}^{(p)}\!j_0(\vec r) = {}^{(b)}\!k_0(\vec r) \ .
\end{equation}
As a consequence, the electric fields $\vec E_a(\vec r)$ will also differ in sign
\begin{equation}\label{3.13}
  ortho \textnormal{ \& } para\textnormal - positronium: \quad \vec E_1(\vec r) \equiv
  -\vec E_2(\vec r) 
\doteqdot \vec E_p(\vec r) \ .
\end{equation}
Furthermore, the missing of exchange interactions lets the electric fields $\vec E_a(\vec
r)$ appear as pure gradient fields, cf. (\ref{2.31c})
\begin{equation}\label{3.14}
\vec E_a(\vec r) = - \vec \nabla {}^{(a)}\!A_0(\vec r) \ ,
\end{equation}
with the {\it electrostatic potentials} ${}^{(a)}\!A_0(\vec r)$ to be identified in the following way
\begin{equation}\label{3.15}
ortho \textnormal{ \& } para\textnormal - positronium: \quad \eAo(\vec r) \equiv -
\zAo(\vec r) 
\doteqdot {}^{(p)}\!A_0(\vec r) \ .
\end{equation}
Indeed, this electrostatic arrangement matches again the (time-component of) Maxwell's
equations (\ref{2.43a})-(\ref{2.43b})
\begin{equation}\label{3.16}
\vec \nabla \cdot \vec E_a = -\vpas \, {}^{(a)}\!j_0(\vec r) \ ,
\end{equation}
i.e. under use of the electrostatic identifications (\ref{3.12})-(\ref{3.15}):
\begin{equation}\label{3.17}
\Delta {}^{(p)}\!A_0(\vec r) = -\vpas \, {}^{(p)}\!j_0(\vec r) \ .
\end{equation}

\subsection{One-Particle Equations}
\indent This specific way of identifications of densities for both positronium
configurations must now necessarily imply a corresponding identification of the Pauli
spinors ${}^{(a)}\!\varphi_\pm (\vec r)$ (\ref{3.4}) since the charge and current
densities are ultimately built up by the wave functions $\psi_a(\vec r)$. The result of
such a spinor identification is the contraction of the four mass eigenvalue equations
(\ref{3.5})-(\ref{3.6}) to only two equations which then formally describe only one
particle. This is the RST analogue to the conventional energy eigenvalue equation being
written in terms of the relative coordinate $\vec r$ ($=\vec r_1 - \vec r_2$) after the
center-of-mass motion $\vec R$ ($=\vec r_1 + \vec r_2$) has been separated-off \cite{Me}.

\indent Turning first to the simpler case of the Dirac densities ${}^{(a)}\!k_0(\vec r)$,
one finds these objects emerging in terms of the Pauli spinors ${}^{(a)}\!\varphi_\pm$
(\ref{3.4}) as
\begin{equation}\label{3.18}
{}^{(a)}\!k_0(\vec r) \doteqdot \bar\psi_a \gamma_0 \, \psi_a =
{}^{(a)}\!\varphi_+^\dagger(\vec r)
 \sdot {}^{(a)}\!\varphi_+(\vec r) + {}^{(a)}\!\varphi_-^\dagger(\vec r) \sdot {}^{(a)}\!\varphi_-(\vec r) \ .
\end{equation}
The somewhat more complicated case of the Dirac three-currents $\vec k_a(\vec r)$ (\ref{3.7'}) looks as follows
\begin{equation}\label{3.19}
\vec k_a(\vec r) = \bar \psi_a(\vec r) \, \vec \gamma \,\psi_a(\vec r) =
{}^{(a)}\!\varphi_+^\dagger(\vec r) \,\vec \sigma \,{}^{(a)}\!\varphi_-(\vec r) +
{}^{(a)}\!\varphi_-^\dagger(\vec r)
 \,\vec \sigma \,{}^{(a)}\!\varphi_+(\vec r) \ ,
\end{equation}
where $\vec \gamma$ denotes the space part of the Dirac matrices $\gamma^\mu$ (standard
form) and $\vec \sigma = (\sigma_x, \sigma_y , \sigma_z)$ refers to the well-known Pauli
matrices.

\indent Now, in order to let the set of four two-particle equations
(\ref{3.5})-(\ref{3.6}) collapse to the desired set of two one-particle equations under
observation of the required density and current identifications, one expresses the second
Pauli spinors ${}^{(2)}\!\varphi_\pm(\vec r)$ in terms of the first Pauli spinors
${}^{(1)}\!\varphi_\pm(\vec r)$ ($\doteqdot \varphi_\pm(\vec r)$) in the following way
\begin{subequations}\label{3.20}
\begin{align}
  ortho \textnormal - positronium: \quad & {}^{(2)}\!\varphi_+(\vec r) = \hat\Omega \,
  {}^{(1)}\!\varphi_+(\vec r)
 \doteqdot \hat\Omega \, \varphi_+(\vec r) \label{3.20a}\\
  \quad & {}^{(2)}\!\varphi_-(\vec r) = -\hat\Omega \, {}^{(1)}\!\varphi_-(\vec r)
  \doteqdot -\hat\Omega \,
 \varphi_-(\vec r) \label{3.20b}\\
  para\textnormal - positronium: \quad & {}^{(2)}\!\varphi_\pm(\vec r) = \hat\Omega \,
  {}^{(1)}\!\varphi_\pm(\vec r) \doteqdot \hat\Omega \, \varphi_\pm(\vec r) \label{3.20c}
  \ .
\end{align}
\end{subequations}
Here, the operator $\hat\Omega$ acts in the two-dimensional Pauli-spinor space and is to
be determined from the identifications of densities and currents. Its first property is
found from the identification (\ref{3.11}) of the Dirac densities ${}^{(a)}\!k_0(\vec r)$
(\ref{3.18}), which holds for the ortho- and para-configurations. Evidently, this density
identification reads explicitly
\begin{equation}\label{3.21}
  \varphi_+^\dagger(\vec r) \Big(\hat\Omega^\dagger \hat\Omega \Big ) \varphi_+ (\vec r) + 
\varphi_-^\dagger(\vec r) \Big(\hat\Omega^\dagger \hat\Omega \Big ) \varphi_- (\vec r) = 
\varphi_+^\dagger(\vec r) \varphi_+ (\vec r) + \varphi_-^\dagger(\vec r) \varphi_- (\vec r) 
\end{equation}
and thus is surely obeyed by adopting the operator $\hat\Omega$ to be unitary ($\hat\Omega^\dagger = \hat\Omega^{-1}$)
\begin{equation}\label{3.22}
\hat\Omega^\dagger \, \hat\Omega = 1 \ .
\end{equation}

\indent The next property of the operator $\hat\Omega$ may be deduced from the current
identifications (\ref{3.8a})-(\ref{3.8b}) which by use of equation (\ref{3.19}) explicitly
read for {\it both} configurations
\begin{multline}\label{3.23}
  \varphi_+^\dagger (\vec r) \Big ( \hat\Omega^\dagger \vec \sigma \, \hat\Omega \Big)
  \varphi_-(\vec r) + \varphi_-^\dagger (\vec r) \Big ( \hat\Omega^\dagger \vec \sigma \,
  \hat\Omega \Big) \varphi_+(\vec r) = \\ = \varphi_+^\dagger (\vec r) \, \vec \sigma \,
  \varphi_-(\vec r) + \varphi_-^\dagger (\vec r) \, \vec \sigma \, \varphi_+(\vec r) \ .
\end{multline}
But this requirement can easily be satisfied by adopting the following additional property
of $\hat\Omega$:
\begin{align}\label{3.24}
  \hat\Omega^\dagger \big( \hat{ \vec k }\sdot \vec \sigma \big ) \hat\Omega &= \hat{ \vec k } \sdot \vec \sigma \\
  (ortho \textnormal{ \& } para\textnormal - &positronium) \nonumber
\end{align}
where the unit three-vector $\hat{\vec k}$ ($\vert\!\vert \hat{\vec k} \vert\!\vert =1$)
points in the direction of the corresponding three-current, i.e. one has (cf.
(\ref{3.8a})-(\ref{3.8b}))
\begin{subequations}\label{3.25}
\begin{align}
ortho \textnormal - positronium: \quad & \vec k_p (\vec r) = k_p (\vec r) \, \hat{\vec k}(\vec r) \label{3.25a} \\
para \textnormal - positronium: \quad & \vec k_b (\vec r) = k_b (\vec r) \, \hat{\vec k}(\vec r) \ . \label{3.25b}
\end{align}
\end{subequations}
But clearly, it is no problem to find unitary solutions (\ref{3.22}) for the equation (\ref{3.24}), namely the following two
\begin{subequations}\label{3.26}
\begin{align}
ortho \textnormal - positronium: \quad & \hat\Omega = i \cdot \mathbf 1 \ , \label{3.26a} \\ 
para \textnormal - positronium: \quad & \hat\Omega = i \, \hat{\vec k} \sdot \vec \sigma \ . \label{3.26b}
\end{align}
\end{subequations}
We shall readily elaborate the reason for this peculiar arrangement of the ortho- and para-cases, resp.

\indent But now that the operator $\hat\Omega$ is known, one can turn to the main problem,
namely the reduction of the two-particle equations (\ref{3.5})-(\ref{3.6}) to one-particle
equations by means of the preceding identifications. Considering first the ortho-case, one
finds the mass eigenvalue equations (\ref{3.6}) of the second particle (electron) adopting
the following form through application of the relations (\ref{3.20a})-(\ref{3.20b}):
\begin{subequations}\label{3.28}
\begin{align}
  i\big(\vec \sigma \sdot \vec \nabla \big ) \hat\Omega \,\varphi_+(\vec r) -
  {}^{(p)}\!A_0 \cdot
 \hat\Omega \,\varphi_-(\vec r) - \vec A_b \sdot \vec \sigma \, \hat\Omega \,
 \varphi_+(\vec r) &= \frac{M_* +M}{\hbar}c\cdot \hat\Omega \, \varphi_-(\vec r) \label{3.28a}\\
  -i\big(\vec \sigma \sdot \vec \nabla \big ) \hat\Omega \,\varphi_-(\vec r) +
  {}^{(p)}\!A_0 \cdot \hat\Omega \,\varphi_+(\vec r) + \vec A_b \sdot \vec \sigma \,
  \hat\Omega \, \varphi_-(\vec r) &= -\frac{M_* -M}{\hbar}c\cdot \hat\Omega \,
  \varphi_+(\vec r) \label{3.28b} \ ,
\end{align}
\end{subequations}
whereas the ortho-identifications recast the eigenvalue equations (\ref{3.5}) of the first
particle (positron) to the following form:
\begin{subequations}\label{3.29}
\begin{align}
  i\big(\vec \sigma \sdot \vec \nabla \big ) \varphi_+(\vec r) - {}^{(p)}\!A_0 \cdot
  \varphi_-(\vec r) - \vec A_b \sdot \vec \sigma \, \varphi_+(\vec r) &= \frac{M_*
    +M}{\hbar}c\cdot
  \varphi_-(\vec r) \label{3.29a}\\
  i\big(\vec \sigma \sdot \vec \nabla \big ) \varphi_-(\vec r) - {}^{(p)}\!A_0 \cdot
  \varphi_+(\vec r) - \vec A_b \sdot \vec \sigma \, \varphi_-(\vec r) &= \frac{M_*
    -M}{\hbar}c\cdot \varphi_+(\vec r) 
\label{3.29b} \\
  (\,ortho \textnormal - positroniu&m\,) \nonumber \ .
\end{align}
\end{subequations}
Hence the identification of both sets of equations (\ref{3.28a})-(\ref{3.28b}) and
(\ref{3.29a})-(\ref{3.29b}) yields a further condition upon the operator $\hat\Omega$,
namely
\begin{subequations}\label{3.30}
\begin{align}
  \hat\Omega^\dagger \big (\vec \sigma \sdot \vec \nabla \big ) \hat\Omega\varphi_\pm(\vec
  r) &= 
\vec \sigma \sdot \vec \nabla \,\varphi_\pm (\vec r) \label{3.30a} \\
  \hat\Omega^\dagger \big ( \vec A_b \sdot \vec \sigma \big )\, \hat\Omega &= 
\big (\vec A_b \sdot \vec \sigma \big ) \label{3.30b} \\
  (\,ortho \textnormal - pos&itronium\,) \nonumber \ .
\end{align}
\end{subequations}
Evidently, these two requirements for ortho-positronium do admit the solution
(\ref{3.26a}) for the operator $\hat\Omega$ with no further constraints which is in
contrast to the corresponding situation with the para-configuration.

\indent The treatment of the para-configuration runs in a quite analogous way: first,
apply the para- transformation (\ref{3.20c}) again to the eigenvalue equations (\ref{3.6})
for the electron and find in this case
\begin{subequations}\label{3.31}
\begin{align}
  i \big (\vec \sigma \sdot \vec \nabla \big ) \hat\Omega \, \varphi_+(\vec r )
  +{}^{(p)}\!A_0 \cdot 
\hat\Omega \, \varphi_-(\vec r) - \big ( \vec A_p \sdot \vec \sigma \big ) \hat\Omega \,
\varphi_+(\vec r) = 
-\frac {M_* + M}{\hbar} c \cdot \hat\Omega \, \varphi_-(\vec r) \label{3.31a} \\
  i \big (\vec \sigma \sdot \vec \nabla \big ) \hat\Omega \, \varphi_-(\vec r )
  +{}^{(p)}\!A_0 \cdot \hat\Omega \, \varphi_+(\vec r) - \big ( \vec A_p \sdot \vec \sigma
  \big ) \hat\Omega \, \varphi_-(\vec r) = -\frac {M_* - M}{\hbar} c \cdot \hat\Omega \,
  \varphi_+(\vec r) \label{3.31b} \ ,
\end{align}
\end{subequations}
whereas the eigenvalue equations (\ref{3.5}) of the positron appear under the action of the para-identifications as
\begin{subequations}\label{3.32}
\begin{align}
  i \big (\vec \sigma \sdot \vec \nabla \big ) \varphi_+(\vec r ) - {}^{(p)}\!A_0 \cdot
  \varphi_-(\vec r) &+ 
\big ( \vec A_p \sdot \vec \sigma \big ) \varphi_+(\vec r) = \frac {M_* + M}{\hbar} c
\cdot \varphi_-(\vec r) 
\label{3.32a} \\
  i \big (\vec \sigma \sdot \vec \nabla \big ) \varphi_-(\vec r ) - {}^{(p)}\!A_0 \cdot
  \varphi_+(\vec r) &+ 
\big ( \vec A_p \sdot \vec \sigma \big ) \varphi_-(\vec r) = \frac {M_* - M}{\hbar} c
\cdot \varphi_+(\vec r) 
\label{3.32b} \\
  (\,par&a\textnormal - positronium\,) \nonumber \ .
\end{align}
\end{subequations}
Therefore, if both sets of eigenvalue equations (\ref{3.31a})-(\ref{3.31b}) and
(\ref{3.32a})-(\ref{3.32b}) are required again to be identical, one is led in this case to
the following condition
\begin{subequations}\label{3.33}
\begin{align}
  \hat\Omega^\dagger \big ( \vec \sigma \sdot \vec \nabla \big ) \hat\Omega \, \varphi_\pm
  (\vec r) &= - 
\big ( \vec \sigma \sdot \vec \nabla \big ) \varphi_\pm (\vec r) \label{3.33a} \\
  \hat\Omega^\dagger \big ( \vec A_p \sdot \vec \sigma \big ) \hat\Omega &= \big (\vec A_p
  \sdot \vec \sigma \big ) \label{3.33b} \\
  (\,para\textnormal - pos&itronium\,) \ . \nonumber
\end{align}
\end{subequations}
Observe here that the first condition (\ref{3.33a}) does not admit the first solution
(\ref{3.26a}) for $\hat\Omega$, which therefore can apply only to the
ortho-configurations. But the requirement (\ref{3.33a}) does admit the second solution
(\ref{3.26b}); and in this case the relation (\ref{3.33a}) presents some restrictive
condition upon the form of the Pauli spinors $\varphi_\pm(\vec r)$ (i.e. double
valuedness, see below). Furthermore, the second condition (\ref{3.33b}) with the operator
$\hat\Omega$ being given by (\ref{3.26b}) requires the vector potential $\vec A_p$ for the
para-configurations to be (anti)parallel to the Dirac current $\vec k_b$ (\ref{3.8b})
\begin{equation}\label{3.34}
\vec A_p = A_p \, \hat{\vec k} \ .
\end{equation}

\indent Thus, summarizing the situation with the one-particle eigenvalue equations, one
arrives at two forms thereof: namely, the ortho-form (\ref{3.29a})-(\ref{3.29b}) and the
para-form (\ref{3.32a})-(\ref{3.32b}). Furthermore, the ortho-form is found to admit a
richer spectrum of mass eigensolutions; the reason is that the shape of the
ortho-eigensolutions $\varphi_\pm(\vec r)$ is not constrained by the ortho-condition
(\ref{3.30a}). Naturally, further insight into this peculiarity of the positronium
eigenvalue problem will now be gained by a closer inspection of just those restrictive
conditions (\ref{3.33a})-(\ref{3.33b}) upon the form of the para-eigensolutions
$\varphi_\pm(\vec r)$.

\subsection{Para-Positronium}
The task is now to elaborate the specific form of the Pauli spinors which is admitted by
the condition (\ref{3.33a}) with the operator $\hat\Omega$ being given by equation
(\ref{3.26b}). For the sake of simplicity, we assume a cylindrical symmetry so that both
Dirac currents $\vec k_{b/p}$ (\ref{3.8a}) and (\ref{3.8b}) encircle the z-axis, where the
unit vector $\hat{\vec k}$ coincides with the azimuthal basis vector $\vec e_\phi$ of an
orthonormal triad $\{ \vec e_r, \vec e_\vartheta, \vec e_\phi \}$ being due to the
spherical polar coordinates $\{ r, \vartheta, \phi\}$, i.e. we put for the vector
potentials
\begin{equation}\label{3.35}
\vec A_{b/p}(\vec r) = {}^{(b/p)}\!A_\phi(r,\vartheta) \vec e_\phi \ ,
\end{equation}
and similarly for the Dirac currents
\begin{equation}\label{3.36}
\vec k_{b/p}(\vec r) = {}^{(b/p)}\!k_\phi(r,\vartheta) \vec e_\phi \ ,
\end{equation}
cf. (\ref{3.25a})-(\ref{3.25b}). By this assumption of cylindrical symmetry (for a
toroidal symmetry, see ref.~\cite{GrBeMa}), the para-condition (\ref{3.33a}) adopts the
following anticommutator form
\begin{align}\label{3.37}
\big\{ \big(\vec \sigma\sdot \vec \nabla\big ),\vec e_\phi \sdot &\vec \sigma \big\} \varphi_\pm (\vec r) = 0 \\
(\,para\textnormal - pos&itronium\,) \nonumber \ .
\end{align}

\indent But here one can show by means of some elementary mathematics, that this condition
upon the Pauli spinors $\varphi_\pm(\vec r)$ can be recast to the following form
\begin{align}\label{3.38}
  \big\{ i \,\vec \sigma \sdot \big(\vec \nabla \times \vec e_\phi \big )+ \vec \nabla
  \sdot 
\vec e_\phi + &2 \big(\vec e_\phi \sdot \vec \nabla \big ) \big\} \varphi_\pm (\vec r) = 0 \\
  (\,para\textnormal - pos&itronium\,) \nonumber \ ,
\end{align}
where of course the divergence of the azimuthal basis vector $\vec e_\phi$ vanishes
\begin{equation}\label{3.39}
\vec \nabla \sdot \vec e_\phi \equiv 0 \ ,
\end{equation}
and the curl of $\vec e_\phi$ is given by
\begin{equation}\label{3.40}
\vec \nabla \times \vec e_\phi = \frac{\vec e_z}{r \sin \vartheta} \ .
\end{equation}
Thus the restriction (\ref{3.38}) upon the Pauli spinors $\varphi_\pm(\vec r)$ adopts its final form as
\begin{align}\label{3.41}
\hat J_z \, \varphi_\pm(\vec r) &= 0 \\
(\,para\textnormal - positr&onium\,) \nonumber \ ,
\end{align}
where the $z$-component of the {\it total angular momentum operator} $\hat{\vec J} (=
\hat{\vec S}+\hat{\vec L} )$ is defined as usual through
\begin{equation}\label{3.42x}
\hat J_z \doteqdot \hbar \, \big (\frac{1}{2} \sigma_z + \frac{1}{i} \frac{\partial}{\partial \phi}\big) \ .
\end{equation}

\indent Consequently, the physical content of the para-restriction (\ref{3.33a}) is
nothing else than the fact that the para-eigensolutions must have vanishing $z$-component
of total angular momentum. Observe that this requirement refers to a one-particle spinor
state $\psi(\vec r) = \varphi_+(\vec r) \oplus \varphi_-(\vec r)$ which is the direct sum
of both Pauli spinors $\varphi _+(\vec r)$ and $\varphi_-(\vec r)$. Such {\it exotic
  states} do not occur in the conventional theory of a single spin-$\frac{1}{2}$ Dirac
particle just on behalf of the required uniqueness of the wave function; but in the
present positronium description, these exotic states must necessarily emerge as a
consequence of the {\it physical equivalence} of both constituents, cf. the preceeding
identification process leading to (\ref{3.33a}).

\indent It is also interesting to remark that para-positronium is somewhat exceptional
also in other respects: since the {\it total charge density} $j_0$ is zero according to
equation (\ref{3.12})
\begin{equation}\label{3.42'}
j_0 \doteqdot {}^{(1)}\!j_0 + {}^{(2)}\!j_0 = 0 \ ,
\end{equation}
as well as the {\it total electric potential} $A_0$, cf. (\ref{3.15})
\begin{equation}\label{3.43'}
A_0 \doteqdot {}^{(1)}\!A_0 + {}^{(2)}\!A_0 = 0 \ ,
\end{equation}
the total electric field strength $\vec E$ must also be zero:
\begin{equation}\label{3.44'}
\vec E(\vec r) = \vec E_1(\vec r) + \vec E_2(\vec r) = -\vec \nabla A_0(\vec r) = 0 \ .
\end{equation}
Similar conclusions do hold also for the total magnetic objects, i.e.
\begin{subequations}\label{3.45'}
\begin{align}
\vec j(\vec r) &\doteqdot \vec j_1(\vec r) + \vec j_2(\vec r) \equiv 0 \label{3.45'a} \\
\vec A(\vec r) &\doteqdot \vec A_1(\vec r) + \vec A_2(\vec r) \equiv 0 \label{3.45'b} \\
\vec H(\vec r) &\doteqdot \vec H_1(\vec r) + \vec H_2(\vec r) \equiv 0 \label{3.45'c} \ ,
\end{align}
\end{subequations}
see equation (\ref{3.7b}) and (\ref{3.9b}). On the other hand, since the composite RST
systems do interact with external sources just via their total objects (see the discussion
of this point in the preceding papers \cite{SSMS}-\cite{BeSo2},~\cite{GrBeMa}), the
para-positronium appears to be reluctant to undergo interactions with the outside world.

\indent Concerning now the construction of appropriate Pauli spinors $\varphi(\vec r)$
obeying the angular-momentum condition (\ref{3.41}), one first constructs four basis
spinors $\{\omega_0^{(+)}, \omega_0^{(-)}; \omega_1^{(+)}, \omega_1^{(-)}\}$:
\begin{subequations}\label{3.42}
\begin{align}
\omega_0^{(+)} &= e^{-i\phi/2} \, \zetappn \label{3.42a}\\
\omega_0^{(-)} &= e^{i\phi/2} \, \zetapmn \label{3.42b}\\
\omega_1^{(+)} &= e^{-i\phi/2} \, \zetappe \label{3.42c}\\
\omega_1^{(-)} &= e^{i\phi/2} \, \zetapme \label{3.42d} \ ,
\end{align}
\end{subequations}
which are required to have vanishing z-component of total angular momentum $\hat J_z$:
\begin{equation}\label{3.43}
\hat J_z \, \omega_0^{(\pm)} = \hat J_z \, \omega_1^{(\pm)} = 0 \ .
\end{equation}
This goal can be achieved by taking for the $\zeta$-basis the conventional choice \cite{Me2}
\begin{subequations}\label{3.44}
\begin{align}
\hat{\vec J}^{\,2} \, \zetajml &= \hbar^2 \, j(j+1) \, \zetajml \label{3.44a} \\
\hat J_z \, \zetajml &= m \hbar \, \zetajml \label{3.44b} \\
\hat{\vec L}^2 \, \zetajml &= \hbar^2 \, l(l+1) \, \zetajml \label{3.44c} \\
\hat{\vec S}^2 \, \zetajml &= \hbar^2 \, s(s+1) \, \zetajml \label{3.44d} \ ,
\end{align}
\end{subequations}
with the following spin-$\frac{1}{2}$ combination of quantum numbers for the positronium {\it groundstate}
\begin{equation}\label{3.45}
s=j=\frac{1}{2}; \ \ m=\pm\frac{1}{2}; \ \ l=0,1 \ .
\end{equation}
And then one decomposes the Pauli spinors $\varphi_\pm(\vec r)$ with respect to this $\omega$-basis as
\begin{subequations}\label{3.46}
\begin{align}
\varphi_+(\vec r) &= R_+(r,\vartheta) \cdot \omega_0^{(+)} + S_+(r,\vartheta)\cdot \omega_0^{(-)} \label{3.46a} \\
\varphi_-(\vec r) &= - i \, R_-(r,\vartheta) \cdot \omega_1^{(+)} - i \, S_-(r,\vartheta)\cdot \omega_1^{(-)} \label{3.46b} \ .
\end{align}
\end{subequations}
Finally, one inserts this form of the Pauli spinors into the eigenvalue equations
(\ref{3.32a})-(\ref{3.32b}) for para-positronium in order to deduce the corresponding
eigenvalue equations for the wave amplitudes $R_\pm(r,\vartheta)$ and $S_\pm
(r,\vartheta)$ as \cite{BeSo2}:
\begin{subequations}\label{3.47}
  \begin{align}
    &\frac{\partial \tilde R_+}{\partial r} + \frac{1}{r} \frac{\partial \tilde
      S_+}{\partial \vartheta} - {}^{(p)}\!A_0 \cdot \tilde R_- + {}^{(p)}\!A_\phi [\sin
    \vartheta \cdot \tilde R_+ - \cos \vartheta \cdot
    \tilde S_+ ] =\frac{M+M_*}{\hbar}c\cdot \tilde R_- \label{3.47a} \\
    &\frac{1}{r} \frac{\partial (r\tilde R_-)}{\partial r} - \frac{1}{r} \frac{\partial
      \tilde S_-}{\partial \vartheta} + {}^{(p)}\!A_0 \cdot \tilde R_+ - {}^{(p)}\!A_\phi
    [\sin \vartheta \cdot \tilde R_- - \cos \vartheta \cdot \tilde S_- ]
    =\frac{M-M_*}{\hbar}c\cdot \tilde R_+ \label{3.47b}\\
    &\frac{\partial \tilde S_+}{\partial r} - \frac{1}{r} \frac{\partial \tilde
      R_+}{\partial \vartheta} - {}^{(p)}\!A_0 \cdot \tilde S_- - {}^{(p)}\!A_\phi [\sin
    \vartheta \cdot \tilde S_+ + \cos \vartheta \cdot
    \tilde R_+ ] =\frac{M+M_*}{\hbar}c\cdot \tilde S_- \label{3.47c}\\
    &\frac{1}{r} \frac{\partial (r\tilde S_-)}{\partial r} + \frac{1}{r} \frac{\partial
      \tilde R_-}{\partial \vartheta} + {}^{(p)}\!A_0 \cdot \tilde S_+ + {}^{(p)}\!A_\phi
    [\sin \vartheta \cdot \tilde S_- + \cos \vartheta \cdot \tilde R_- ]
    =\frac{M-M_*}{\hbar}c\cdot \tilde S_+ \ . \label{3.47d}
\end{align}
\end{subequations}
Here, the original wave amplitudes $R_\pm(r, \vartheta)$ and $S_\pm(r,\vartheta)$
(\ref{3.46a})-(\ref{3.46b}) have been slightly changed to
\begin{subequations}\label{3.48}
\begin{align}
\tilde R_\pm(r,\vartheta) &\doteqdot \sqrt{r\sin\vartheta} \cdot R_\pm(r,\vartheta) \label{3.48a}\\
\tilde S_\pm(r,\vartheta) &\doteqdot \sqrt{r\sin\vartheta} \cdot S_\pm(r,\vartheta) \label{3.48b} \ ;
\end{align}
\end{subequations}
and furthermore, since the vector potential $\vec A_p$ (\ref{3.7b}) must always be
(anti)parallel to the unit vector $\hat{\vec k} \equiv \vec e_\phi$ (\ref{3.34}), we have
put also for its azimuthal component:
\begin{equation}\label{3.49}
\vec A_p = {}^{(p)}\!A_\phi \vec e_\phi \ .
\end{equation}

\indent Notice also that the use of the $\omega$-basis (\ref{3.42a})-(\ref{3.42d}) induces
a certain non-uniqueness of the Pauli spinors $\varphi_\pm (\vec r)$
(\ref{3.46a})-(\ref{3.46b}); indeed it is obvious that they are unique over space-time
only up to sign:
\begin{equation}\label{3.49'}
\varphi_\pm (r,\vartheta,\phi+2\pi ) = -\varphi_\pm(r, \vartheta,\phi) \ .
\end{equation}
However, it is also obvious that the physical densities, e.g. charge density
${}^{(a)}\!k_0(\vec r)$ (\ref{3.18}) and current $\vec k_a(\vec r)$ (\ref{3.19}), are
nevertheless unique!

\subsection{Ortho-Positronium}
Naturally, the ortho-configuration must appear somewhat simpler because the corresponding
operator $\hat\Omega$ (\ref{3.26a}) does trivially obey the ortho-constraint (\ref{3.30a})
and therefore does not impose any restrictive condition upon the Pauli spinors
$\varphi_\pm(\vec r)$, as is the case with the para-configuration (\ref{3.33a}).
Nevertheless, it is reasonable to assume that also the stationary ortho-configurations are
associated with a well-defined z-component $\hat J_z$ of total angular-momentum; albeit
this component is not necessarily zero as for the para-configurations. But for the special
case of the {\it groundstate} it is self-suggestive to adopt zero angular-momentum also
for ortho-positronium. This means that we can resort to the above mentioned $\omega$-basis
(\ref{3.42a})-(\ref{3.42d}) and thus the para-ansatz (\ref{3.46a})-(\ref{3.46b}) may be
applied also for the ortho-configurations! Consequently, one inserts this ansatz into the
ortho-system (\ref{3.29a})-(\ref{3.29b}) and then arrives at the following eigenvalue
system for the two wave amplitudes $\tilde R_\pm(r,\vartheta)$ and $\tilde
S_\pm(r,\vartheta)$:
\begin{subequations}\label{3.50}
\begin{align}
  \frac{\partial \tilde R_+}{\partial r} + \frac{1}{r} \frac{\partial \tilde S_+}{\partial
    \vartheta} - {}^{(p)}\!A_0 \cdot \tilde R_- - {}^{(b)}\!A_\phi [\sin \vartheta \cdot
  \tilde R_+ - \cos \vartheta \cdot
  \tilde S_+ ] &=\frac{M+M_*}{\hbar}c\cdot \tilde R_- \label{3.50a} \\
  \frac{1}{r} \frac{\partial (r\tilde R_-)}{\partial r} - \frac{1}{r} \frac{\partial
    \tilde S_-}{\partial \vartheta} + {}^{(p)}\!A_0 \cdot \tilde R_+ + {}^{(b)}\!A_\phi
  [\sin \vartheta \cdot \tilde R_- - \cos \vartheta \cdot \tilde S_- ]
  &=\frac{M-M_*}{\hbar}c\cdot \tilde R_+ \label{3.50b}\\
  \frac{\partial \tilde S_+}{\partial r} - \frac{1}{r} \frac{\partial \tilde R_+}{\partial
    \vartheta} - {}^{(p)}\!A_0 \cdot \tilde S_- + {}^{(b)}\!A_\phi [\sin \vartheta \cdot
  \tilde S_+ + \cos \vartheta \cdot \tilde R_+ ]
  &=\frac{M+M_*}{\hbar}c\cdot \tilde S_- \label{3.50c}\\
  \frac{1}{r} \frac{\partial (r\tilde S_-)}{\partial r} + \frac{1}{r} \frac{\partial
    \tilde R_-}{\partial \vartheta} + {}^{(p)}\!A_0 \cdot \tilde S_+ - {}^{(b)}\!A_\phi
  [\sin \vartheta \cdot \tilde S_- + \cos \vartheta \cdot \tilde R_- ]
  &=\frac{M-M_*}{\hbar}c\cdot \tilde S_+ \ . \label{3.50d}
\end{align}
\end{subequations}

\indent Evidently, this ortho-system differs from its para-counterpart
(\ref{3.47a})-(\ref{3.47d}) merely in the sign of the magnetic term (${}^{(b)}\!A_\phi
\rightarrow - {}^{(p)}\!A_\phi$); and if the magnetic (i.e. spin-spin) interactions are
neglected, both systems collapse into the same form ("{\it electrostatic approximation}").
This is reasonable because the ortho-para dichotomy of positronium is just due to the
different magnetic interactions and therefore must disappear when the latter type of
interaction is neglected. This circumstance will readily be exploited for a lowest-order
approximation.

\subsection{Approximations}
\indent It should be obvious that both eigenvalue problems (\ref{3.47a})-(\ref{3.47d}) and
(\ref{3.50a})-(\ref{3.50d}) are too complicated in order to find exact solutions
(which is mostly the case with modern gauge field theories). Therefore it is important to
look for simple approximative solutions which, however, are
still well-suited to demonstrate the essential physical effects. Surely, the binding
phenomenon can be understood already by reference to the electric interactions alone,
which mostly in atomic physics are much stronger than their magnetic counterparts.
Therefore one may expect to obtain a relatively good approximation to the binding energy
by neglecting the magnetic terms in the eigenvalue equations, with preservation of their
relativistic form. Furthermore it seems reasonable to assume that magnetism is closely
related to a non-trivial angular momentum, which however is known to be in conflict with
the spherical symmetry of the field configuration. Therefore, if one wishes to
consider first a {\it "spherically symmetric approximation"} of the positronium groundstate, one
will skip the $\vartheta$-dependence of the wave amplitudes $\tilde R_\pm, \tilde S_\pm$
together with the magnetic (spin-spin) interactions. By these approximative assumptions
both eigenvalue systems (\ref{3.47a})-(\ref{3.47d}) and (\ref{3.50a})-(\ref{3.50d})
coincide to the same form:
\begin{subequations}\label{3.51}
\begin{align}
  \frac{d \tilde R_+(r) }{dr} - {}^{(p)}\!A_0(r) \cdot \tilde R_-(r)
  &=\frac{M+M_*}{\hbar}c\cdot
  \tilde R_-(r) \label{3.51a} \\
  \frac{1}{r}\frac{d(r\tilde R_-(r))}{dr}+{}^{(p)}\!A_0(r)\cdot\tilde
  R_+(r)&=\frac{M-M_*}{\hbar}c\cdot\tilde R_+(r) 
\label{3.51b}\\
  \frac{d \tilde S_+(r)}{dr} - {}^{(p)}\!A_0(r) \cdot \tilde
  S_-(r)&=\frac{M+M_*}{\hbar}c\cdot\tilde S_-(r) 
  \label{3.51c}\\
  \frac{1}{r} \frac{d(r\tilde S_-(r))}{dr} + {}^{(p)}\!A_0(r) \cdot \tilde
  S_+(r)&=\frac{M-M_*}{\hbar}c\cdot \tilde S_+(r) \ . \label{3.51d}
\end{align}
\end{subequations}

\indent The striking feature of this simplification evidently refers to the circumstance
that the spin-up fields $\tilde R_\pm(r)$ become decoupled from the spin-down fields
$\tilde S_\pm(r)$; and therefore it is sufficient to consider the spin-up configurations
$\tilde R_\pm(r)$ alone, which are solutions of the coupled pair of equations
(\ref{3.51a})-(\ref{3.51b}). Clearly, the spin-down solutions $\tilde S_\pm(r)$ of
(\ref{3.51c})-(\ref{3.51d}) are equally well possible and are identical (up to
spin-orientation) with the spin-up solutions. Consequently we can concentrate exclusively
upon the spin-up solutions $\tilde R_\pm(r)$.

\indent The next problem refers to the electrostatic potential ${}^{(p)}\!A_0(r)$ which
has also been assumed to be spherically symmetric in the simplified system
(\ref{3.51a})-(\ref{3.51d}). Properly speaking, this potential must be determined from its
own Poisson equation, cf. (\ref{3.17}), because it is generated by the charge distribution
${}^{(p)}\!j_0$ (\ref{3.12}) of the other particle. However, it is a well-known fact in
classical electrodynamics that the potential ${}^{(p)}\!A_0(\vec r)$ as a solution of the
Poisson equation (\ref{3.17}) must be of the following form
\begin{equation}\label{3.52} {}^{(p)}\!A_0(\vec r)=\alpha_{\rm S} \int d^3\vec r\,' \;
  \frac{{}^{(p)}\!j_0(\vec r\,')}{\vert\vert \vec r - \vec r\,'\vert\vert}
  \mathop{\Longrightarrow}\limits_{\mathrm{(r\rightarrow\infty)}} \frac{\alpha_{\rm S}}{r}
  \ ,
\end{equation}
which in the asymptotic region ($r\!\rightarrow\!\infty$) adopts the Coulomb form ($\sim
r^{-1}$). This fact may now be exploited in order to briefly demonstrate that the residual
eigenvalue equations
\begin{subequations}\label{3.53}
\begin{align}
  \frac{d\tilde R_+(r)}{dr}-\frac{\alpha_{\rm S}}{r}\cdot\tilde
  R_-(r)&=\frac{M+M_*}{\hbar}c\cdot\tilde R_-(r)
  \label{3.53a}\\
  \frac{1}{r}\frac{d \big(r\,\tilde R_-(r)\big)}{dr} + \frac{\alpha_{\rm S}}{r} \cdot
  \tilde R_+(r) &= \frac{M-M_*}{\hbar} c \cdot \tilde R_+(r) \label{3.53b}
\end{align}
\end{subequations}
do not agree with the well-known solutions of the conventional Coulomb force problem,
despite the use of the asymptotic Coulomb potential (\ref{3.52}).  Indeed, trying here the
usual ansatz for the wave amplitudes $\tilde R_\pm(r)$
\begin{equation}\label{3.54}
\tilde R_\pm(r) = \tilde N_\pm \cdot r^\nu \exp [-\frac{r}{r_*}]
\end{equation}
yields by insertion into the simplified eigenvalue equations (\ref{3.53a})-(\ref{3.53b})
for the scale parameter $r_*$, power $\nu$, mass eigenvalue $M_*$ and normalization
constants $\tilde N_\pm$ the following results:
\begin{subequations}\label{3.55}
\begin{align}
r_* &= \frac{1}{2} a_{\rm B} \doteqdot \frac{\hbar^2}{2M e^2} \ \dots \ \textnormal{Bohr radius} \label{3.55a} \\
\nu &=-\frac{1}{2} \big(1- \sqrt{(1-(2\alpha_{\rm S})^2} \big) \label{3.55b}\\
M_*&=M\sqrt{1-(2\alpha_{\rm S})^2} \label{3.55c} \\
\frac{\tilde N_-}{\tilde N_+} &= - \frac{1-\sqrt{1-(2\alpha_{\rm S})^2}}{2\alpha_{\rm S}} \label{3.55d} \ .
\end{align}
\end{subequations}

\indent But these results do {\it not} agree with the corresponding ones of the
conventional Coulomb force problem (see, e.g., ref.\cite{Gr}), which may be seen even more
clearly through passing over to the non-relativistic approximation. For this
approximation, the scale parameter $r_*$ (\ref{3.55a}) remains unchanged whereas the power
$\nu$~(\ref{3.55b}) is put to zero on account of the smallness of the fine structure constant
$\alpha_{\rm S}=e^2/\hbar c \ (\approx 1/137)$. Furthermore, the negative Pauli component
$\tilde R_-$ is put to zero for the same reason (see the ratio (\ref{3.55d}) of the
normalization constants $\tilde N_\pm$), and finally the mass eigenvalue $M_*$ is expanded
with respect to $\alpha_{\rm S}$
\begin{equation}\label{3.56}
M_* \cong M(1-2 \alpha_{\rm S}^2) = M-2\,\frac{Me^4}{\hbar^2c^2} \ .
\end{equation}
Defining here the non-relativistic eigenvalue $E_S$ through
\begin{equation}\label{3.57}
E_S \doteqdot M_*c^2-Mc^2
\end{equation}
yields finally
\begin{equation}\label{3.58}
E_S = -2\,\frac{Me^4}{\hbar^2}=-2\, \frac{e^2}{a_{\rm B}} \ .
\end{equation}
Thus the groundstate binding energy $E_B$ ($\doteqdot -E_S$) of a point particle of mass
$M$ and charge $e$ in the Coulomb potential (\ref{3.52}) is found here to be four times
the conventional hydrogen result
\begin{equation}\label{3.59}
E_B\Big \vert_{RST} = 2\,\frac{e^2}{a_{\rm B}} = 4\cdot E_B \Big \vert_{conv.} = 4\cdot\Big(\frac{e^2}{2\, a_{\rm B}}\Big) \ .
\end{equation}
(For the conventional treatment of the hydrogen atom see, e.g., ref.\cite{Gr}). The
non-conventional RST result (\ref{3.59}) does not mean that RST is wrong, but simply says
that RST treats the {\it internal} motion of a compound system in a way different from
to the {\it external} motion of a pointlike particle!

\indent In order to complete this RST picture of the internal motion of the positronium
constituents, one wishes to see the non-relativistic approximation ($\tilde R(r)$, say) of
the wave function $\tilde R_\pm(r)$ (\ref{3.54})
\begin{subequations}\label{3.60}
\begin{align}
\tilde R_+(r) &\Rightarrow \tilde N \exp\Big[-\frac{r}{r_*}\Big] \doteqdot \tilde R(r) \label{3.60a}\\
\tilde R_-(r) &\Rightarrow 0 \ .
\end{align}
\end{subequations}
Naturally, one expects that this non-relativistic form of the solution will satisfy some
non-relativistic wave equation (for the internal Coulomb force problem); but this wave
equation cannot coincide with the conventional Schr\"odinger equation because of the
non-conventional eigenvalue $E_S$ (\ref{3.58}). Rather, the desired non-relativistic form
of the original RST eigenvalue equations (\ref{3.53a})-(\ref{3.53b}), with arbitrary
electric potential $A_0(r)$, is obtained by approximating the first one (\ref{3.53a})
through
\begin{equation}\label{3.61}
\tilde R_-(r) \approx \frac{\hbar}{2Mc}\cdot \frac{d\tilde R_+(r)}{dr}
\end{equation}
and substituting this into the second one (\ref{3.53b}) in order to find
\begin{equation}\label{3.62}
  -\frac{\hbar^2}{2M} \Big( \frac{d^2}{dr^2} + \frac{1}{r}\frac{d}{dr} \Big) \tilde R(r) -
  \hbar c \, {}^{(p)}\!A_0(r) = E_S \cdot \tilde R(r) \ .
\end{equation}
Clearly, the non-relativistic form (\ref{3.60a}) of the relativistic solution
(\ref{3.54})-(\ref{3.55d}) together with the Schr\"odinger eigenvalue $E_S$ (\ref{3.58})
does obey this non-relativistic wave equation~(\ref{3.62}) with ${}^{(p)}\!A_0$ being
identified as the Coulomb potential (\ref{3.52}); but nevertheless this non-relativistic
eigenvalue equation (\ref{3.62}) is {\it not} the ordinary Schr\"odinger equation for the
Coulomb force problem! The reason is that it is not the Laplacean $\Delta(r) = \vec
\nabla^2$, in spherical polar coordinates
\begin{equation}\label{3.63}
\Delta(r) = \frac{d^2}{dr^2} + \frac{2}{r} \frac{d}{dr}\ ,
\end{equation}
which acts upon the wave function $\tilde R(r)$. But rather in equation (\ref{3.62}),
there emerges the Laplacean written in \emph{cylindrical} coordinates, so that the
variable $r$ changes its geometric meaning from a spherical polar coordinate to a
cylindrical coordinate. This change then results in the unconventional eigenvalue $E_S$
(\ref{3.58}), in contrast to its conventional counterpart (\ref{3.59}).

\indent Summarizing, the requirement of physical equivalence of the electron and positron
has lead us to the introduction of those exotic states with vanishing $z$-component $\hat
J_z$ (\ref{3.41}) of the total angular momentum; and these exotic states were then
revealed to obey a wave equation whose non-relativistic approximation does not exactly
agree with the conventional Schr\"odinger form. Therefore, when the exotic states are used,
the non-relativistic limit of RST does not agree with the Hartree-Fock approach which is
based upon the Schr\"odinger type of wave equation. This opens the possibility for RST to
overcome the deficiencies of the HF approach in the sense that the RST predictions may
come closer to both the experimental data and the conventional Schr\"odinger predictions
than it is possible for the HF approach. (For the emergence of the HF approach as the
non-relativistic limit of RST by using the conventional states, see ref.~\cite{PrSo}).

\indent However there is a second peculiarity of RST by which its non-relativistic limit
differs from the conventional HF approach: this refers to the specific form of the
electric potential ${}^{(p)}\!A_0(r)$, by means of which the positronium constituents are
attracting each other. For our preceding heuristic demonstrations in connection with the
approximative wave equations of relativistic (\ref{3.53a})-(\ref{3.53b}) and
non-relativistic form (\ref{3.62}) we preferred to work with the (exact) Coulomb potential
(${}^{(p)}\!A_0 \Rightarrow \frac{\alpha_{\rm S}}{r}$; see equation (\ref{3.52})) because
this is used also in the conventional Schr\"odinger theory, cf. the non-relativistic
positronium Hamiltonian $\hat H$ (\ref{2.19''}). However the {\it exact} RST potential
${}^{(p)}\!A_0$ (\ref{3.52}) surely differs from its asymptotic Coulomb form, but {\it
  not} in that way as is the case in the HF approach. Whereas, in the latter theory, the
singularity of the Coulomb potential (for $r\rightarrow0$) is completely {\it
  regularized}, the RST modification of the Coulomb potential {\it remains singular}, but
in a less pathological manner. Since the specific form of the interaction potential
${}^{(p)}\!A_0$ influences the mass/energy eigenvalues not less than the general form of
the eigenvalue equations themselves, it is now necessary to inspect thoroughly that new
form of the electrostatic potential.

\section{Poisson Equations}
\indent As the preceding discussion is mainly concentrated upon the eigenvalue equations
for the matter fields, it must be complemented by a similar investigation of the gauge
fields, i.e. the electrostatic potential ${}^{(p)}\!A_0(\vec r)$. This object mediates the
electric interaction of the positron and electron and thereby dominates the magnetic
interaction (which is mediated by the azimuthal component ${}^{(p/b)}\!A_\phi$ of the
vector potential $\vec A_{b/p}(\vec r)$ (\ref{3.35})). Especially, the electric potential
is responsible for the occurence of binding while the magnetic interactions do at most
contribute certain small corrections to the binding energy. We therefore neglect now the
magnetic forces altogether (i.e. putting $\vec A_{p/b}$ to zero) and thus work with the
truncated eigenvalue systems (\ref{3.51a})-(\ref{3.51d}) or (\ref{3.53a})-(\ref{3.53b}),
resp. These equations demonstrate clearly the specific way in which both particles do
organize their mutual coupling: first, the wave function $\psi_a$ of either particle
builds up the corresponding Dirac current $k_{a\mu}$ (\ref{2.44}), then its Maxwell
counterpart $\jaomu$ (\ref{2.47a})-(\ref{2.47b}) acts as the source of the electromagnetic
vector potentials $\Aamu$ via the Maxwell equations (\ref{2.43a})-(\ref{2.43b}), and
finally these four-vector potentials influence the wave function $\psi_a$ of either
particle via the mechanism of minimal coupling (\ref{2.22a}-\ref{2.22b}). For the present
stationary situation, the Maxwell equations (\ref{2.43a})-(\ref{2.43b}) are transcribed to
the electrostatic Poisson equation (\ref{3.17}) for the electric interaction potential
${}^{(p)}\!A_0(\vec r)$. It is true, this interaction mechanism is of the conventional
electromagnetic type; but the new aspect is now that the potential ${}^{(p)}\!A_0(\vec r)$
does adopt a very unusual form when it is generated by those exotic one-particle states
with vanishing $z$-component $\hat{J}_z$ of their total angular momentum $\hat{\vec J}$,
see the discussion of equation (\ref{3.41}).

\subsection{Non-relativistic Approximation}
\indent In order to elaborate this point in some detail, one will not be satisfied with
the asymptotic Coulomb form of the solution (\ref{3.52}) to the Poisson equation
(\ref{3.17}), but one will try to compute that integral (\ref{3.52}) more rigorously. To
this end, one makes use of the fact that the Maxwell charge density ${}^{(p)}\!j_0(\vec
r)$ (\ref{3.12}) coincides with the Dirac density ${}^{(b)}\!k_0(\vec r)$, where the
latter reads in terms of the Pauli spinors $\varphi_\pm(\vec r)$ for both positronium
configurations, cf. (\ref{3.18}):
\begin{equation}\label{4.1} {}^{(p)}\!j_0(\vec r)\equiv {}^{(b)}\!k_0(\vec r) =
  \varphi_+^\dagger(\vec r)\sdot \varphi_+(\vec r) + \varphi_-^\dagger(\vec r)\sdot
  \varphi_-(\vec r) \ .
\end{equation}
Furthermore, if the decompositions (\ref{3.46a})-(\ref{3.46b}) of the Pauli spinors with
respect to the ({\it double valued} !) $\omega$-Basis are inserted here, one finds the
desired Maxwell density ${}^{(p)}\!j_0(\vec r)$ appearing in terms of the ({\it unique})
wave amplitudes $\tilde R_\pm(r,\vartheta)$ and $\tilde S_\pm(r,\vartheta)$
(\ref{3.48a})-(\ref{3.48b}) in the following form
\begin{equation}\label{4.2} {}^{(p)}\!j_0(\vec r) = \frac{\tilde R_+^{\; 2} + \tilde
    R_-^{\; 2} + \tilde S_+^{\; 2} + \tilde S_-^{\; 2}}{4\pi\, r \sin\vartheta} \ .
\end{equation}

\indent However, it is not necessary to work with this exact form of the Maxwell charge
density; but for our {\it non-relativistic approximation} we can make use of the fact that
the spin-up and spin-down components of the time-independent Pauli spinors
(\ref{3.46a})-(\ref{3.46b}) become decoupled (see the discussion of this point below
equation (\ref{3.51d})). This enables us to restrict ourselves to the spin-up
configurations alone (i.e. $\tilde S_\pm\Rightarrow 0$), for which the charge density
${}^{(p)}\!j_0(\vec r)$ (\ref{4.2}) reduces to
\begin{equation}\label{4.3}
{}^{(p)}\!j_0(\vec r) \Rightarrow \frac{\tilde R_+^{\; 2} + \tilde R_-^{\; 2}}{4\pi\, r\sin\vartheta} \ .
\end{equation}
But here one may go one step further and restrict oneself to the non-relativistic limit
which means putting $\tilde R_-$ to zero (see the discussion of this point below equation
(\ref{3.55d})). Thus we ultimately arrive at the following approximative form of the
one-particle charge density
\begin{equation}\label{4.4}
{}^{(p)}\!j_0(\vec r) \cong \frac{\tilde R_+^{\;2}(r)}{4\pi\, r\sin\vartheta} \ ,
\end{equation}
where the non-relativistic wave amplitude $\tilde R_+(r) \Rightarrow \tilde R(r)$ has
already been specified by equation (\ref{3.60a}). The normalization of the
non-relativistic density ${}^{(p)}\!j_0(\vec r)$ (\ref{4.4}) is the following, in view of
the fact that both matter fields $\psi_a(r)$ ($a=1,2$) carry just one charge unit:
\begin{equation}\label{4.5}
  1=\int d^3\vec r \; {}^{(p)}\!j_0(\vec r)=\frac{1}{2} \int d^2\vec r \; \tilde
  R^{\;2}(r) =  \frac{\pi}{2} \int dr \, r\tilde R^{\;2}(r) \ ,
\end{equation}
with the two- and three-dimensional volume elements being given by
\begin{subequations}\label{4.6}
\begin{align}
d^3\vec r &= r^2 \sin\vartheta \, dr\, d\vartheta \label{4.6a} \\
d^2\vec r &= r \, dr\, d\vartheta \label{4.6b} \ .
\end{align}
\end{subequations}
Here, the non-relativistic wave amplitude $\tilde R(r)$ to be used hereafter for the
calculation of the positronium groundstate data is the following, cf. (\ref{3.60a})
\begin{equation}\label{4.7}
\tilde R(r) \doteqdot \sqrt{\frac{8}{\pi\, r_*^{\, 2}}} \, \exp\Big[-\frac{r}{r_*} \Big ] \ .
\end{equation}
The length parameter $r_*$ will subsequently be considered as a variational parameter for
the minimalization of the groundstate energy.

\subsection{Spherically Symmetric Approximation}
\indent But now that a suitable form of non-relativistic wave amplitude $\tilde R(r)$
(\ref{4.7}) is fixed, one can substitute this into the integral (\ref{3.52}) for the
determination of the electric potential ${}^{(p)}\!A_0(\vec r)$ which then appears in the
following form:
\begin{equation}\label{4.8} {}^{(p)}\!A_0(\vec r) = \frac{\alpha_{\rm S}}{4\pi}
  \int\!\!\!\!\int d^2\vec r\,'\,d\varphi' \, \frac{\tilde R(r')^{\, 2}}{\vert\vert \vec
    r- \vec r\,'\vert\vert} \ .
\end{equation}
Observe that the asymptotic Coulomb form is guaranteed here by just the normalization
condition (\ref{4.5}). However, the present result (\ref{4.8}) is not yet the desired form
of the electric potential because it is not spherically symmetric. This obliges us to look
for its spherically symmetric approximation which will be found by expanding the
denominater of the integrand (\ref{4.8}) as follows:
\begin{equation}\label{4.9}
  \frac{1}{\vert\vert \vec r- \vec r\,'\vert\vert} = \frac{1}{\sqrt{r^2+r'^2}} \cdot
  \Big\{ 1 + \frac{\vec r\sdot
    \vec r\,'}{r^2+r'^2} + \frac{3}{2} \Big( \frac{\vec r\sdot \vec r\,'}{r^2+r'^2} \Big)^2 + \dots \Big\} \ .
\end{equation}
Clearly, such an expansion induces an analogous expansion of the electric potential
${}^{(p)}\!A_0(\vec r)$ (\ref{4.8}), i.e. one finds
\begin{equation}\label{4.10}
 {}^{(p)}\!A_0(\vec r) = {}^{(I)}\!A_0(r) + {}^{(I\!I)}\!A_0(r,\vartheta) + \dots \ ,
\end{equation}
with the {\it spherically symmetric contribution} ${}^{(I)}\!A_0(r)$ being given by
\begin{equation}\label{4.11}
{}^{(I)}\!A_0(r)=\frac{\alpha_{\rm S}}{2} \int d^2\vec r\,' \; \frac{\tilde R(r')^2}{\sqrt{r^2+r'^2}} \ ,
\end{equation}
and correspondingly the first anisotropic correction ${}^{(I\!I)}\!A_0(r,\vartheta)$ by
\begin{equation}\label{4.12} 
{}^{(I\!I)}\!A_0(r,\vartheta)=\frac{3\,\alpha_{\rm S}}{8\pi}
\int d^2\vec r\,' \, d\varphi'\; \frac{\big(\vec r \cdot \vec r\,'\big)^2\cdot \tilde
  R(r')^2}{\big(r^2+r'^2\big)^\frac{5}{2}} \ .
\end{equation}

\indent Naturally, one would like to neglect here the first anisotropic correction
${}^{(I\!I)}\!A_0(r,\vartheta)$ (and all higher-order terms) and retain only the
spherically symmetric contribution ${}^{(I)}\!A_0(r)$ (\ref{4.11}). But this is allowed
only if the anisotropic correction terms are actually much smaller than the spherically
symmetric term. More concretely, it is easy to see that the first anisotropic correction
(\ref{4.12}) is factorized according to
\begin{equation}\label{4.13}
{}^{(I\!I)}\!A_0(r,\vartheta) = {}^{(I\!I)}\!A_0(r)\cdot {}^{(I\!I)}\!A_0(\vartheta)
\end{equation}
with the radial factor ${}^{(I\!I)}\!A_0(r)$ being given by
\begin{equation}\label{4.14}
{}^{(I\!I)}\!A_0(r) =\alpha_{\rm S} \, r^2 \int dr'\; r'^3\cdot \frac{\tilde R(r')^2}{(r^2+r'^2)^\frac{5}{2}} \ ,
\end{equation}
and analogously the angular factor ${}^{(I\!I)}\!A_0(\vartheta)$ by
\begin{align}\label{4.15}
{}^{(I\!I)}\!A_0(\vartheta) = \int\!\!\!\!\int d\vartheta'd\varphi'\; (\hat{\vec  r}\,'\cdot \hat{\vec r}\,)^2
&=\frac{3\pi}{16}(1+\cos^2\vartheta) \ .\\
(\hat{\vec r}\doteqdot \vec r/\vert\!\vert\vec r\vert\!\vert \ ,\
\textnormal{etc.})&\nonumber
\end{align}
Consequently, since this angular factor is of the order of unity, the anisotropic
corrections could be safely neglected if the radial factor ${}^{(I\!I)}\!A_0(r)$
(\ref{4.14}) is much smaller than the spherically symmetric potential ${}^{(I)}\!A_0(r)$
(\ref{4.11}). It may be somewhat difficult to show this for the most general situation,
but a rough estimate can be carried out for a concrete model density $\tilde R(r)^2$, i.e.
the step function (see Appendix). Indeed for this special demonstration, the magnitude of
the radial factor ${}^{(I\!I)}\!A_0(r) $ amounts to (at most) a few percent of the
spherically symmetric potential ${}^{(I)}\!A_0(r)$ (\ref{4.11}), see fig.1.

\indent Furthermore, the radial factor ${}^{(I\!I)}\!A_0(r)$ (\ref{4.14}) vanishes at the
origin ($r=0$) so that the exact potential ${}^{(p)}\!A_0(r,\vartheta)$ (\ref{4.8}) looks
in any case spherically symmetric around the origin. And additionally it looks spherically
symmetric at infinity ($r\rightarrow \infty$) since it approaches the (spherically
symmetric) Coulomb potential in that asymptotic region, cf. (\ref{3.52}). Thus the
anisotropy of the electric potential ${}^{(p)}\!A_0(\vec r)$ can be of some (minor)
relevance only in the intermediate region (fig.1). In any case, the responsibility for the
occurrence of binding is surely due to the spherically symmetric part ${}^{(I)}\!A_0(r)$
alone; and the effect of anisotropy will merely consist in some minor shift of the energy
levels whose magnitude will be determined mainly by the spherically symmetric part
${}^{(I)}\!A_0(r)$. For this reason, we will subsequently restrict ourselves to the use of
that spherically symmetric approximation ${}^{(I)}\!A_0(r)$ of the electric potential.

\subsection{Struve-Neumann Potential}
Clearly, even if one has decided to be satisfied with the spherically symmetric
approximation ${}^{(I)}\!A_0(r)$ (\ref{4.11}) of the potential, one nevertheless wants to
have this approximation as realistic as possible, i.e. one becomes faced now with the
problem of an appropriate choice of the wave amplitude $\tilde R(r)$ in equation
(\ref{4.11}). Since in the asymptotic region ($r\rightarrow\infty$) the approximative
potential ${}^{(I)}\!A_0(r)$ approaches the Coulomb potential, cf. (\ref{3.52}), and since
in this case the non-relativistic wave equation (\ref{3.62}) has the normalized wave
amplitude $\tilde R(r)$ (\ref{4.7}) as its exact solution, one will now take this function
(\ref{4.7}) as a trial function and will calculate with its help the associated trial
potential ${}^{(I)}\!A_0(r)$ (\ref{4.11})
\begin{align}\label{4.16}
{}^{(I)}\!A_0(r) \Rightarrow {}^{[1]}\!A_0(r)\doteqdot \frac{\alpha_{\rm S}}{\pi} \Big(\frac{2}{r_*}\Big)^2 \int d^2&\vec r\,' \; \frac{\exp\big[-2\, \frac{r'}{r_*}\big]}{\sqrt{r'^2+r^2}} \\
(Struve\textnormal{-}Neumann\textnormal{ } &potential) \nonumber \ .
\end{align}

\indent Indeed if one lets the length parameter $r_*$ tend to zero ($r_*\rightarrow 0$),
the trial charge distribution $\tilde R(r)^2$ shrinks to a point-like distribution whose
associated potential ${}^{[I]}\!A_0(r)$ (\ref{4.16}) actually tends then to the Coulomb
potential. However, the solution of the (non-relativistic) wave equation (\ref{3.62}) is
by no means point-like, but one rather expects that the realistic value of the ansatz
parameter $r_*$ will fall in the vicinity of the Bohr radius $a_{\rm B}$ (\ref{3.55a});
and the tentative hypothesis is now that for this order of magnitude of $r_*$, there will
arise some potential ${}^{[1]}\!A_0(r)$ from the prescription (\ref{4.16}) so that its
associated solution $\tilde R(r)$ of the wave equation(\ref{3.62}) has an extension again
of the order of magnitude of the starting value $r_*\approx a_{\rm B}$. The question is
now, how does such a potential ${}^{[1]}\!A_0(r)$ look like? Especially, is there some
similarity to the original Coulomb potential?

\indent In order to answer these questions, one simply calculates exactly the integral in
(\ref{4.16}) (see ref.\cite{GrRy}) and finds the following form:
\begin{equation}\label{4.17} {}^{[1]}\!A_0(r) = \alpha_{\rm S} \Big(\frac{2}{r_*}\Big)^2
  \int\limits_{r'=0}^{\infty} dr' \;\frac{r'\cdot
    \exp\big[-2\frac{r'}{r_*}\big]}{\sqrt{r'^2+r^2}} = f_{SN}\Big(\frac{2r}{r_*}\Big)\cdot
  \frac{\alpha_{\rm S}}{r} \ ,
\end{equation}
where the Struve-Neumann {\it screening factor} $f_{SN}$ is given by
\begin{equation}\label{4.18}
  f_{SN}\Big(\frac{2r}{r_*}\Big) = \Big(\frac{2r}{r_*}\Big)^2 \cdot \Big\{\frac{\pi}{2}
  \Big[I\!\!H_1  \Big (\frac{2r}{r_*}\Big ) - N_1\Big(\frac{2r}{r_*}\Big)\Big ]-1\Big\} \ .
\end{equation}
Here $I\!\!H_1(z)$ is the first {\it Struve function} \cite{GrRy}
\begin{equation}\label{4.19}
  I\!\!H_1(z) = \frac{2}{\pi} - \sum\limits_{n=0}^{\infty} \frac{(-1)^n
    \big(\frac{z}{2}\big)^n}{\big(n+\frac{1}{2}\big)
    \cdot \Gamma\big(n+\frac{1}{2}\big)^2} \doteqdot \frac{2}{\pi}-E_1(z)
\end{equation}
and $N_1(z)$ is the first {\it Neumann function} which admits a similar expansion as the
Struve function. But the physically important point with the Struve-Neumann potential
(\ref{4.17}) is its asymptotic behaviour for $r\rightarrow 0$ (origin) and $r\rightarrow
\infty$. For the latter case, one has the following expansion for the first {\it Weber
  function} $E_1(z)$ \cite{GrRy}
\begin{equation}\label{4.20}
E_1(z) = -N_1(z) - \frac{2}{\pi z^2} + \mathcal O\Big(\frac{1}{\vert z \vert^4}\Big) \ .
\end{equation}
Therefore the screening factor $f_{SN}$ (\ref{4.18}) becomes in the asymptotic region ($\vert z \vert \rightarrow \infty$)
\begin{multline}\label{4.21}
f_{SN}(z)=z^2 \bigg\{\frac{\pi}{2}\Big[\Big(\frac{2}{\pi}-E_1(z)\Big)-N_1(z)\Big]-1\bigg\} \\
=-\frac{\pi}{2}z^2\Big[E_1(z) + N_1(z)\Big] =1+\mathcal O\Big(\frac{1}{\vert z\vert^2}\Big)\ .
\end{multline}
But since the screening factor becomes thus unity far away from the origin, the
Struve-Neumann potential ${}^{[1]}\!A_0(r)$ (\ref{4.17}) actually adopts the exact Coulomb
form (\ref{3.52}) for $r\rightarrow\infty$. Clearly, this result is just what must be
expected if the integral in (\ref{4.17}) is correctly calculated; but the important point
refers of course to the origin ($r\rightarrow 0$).

\indent For $\vert z \vert \rightarrow 0$, the Struve and Neumann functions have the following behaviour \cite{GrRy}
\begin{subequations}\label{4.22}
\begin{align}
I\!\!H_1(z) &= o(\vert z\vert^2)\\
N_1(z)&=-\frac{2}{\pi z} + \frac{z}{\pi} \ln \frac{z}{2} + o(\vert z \vert) \ ,
\end{align}
\end{subequations}
and therefore the Struve-Neumann screening factor $f_{SN}(z)$ (\ref{4.18}) becomes in the vicinity of the origin
\begin{equation}\label{4.23}
f_{SN}(z) = z-z^2-\frac{z^3}{2}\ln \frac{z}{2} + o(\vert z\vert^3) \ .
\end{equation}
Thus the screening factor weakens somewhat the Coulomb singularity at the origin, but does
not eliminate it completely: The Struve-Neumann potential ${}^{[1]}\!A_0(r)$ (\ref{4.17})
behaves at the origin in the following way ($r \ll r_*$)
\begin{equation}
  \label{4.24} {}^{[1]}\!A_0(r)=\frac{2\alpha_{\rm S}}{r_*} \Big( 1-
  \frac{2r}{r_*} - \frac{1}{2}\Big(\frac{2r}{r_*}\Big)^2\ln\Big(\frac{r}{r_*}\Big) +
  o(r^2)\Big) \ .
\end{equation}
This result says that the potential remains finite at the origin
\begin{equation}\label{4.25}
{}^{[1]}\!A_0(0) = \frac{2\alpha_{\rm S}}{r_*} \ ,
\end{equation}
and this holds also for the radial component ${}^{[1]}\!E_r$ of the corresponding electric
field ${}^{[1]}\!\vec E_(\vec r) = -\vec \nabla {}^{[1]}\!A_0(r)$, cf. (\ref{3.14}):
\begin{equation}\label{4.26}
\lim\limits_{r\rightarrow 0} {}^{[1]}\!E_r(r) = \alpha_{\rm S}\Big(\frac{2}{r_*}\Big)^2 \ ,
\end{equation}
see fig.1.

\indent Such a less singular behaviour at the origin ($r=0$) has a very important
consequence. The field energy $\hat E_R^{(e)}$ being carried by the electrostatic field
${}^{[1]}\!A_0(r)$ is given by the integral of the corresponding energy density as
\cite{BeSo,BeSo2}
\begin{equation}\label{4.27}
  \hat E_R^{(e)} = - \frac{\hbar c}{\vpas} \int d^3\vec r \; \big \vert\!\big\vert \vec
  \nabla \, {}^{[1]}\!A_0(r) \big\vert\!\big\vert^2 \ ,
\end{equation}
see also the energy functional $E_T$ below. However, if the present interaction potential
${}^{[1]}\!A_0(r)$ would exactly obey the Coulomb form with its singularity ($\approx
1/r$) at the origin, the energy integral (\ref{4.27}) would diverge and thus the
positronium groundstate energy would be infinite. Therefore, the partial regularization of
the Coulomb interaction potential is necessary in order to equip the RST field
configuration with a well-defined energy content. But this point requires now a more
thorough inspection.

\section{Energy Functional}
\indent Quite generally speaking, the energy carried by a bound field configuration
provides an ideal handle for testing the physical relevance of any theory because the
atomic and molecular energy levels are immediately accessible to spectroscopic
observation. For the present situation, the {\it total energy} $E_T$ of a bound RST field
configuration appears as the sum of a {\it matter part} ($E_D$, say) and of a {\it gauge
  field part} ($E_G$), i.e.
\begin{equation}\label{5.1}
E_T=E_D+E_G \ .
\end{equation}
All three kinds of energy are calculated as the spatial integrals of the corresponding
energy densities $T_{00}(\vec r)$, i.e.
\begin{subequations}\label{5.2}
\begin{align}
E_T &\doteqdot \int d^3\vec r \; \TToo(\vec r) \label{5.2a}\\
E_D &\doteqdot \int d^3\vec r \; \DToo \label{5.2b}\\
E_G &\doteqdot \int d^3\vec r \; \GToo \label{5.2c}\ ,
\end{align}
\end{subequations}
where these energy densities are nothing else than the time components of the
corresponding energy-momentum densities $\TTmunu$, $\DTmunu$ and $\GTmunu$, resp. (see the
extended discussions of this in the preceding papers).

\subsection{Gauge Field Energy $E_G$}
\indent On the other hand, these energy-momentum densities $\Tmunu$ are built up by the
corresponding matter field ($\Psi$) and gauge field ($\MFmunu$) so that, e.g., the
electromagnetic gauge field energy $E_G$ will appear as a sum of the energy content $\hat
E_R^{(e)}$ due to the {\it electric} fields $\vec E_a(\vec r)$ ($a=1,2$) and of the energy
content $\hat E_R^{(m)}$ due to the {\it magnetic} fields $\vec H_a(\vec r)$:
\begin{equation}\label{5.3}
  E_G \Rightarrow \hat E_R = \hat E_R^{(e)} + \hat E_R^{(m)} = \frac{\hbar c}{\vpas} \int
  d^3\vec r \; 
\big\{ \vec E_1(\vec r)\sdot \vec E_2(\vec r) + \vec H_1(\vec r) \sdot \vec H_2(\vec r) \big\} \ ,
\end{equation}
cf. (\ref{4.27}). (For {\it identical} particles, the gauge field energy $E_G$ contains
also the exchange energy $E_C$, which however is missing here because positronium consists
of two {\it non-identical} particles).

\indent An important point with the electric type of gauge field energy $\hat E_R^{(e)}$
(\ref{5.3}) refers now to the fact that this quantity can be recast also in an alternative
form in terms of currents and potentials. To this end, one reconsiders the electrostatic
Poisson equation (\ref{3.17}) from which the following identity can be deduced, namely by
multiplying through with the potential ${}^{(p)}\!A_0(\vec r)$ and integrating over, under
use of Gau\ss' integral theorem:
\begin{equation}\label{5.4}
  N_G\doteqdot \int d^2\vec r \; \Big\{ r \sin\vartheta \,\big\vert\!\big\vert \vec\nabla 
  \,{}^{(p)}\!A_0(r,\vartheta) \big\vert\!\big\vert^2 - \alpha_{\rm S} {}^{(p)}\!A_0(\vec
  r)\cdot (\tilde R_+^{\;2}+\tilde R_-^{\;2}+\tilde S_+^{\;2}+\tilde S_-^{\;2})\Big\} \equiv 0 \ .
\end{equation}
Here, the specific relativistic form of the charge density ${}^{(p)}\!j_0(\vec r)$
(\ref{4.2}) has been used, which of course for our present non-relativistic purposes may
be approximated as shown by equation (\ref{4.4}). As we shall readily see, this {\it
  Poisson identity} (\ref{5.4}) turns out to be crucial for the subsequent principle of
minimal energy.

\subsection{Matter Energy $E_D$}
\indent The matter energy $E_D$ is a somewhat more complicated object than the gauge field
energy $E_G$. First, since the matter energy density $\DToo$ is built up by the time
derivatives of the matter field $\Psi$, cf. (\ref{2.19'})
\begin{equation}\label{5.5}
  \DToo = \frac{i\hbar c}{2} \Big[ \bar \Psi \textnormal I \! \Gamma_0 \big(\mathcal D_0
  \Psi \big) - \big(\mathcal D_0 \bar \Psi \big) \textnormal I \! \Gamma_0 \Psi \Big] \ ,
\end{equation}
and since furthermore the time derivatives of $\Psi$ (\ref{3.1a})-(\ref{3.1b}) must
explicitly contain the mass eigenvalues $M_a$ ($\Rightarrow \pm M_*$), the matter energy
$E_D$ (\ref{5.2b}) of the two-particle system must necessarily be built up by twice the
one-particle mass eigenvalue $M_*$ \cite{BeSo,BeSo2}:
\begin{equation}\label{5.6}
E_D=2\, M_* c^2 -2\, M_R^{(e)}c^2 \ ,
\end{equation}
where the mass equivalent ($M_R^{(e)} c^2$) of the electrostatic gauge field energy $\hat
E_R^{(e)}$ (\ref{5.3}) is defined through
\begin{equation}\label{5.7}
  M_R^{(e)}c^2 = - \frac{\hbar c}{2} \int d^2\vec r\; {}^{(p)}\!A_0(\vec r) \big\{\tilde
  R_+^{\; 2} + \tilde R_-^{\; 2} + \tilde S_+^{\; 2} + \tilde S_-^{\; 2}\big\} \ .
\end{equation}
Indeed, the Poisson identity (\ref{5.4}) ensures just the identity of the electrostatic
gauge field energy $\hat E_R^{(e)}$ (\ref{5.3}) and its mass equivalent $M_R^{(e)} c^2$
(\ref{5.7})! Now the reason, why the electrostatic interaction energy $\hat E_R^{(e)}$ (or
its mass equivalent, resp.) must be subtracted from the mass eigenvalue $M_*c^2$ (for any
particle) becomes immediately clear: since the mass functional $M_*c^2$ (see below) is
composed of the rest mass energy ($Mc^2$), the kinetic energy ($T_{kin}$) and the
electrostatic interaction energy $M_R^{(e)}c^2$ one has to subtract the latter (as a
two-particle effect) from the mass functional $M_*c^2$ in order to find the {\it
  two-particle} matter energy $E_D$ (\ref{5.6}) as the sum of pure {\it one-particle}
contributions! Clearly it is very pleasant to see this subtraction mechanism working
automatically by the right choice of the energy-momentum density $\DTmunu$ (\ref{2.19'})
for the matter field $\Psi$.

\indent But with both constituents $E_D$ and $E_G$ of the total energy $E_T$ (\ref{5.1})
being elaborated, one may now face the question whether perhaps this energy functional
$E_T$ adopts its minimally possible value just for the solutions of the RST eigenvalue
equations (\ref{3.47a})-(\ref{3.47d})?

\subsection{Action Principle}
\indent The preceding RST dynamics, as a coupled set of matter and gauge field equations,
can also be deduced from a variational principle. The existence of such a principle is
important for finding approximative (variational) solutions to the eigenvalue equations
for those situations where it is hard to find the exact solutions. Indeed, we will
subsequently present such a variational solution for the positronium groundstate. This
will not only support the claim that exact solutions of the RST eigenvalue system do
really exist, but it yields also valuable hints on further improvements of those
variational techniques.

\indent Mostly, the action principles are based upon a {\it Lagrangean density}, here
$\LRST[\Psi,\MAmu]$, which is integrated over some space-time region in order to yield the
action integral ($W_{\rm RST}$)
\begin{equation}\label{5.8}
W_{{\rm RST}}=\int d^4x\;\LRST[\Psi,\MAmu] \ .
\end{equation}
The extremalization of this action integral with respect to the matter field $\Psi(x)$
then reproduces the matter field equations (here the two-particle Dirac equation
(\ref{2.17})); and similarly the variation of $W_{RST}$ with respect to the bundle
connection $\MAmu(x)$ yields the gauge field dynamics (i.e. the non-Abelian Maxwell
equations (\ref{2.35})). For the present case of RST, the Lagrangean $L_{RST}$ splits up
into two parts, namely the matter part ($\LD$) and the gauge field part ($\LG$):
\begin{equation}\label{5.9}
\LRST[\Psi, \MAmu] = \LD[\Psi] + \LG[\MAmu] \ .
\end{equation}
Here the matter part is given by \cite{SSMS}
\begin{equation}\label{5.10}
\LD[\Psi] =\frac{i \hbar c}{2} \Big[\bar\Psi \DGmo \big(\MDmu \Psi\big) - \big(\MDmu\bar
\Psi \big ) \DGmo \Psi\Big] \ ,
\end{equation}
and the gauge field part looks as follows
\begin{align}\label{5.11}
\LG[\MAmu] &=\frac{\hbar c}{16\pi\alpha_{\rm S}} K_{\alpha\beta} \Falomunu \Fbomono \\
&\Rightarrow \frac{\hbar c}{16\pi\alpha_{\rm S}} \sum\limits^2_{a,b=1} K_{ab} F^a_{\;\;
\mu\nu} F^{b\mu\nu} \nonumber \ .
\end{align}
Observe here that the four-dimensional structure algebra $\mathfrak u(2)$ is not sweeped
out completely by the positronium curvature $\MFmunu$ because the two-particle structure
group $U(2)$ becomes reduced to $U'(2) = U(1) \times U(1)$, see the discussion of the
reduced Maxwell equations (\ref{2.43a})-(\ref{2.43b}).

\indent In order to be convinced that the present positronium eigenvalue systems
(\ref{3.47a})-(\ref{3.47d}) and (\ref{3.50a})-(\ref{3.50d}) actually do emerge as the
Euler-Lagrange variational equations from the claimed action integral
(\ref{5.8})-(\ref{5.11}), one substitutes the stationary ansatz (\ref{3.1a})-(\ref{3.1b})
into the matter Lagrangean $\LD$ (\ref{5.10}), uses the decomposition (\ref{3.4}) of the
Dirac spinors into Pauli spinors together with the ansatz (\ref{3.46a})-(\ref{3.46b}) for
those Pauli spinors, and thus finds the matter Lagrangean $\LD[\Psi]$ being splitted up
into three parts: kinetic term ($\LD^{(kin)}$) and electric ($\LD^{(e)}$) plus magnetic
($\LD^{(m)}$) interaction term, i.e.
\begin{equation}\label{5.12}
L_{D}[\Psi] = \LD^{(kin)}+ \LD^{(e)} + \LD^{(m)} \ .
\end{equation}
Here the electromagnetic interaction terms are the simpler ones and are given by
\begin{subequations}\label{5.13}
\begin{align}
\LD^{(e)} &= 2\,[M_*c^2 + \hbar c \, {}^{(p)}\!A_0(\vec r) ] \cdot {}^{(b)}\!k_0(\vec r) \label{5.13a} \\
\LD^{(m)} &= \pm 2\, \hbar c \,{}^{(b/p)}\!A_{\phi} \cdot {}^{(p/b)}\!k_\phi(\vec r) \label{5.13b} \ ,
\end{align}
\end{subequations}
where the common mass eigenvalue $M_*$ is included in the electric part (\ref{5.13a});
furthermore the upper/lower sign refers to the ortho/para case, and the common charge and
current densities ${}^{(b)}\!k_0(\vec r)$ (\ref{3.18}) and ${}^{(p/b)}\!k_\phi(\vec r)$
(\ref{3.19}), (\ref{3.36}) are given in terms of the wave amplitudes $\tilde R_\pm$,
$\tilde S_\pm$ by
\begin{subequations}\label{5.14}
  \begin{align}
    {}^{(b)}\!k_0(\vec r) &= \frac{\tilde R_+^2 + \tilde R_-^2 + \tilde S_+^2 + \tilde
      S_-^2}{4\pi\, r\,\sin\vartheta} \label{5.14a}\\[1ex]
    {}^{(b/p)}\!k_\phi(\vec r) &= \frac{\sin\vartheta\, [\tilde R_+ \tilde R_- - \tilde
      S_+ \tilde S_-] - \cos\vartheta\, [\tilde S_+ \tilde R_- + \tilde R_+ \tilde
      S_-]}{2\pi\, r\,\sin\vartheta} \label{5.14b} \ .
\end{align}
\end{subequations}
In contrast to these interaction terms of the matter part $\LD$ (\ref{5.12}), its kinetic
contribution $\LD^{(kin)}$ looks somewhat more complicated but is the same for the ortho-
and para-configurations:
\begin{multline}\label{5.15}
  L_D^{(kin)}=\frac{\hbar c}{2\pi\, r\,\sin\vartheta} \Big[\tilde R_+
  \cdot\big(\frac{\partial\tilde R_-}{\partial r}-\frac{1}{r}\frac{\partial\tilde
    S_-}{\partial \vartheta}\big) - \tilde R_-\cdot \big(\frac{\partial\tilde
    R_+}{\partial r}+\frac{1}{r}\frac{\partial\tilde S_+} {\partial \vartheta}\big) \\
  +\tilde S_+\cdot\big(\frac{\partial\tilde S_-}{\partial
    r}+\frac{1}{r}\frac{\partial\tilde R_-}{\partial \vartheta}\big) - \tilde S_-\cdot
  \big(\frac{\partial\tilde S_+}{\partial r}-\frac{1}{r}\frac{\partial\tilde R_+}{\partial \vartheta}\big)\\
  +\frac{1}{r}\big(\tilde R_+\cdot \tilde R_- + \tilde S_+\cdot \tilde S_- \big) -
  \frac{Mc}{\hbar} \big(\tilde R_+^{\; 2} - \tilde R_-^{\; 2} + \tilde S_+^{\; 2} - \tilde
  S_-^{\; 2}) \Big] \ .
\end{multline}

\indent But now that the matter Lagrangean $\LD$ (\ref{5.12}) is explicitly known, one
substitutes this into the action integral $W_{{\rm RST}}$ (\ref{5.8}), which itself splits
up into two parts according to the splitting of the Lagrangean $\LRST$ (\ref{5.9})
\begin{subequations}\label{5.16}
\begin{align}
W_{{\rm RST}} &= W_D + W_G \label{5.16a}\\
W_{{\rm D}} &= \int d^4x \; \LD[\Psi] \label{5.16b}\\
W_{{\rm G}} &= \int d^4x \; \LG[\MAmu] \label{5.16c}\ ,
\end{align}
\end{subequations}
and then the variational procedure for the matter action $W_{\rm D}$ (\ref{5.16b}) with
respect to the wave amplitudes $\tilde R_\pm, \tilde S_\pm$ just yields the mass
eigenvalue equations (\ref{3.47a})-(\ref{3.47d}) for the para-case and
(\ref{3.50a})-(\ref{3.50d}) for the ortho-case. Here, there is a nice consistency check
which, however, has a deeper meaning. Namely, the value of the matter action functional
$W_{\rm D}$ (\ref{5.16b}) upon the solutions of the mass eigenvalue equations must vanish
because the latter are linear and homogeneous. Indeed, for any solution of such type of
equations one can easily generate new solutions by simple muliplication by some constant
$c_*$ (i.e.  $\tilde R_\pm \Rightarrow c_*\cdot \tilde R_\pm$, $\tilde S_\pm \Rightarrow
c_* \cdot \tilde S_\pm$). But this process of multiplication cannot change the value of
the matter functional $W_{\rm D}$ (\ref{5.16b}) since the multiplied wave amplitudes are
also solutions of the variational procedure and do continously approach the original
solution for the constant $c_*$ tending to unity ($c_* \rightarrow 1$). And actually,
substituting the derivatives of the wave amplitudes from the mass eigenvalue equations
into the matter functional $W_{\rm D}$ (\ref{5.16b}) lets this functional adopt the value
of zero! This is the reason why the present action principle ($\delta W_{\rm D}=0$) for
the matter fields $\tilde R_\pm, \tilde S_\pm$ does {\it not} require some constraint upon
the latter fields (e.g. normalization condition)!

\indent This situation changes if one considers now the gauge fields, i.e. the
electrostatic potential ${}^{(p)}\!A_0(\vec r)$ (its magnetic counterpart is not
considered here because we restrict ourselves to the electrostatic approximation). The
field equation for ${}^{(p)}\!A_0(\vec r)$ is the Poisson equation (\ref{3.17}) which is
to be deduced from the variational principle for the RST action integral (\ref{5.8}). Here
it is sufficient to take into account only those parts $W_{{\rm RST}}^{(e)}$ of $W_{{\rm
    RST}}$ which contain the electric potential ${}^{(p)}\!A_0(\vec r)$, i.e.
\begin{equation}\label{5.17}
W_{{\rm RST}}^{(e)} = W_{{\rm D}}^{(e)} + W_{{\rm G}}^{(e)}
\end{equation}
with
\begin{subequations}\label{5.18}
\begin{align}
W_{{\rm D}}^{(e)}&=\int d^4x\; \LD^{(e)} \label{5.18a} \\
W_{{\rm G}}^{(e)}&=\int d^4x\; \LG^{(e)} \label{5.18b} \ ,
\end{align}
\end{subequations}
where the electric part $\LD^{(e)}$ of the matter Lagrangian $\LD$ is given by equation
(\ref{5.13a}); and similarly the electric part of the gauge field Lagrangean $L_G$
(\ref{5.11}) is given by
\begin{equation}\label{5.19}
\LG^{(e)} = - \frac{\hbar c}{\vpas} \big\vert\!\big\vert\vec\nabla{}^{(p)}\!A_0(\vec r) \big\vert\!\big\vert^2 \ .
\end{equation}

\indent The important difference between the matter and gauge fields is now that the
Poisson equation (\ref{3.17}) is not homogeneous, and therefore one cannot multiply its
solution ${}^{(p)}\!A_0(\vec r)$ by some constant $c_*$ in order to obtain a further
solution $c_* \cdot {}^{(p)}\!A_0(\vec r)$. It is just this peculiarity of the gauge
potential ${}^{(p)}\!A_0(\vec r)$ which generates the former Poisson constraint
(\ref{5.4})! In order to become convinced of this, substitute the modified solution
$c_*\cdot {}^{(p)}\!A_0(\vec r)$ into the electric part $W^{(e)}_{{\rm RST}}$ (\ref{5.17})
of the action integral so that this quantity becomes an ordinary function of the constant
$c_*$: $W_{{\rm RST}}^{(e)}(c_*)$. Since the original potential ${}^{(p)}\!A_0(\vec r)$ is
presumed to extremalize the electric action $W_{{\rm RST}}^{(e)}$ (\ref{5.17}), one is led
to the following conclusion
\begin{equation}\label{5.20}
\frac{dW_{{\rm RST}}^{(e)}(c_*)}{dc_*}\Big\vert_{c_*=1}=0 \ ,
\end{equation}
and this is found to be nothing else than the Poisson identity (\ref{5.4}). Thus, through
the present deduction, the Poisson identity may be interpreted to be a {\it constraint}
for the variational procedure (see below for the use of this constraint in connection with
the extremalization of the binding energy).

\subsection{Mass Functional}
\indent Surely, it is a very pleasant feature of a theory if it can be based upon an
action principle (here: $\delta W_{{\rm RST}} = 0$); but with reference to the treatment
of bound systems the action integral $W$ is not immediately linked to the binding energy
which itself, on the other hand, is the quantity of physical interest. Therefore the
question must necessarily arise whether, besides the action $W_{{\rm RST}}$, the binding
energy perhaps does also adopt an extremal value for the bound states ($\delta E_T = 0$)?
If so, how does the corresponding {\it energy principle} look like and what is its
relation to the {\it action principle}? The answer to these questions is most conveniently
obtained by first considering the {\it mass functional} ($M_T c^2$, say).

\indent For any solution of the mass eigenvalue equations (\ref{3.47a})-(\ref{3.47d}) or
(\ref{3.50a})-(\ref{3.50d}), resp., the mass eigenvalue $M_*c^2$ adopts a well-defined
real number, and this number can be represented in terms of the wave amplitudes $\tilde
R_\pm$, $\tilde S_\pm$ by multiplying through the eigenvalue equations by the associate
amplitudes, integrating over whole three-space and adding up all four contributions. After
the mass eigenvalue $M_*c^2$ has been represented in this way, it may be reinterpreted
also as the value of the corresponding functional $M_*c^2 \Rightarrow M_Tc^2 [\tilde
R_\pm, \tilde S_\pm ]$ upon the solutions $\tilde R_\pm, \tilde S_\pm$ of the mass
eigenvalue problem.  Thus the desired mass functional $M_Tc^2$ is found to appear in the
following general form \cite{BeSo,BeSo2}
\begin{equation}\label{5.21}
M_Tc^2 = \mathcal Z^2\cdot Mc^2 + 2\,(T_r+T_\vartheta) + M_R^{(e)} c^2 + M_R^{(m)}c^2 \ .
\end{equation}

\indent Here, the meaning of the four constituents is the following: First, it must be
mentioned that the constraint of wave function normalization has to be applied in its fully
relativistic form
\begin{equation}\label{5.22}
  1=\int d^3\vec r\; {}^{(p)}\!j_0(\vec r) \equiv \frac{1}{2}\int d^2\vec r \;\{ \tilde
  R_+^{\; 2} + 
\tilde R_-^{\; 2} + \tilde S_+^{\; 2} + \tilde S_-^{\; 2} \} \ ,
\end{equation}
whose non-relativistic approximation has already been specified by the preceding equation
(\ref{4.5}). Next, the first term on the right-hand-side of equation (\ref{5.21}) is a
kind of modified rest mass energy ($Mc^2$) where the {\it renormalization factor}
$\mathcal Z^2$ is given by
\begin{equation}\label{5.23}
\mathcal Z^2 =\frac{1}{2}\int d^2\vec r \;\{ \tilde R_+^{\; 2} - \tilde R_-^{\; 2} +
\tilde S_+^{\; 2} - \tilde S_-^{\; 2} \} \ .
\end{equation}
Furthermore, the relativistic form of the {\it kinetic energy} $T_{kin} = T_r +
T_\vartheta$ reads in terms of the wave amplitudes $\tilde R_\pm, \tilde S_\pm$
\begin{subequations}\label{5.24}
\begin{align}
  T_r &= \frac{\hbar c}{4} \int d^2\vec r \; \Big[\tilde R_-\cdot \frac{\partial \tilde
    R_+}{\partial r} - \frac{\tilde R_+}{r}\cdot \frac{\partial (r \tilde R_-)}{\partial
    r} +\tilde S_-\cdot \frac{\partial \tilde S_+}{\partial r} - \frac{\tilde
    S_+}{r}\cdot \frac{\partial (r \tilde S_-)}{\partial r} \Big ] \label{5.24a} \\
  T_\vartheta &= \frac{\hbar c}{4} \int d^2\vec r \; \frac{1}{r} \Big[\tilde
  R_-\cdot\frac{\partial \tilde S_+}{\partial \vartheta} - \tilde S_+ \cdot \frac{\partial
    (\tilde R_-)}{\partial \vartheta} + \tilde R_+\cdot \frac{\partial \tilde
    S_-}{\partial \vartheta} - \tilde S_- \cdot\frac{\partial (\tilde R_+)}{\partial
    \vartheta} \Big ] \ . \label{5.24b}
\end{align}
\end{subequations}
Finally, the last two terms are the mass-energy equivalents of the energy $\hat E_R$
contained in the electromagnetic field modes, i.e. for the electric type
\begin{align}\label{5.25}
M_R^{(e)}c^2 &= - \frac{\hbar c}{2} \int d^2\vec r\; {}^{(p)}\!A_0 \{ \tilde R_+^{\; 2} + \tilde R_-^{\; 2} + \tilde S_+^{\; 2} + \tilde S_-^{\; 2} \} \nonumber \\
&\equiv -\hbar c \int d^3\vec r\; {}^{(p)}\!A_0(\vec r) \cdot {}^{(b)}\!k_0(\vec r) \ ,
\end{align}
and similarly for the magnetic type
\begin{align}\label{5.26}
  M_R^{(m)}c^2 &= \mp \hbar c \int d^2\vec r\; {}^{(b/p)}\!A_\phi \Big\{ \sin\vartheta
  \big[\tilde R_+\tilde R_- - \tilde S_+ \tilde S_- \big ] -\cos\vartheta \big[\tilde
  S_+\tilde R_- + \tilde R_+ \tilde S_- \big ]\Big\} \nonumber \\
  &\equiv \mp \hbar c \int d^3\vec r\; \vec A_{b/p}(\vec r) \cdot \vec k_{p/b} \ .
\end{align}
Clearly, the upper/lower case of the latter equation refers again to the ortho/para-positronium. 

\indent But with the mass functional $M_T c^2$ being explicitly known, it is a standard
variational procedure to actually deduce thereof the mass eigenvalue equations
(\ref{3.47a})-(\ref{3.47d}) or (\ref{3.50a})-(\ref{3.50d}), resp. Clearly, the
relativistic normalization condition (\ref{5.22}) has to be respected here, but this can
be easily done via the method of Lagrangean multipliers (see below). The mass functional
$M_Tc^2$ (\ref{5.21}) is not identical with the desired energy functional, but its closer
inspection will help to construct the latter one.

\indent It seems very natural that the mass functional $M_Tc^2$ (\ref{5.21}) consists of
an electric and magnetic interaction term, and one expects that the other two terms should
represent the rest mass and kinetic energies, resp. But why does the kinetic energy $T_{kin}$
($=T_r + T_\vartheta$) appear with a pre-factor of two, and why does there emerge a
renormalization factor $\mathcal Z^2$ in front of the rest mass term $Mc^2$ ? These
questions are clarified most transparantly by passing to the non-relativistic limit of the
mass functional. For this purpose, one eliminates the negative Pauli amplitudes $\tilde
R_-, \tilde S_-$ by expressing them approximately in terms of their positive counterparts
$\tilde R_+, \tilde S_+$ via the mass eigenvalue equations, i.e. one puts \cite{BeSo}
\begin{subequations}\label{5.27}
\begin{align}
  \tilde R_- &\approx \frac{\hbar}{2\,Mc} \Big\{ \frac{\partial \tilde R_+}{\partial r} +
  \frac{1}{r}\frac{\partial \tilde S_+}{\partial \vartheta} \Big\} \label{5.27a} \\
  \tilde S_- &\approx \frac{\hbar}{2\,Mc} \Big\{ \frac{\partial \tilde S_+}{\partial r} -
  \frac{1}{r}\frac{\partial \tilde R_+}{\partial \vartheta} \Big\} \label{5.27b} \ ,
\end{align}
\end{subequations}
and if this is substituted into the relativistic kinetic energies $T_r$ (\ref{5.24a}) and
$T_\vartheta$ (\ref{5.24b}) one obtains their non-relativistic approximations
${}^{(r)}\!E_{kin}$ and ${}^{(\vartheta)}\!E_{kin}$ as
\begin{subequations}\label{5.28}
\begin{align}
  T_r \Rightarrow {}^{(r)}\!E_{kin} &= \frac{\hbar^2}{4M} \int d^2\vec
  r\;\Big\{\Big(\frac{\partial \tilde R_+}{\partial r}\Big)^2 + \Big(\frac{\partial \tilde
    S_+}{\partial r}\Big)^2 \Big\} \label{5.28a} \\
  T_\vartheta \Rightarrow {}^{(\vartheta)}\!E_{kin} &= \frac{\hbar^2}{4M} \int d^2\vec
  r\;\Big\{\Big(\frac{1}{r} \frac{\partial \tilde R_+}{\partial \vartheta}\Big)^2 +
  \Big(\frac{1}{r}\frac{\partial
    \tilde S_+} {\partial \vartheta}\Big)^2 \Big\} \label{5.28b} \\
  &\qquad + \frac{\hbar^2}{4M} \int\!\!\!\!\int dr \, d\vartheta \; \Big[\frac{\partial
    \tilde R_+}{\partial r} \cdot \frac{\partial \tilde S_+}{\partial \vartheta} -
  \frac{\partial \tilde S_+}{\partial r}\cdot \frac{\partial \tilde R_+}{\partial
    \vartheta} \Big] \ . \nonumber
\end{align}
\end{subequations}
Thus the non-relativistic approximation $E_{kin}$
($={}^{(r)}\!E_{kin}+{}^{(\vartheta)}\!E_{kin}$) of the relativistic kinetic energy
$T_{kin}$ ($=T_r + T_\vartheta$) is found to appear in the following form
\begin{equation}\label{5.29}
  E_{kin} = \frac{\hbar^2}{4M} \int d^2\vec r \; \Big\{ \big\vert\!\big\vert\vec\nabla
  \tilde R_+\big\vert\!\big\vert^2+\big \vert\!\big\vert\vec\nabla\tilde S_+\big\vert\!\big\vert^2\Big\} + E_W \ .
\end{equation}
Here the {\it "winding energy"} $E_W$ is due to the map from the ($r,\vartheta$)-plane
into the ($\tilde R_+, \tilde S_+$)-space and is defined (as usual for winding numbers) in
terms of the corresponding Jacobian as
\begin{equation}\label{5.30}
  E_W\doteqdot\frac{\hbar^2}{4M} \int\!\!\!\!\int dr\,d\vartheta \; \frac{\partial 
(\tilde R_+,\tilde S_+)}{\partial (r,\vartheta)} \ .
\end{equation}
Indeed, apart from the emergence of this winding energy (to be neglected for the present
purposes), the kinetic energy $E_{kin}$ (\ref{5.29}) is just the result to be expected
from the non-relativistic approximation of the mass eigenvalue equations
\begin{subequations}\label{5.31}
\begin{align}
  -\frac{\hbar^2}{2M} \Big(\frac{\partial^2}{\partial r^2} +
  \frac{1}{r}\frac{\partial}{\partial r} + \frac{1}{r^2} \frac{\partial^2}{\partial
    \vartheta^2} \Big) \tilde R - \hbar c \, {}^{(p)}\!A_0\cdot \tilde R &= E_S\cdot \tilde R \label{5.31a} \\
  -\frac{\hbar^2}{2M} \Big(\frac{\partial^2}{\partial r^2} +
  \frac{1}{r}\frac{\partial}{\partial r} + \frac{1}{r^2} \frac{\partial^2}{\partial
    \vartheta^2} \Big) \tilde S - \hbar c \, {}^{(p)}\!A_0\cdot \tilde S &= E_S\cdot
  \tilde S \label{5.31b}
\end{align}
\end{subequations}

\indent Obviously, the spin-up ($\sim\tilde R$) and spin-down ($\sim\tilde S$) components
of the non-relativistic approximation are decoupled, because the magnetic interactions are
neglected (see equation (\ref{3.62}) for the spherically symmetric approximation hereof).
Concentrating here upon the spin-up solutions (\ref{5.31a}), one finds the Schr\"odinger
energy functional $E_S[\tilde R]$ by simply multiplying through that equation by $\tilde
R$ and integrating by parts as
\begin{equation}\label{5.32}
  E_S = \frac{\hbar^2}{4M}\int d^2\vec r\; \Big\{\Big(\frac{\partial \tilde R}{\partial
    r}\Big)^2 + \frac{1}{r^2}\Big(\frac{\partial \tilde R}{\partial \vartheta}\Big)^2
  \Big\} - \frac{\hbar c}{2}\int d^2\vec r\; {}^{(p)}\!A_0\cdot \tilde R^2 \ ,
\end{equation}
where the non-relativistic normalization condition has been applied, cf. (\ref{4.5})
\begin{equation}\label{5.33}
\int d^2\vec r \; \tilde R(r,\vartheta)^2 = 2 \ .
\end{equation}

\indent But on the other hand, this Schr\"odinger energy functional $E_S$ (\ref{5.32})
must turn out as the non-relativistic electrostatic approximation of the relativistic mass
functional $M_Tc^2$ (\ref{5.21}), where the non-relativistic form of the electric
mass-energy equivalent $M_R^{(e)} c^2$ (\ref{5.25}) in equation (\ref{5.32}) is
self-evident. However, the relativistic kinetic energy $T_{kin}$ ($=T_r+T_\vartheta$)
appears with a factor of two in the {\it relativistic} form (\ref{5.21}) and, on the other
hand, appears in the usual way in the {\it non-relativistic} form (\ref{5.32}), cf.
(\ref{5.29}). In order to clarify this final point, one subtracts the rest-mass energy
$Mc^2$ from the relativistic mass functional $M_Tc^2$ in order to obtain the
non-relativistic Schr\"odinger functional $E_S$
\begin{equation}\label{5.34}
E_S = M_Tc^2 - Mc^2 \ ,
\end{equation}
and furthermore one looks also for the non-relativistic approximation of the
renormalization factor $\mathcal Z^2$ (\ref{5.23}) by use of the approximations
(\ref{5.27a})-(\ref{5.27b}) which yields
\begin{equation}\label{5.35}
\mathcal Z^2 \cong 1 - \frac{E_{kin}}{Mc^2} \ ,
\end{equation}
with the non-relativistic $E_{kin}$ being given by equation (\ref{5.29}) apart from the
winding energy $E_W$. Consequently if this is substituted back into the relativistic mass
functional $M_Tc^2$ (\ref{5.21}), one finally gets by observation of the non-relativistic
form $E_{kin}$ (\ref{5.29}) and with the neglection of the magnetic contribution
$M_R^{(m)} c^2$ (\ref{5.26}) for the Schr\"odinger energy functional $E_S$ (\ref{5.34}):
\begin{align}\label{5.36}
  E_S &= (\mathcal Z^2-1) Mc^2 + 2\, E_{kin} + M_R^{(e)} c^2 = E_{kin} + M_R^{(e)}c^2 \\
  &= \frac{\hbar^2}{4M} \int d^2\vec r \;\big\vert \vec\nabla \tilde R\big\vert^2 -
  \frac{\hbar c}{2} \int d^2\vec r \; {}^{(p)}\!A_0(r) \tilde R^2 \ . \nonumber
\end{align}
Thus the non-relativistic approximation (\ref{5.36}) of the relativistic mass functional
$M_Tc^2$ yields exactly the Schr\"odinger energy functional $E_S$ (\ref{5.32}); and this
of course supports the confidence into the mass functional $M_Tc^2$ as the correct
relativistic generalization of the non-relativistic Schr\"odinger functional $E_S$.

\indent This viewpoint is even further supported by considering the variational equations
due to both energy/mass functionals. Namely, it is well-known that the ordinary
non-relativistic Schr\"odinger equation for a fixed potential ${}^{(p)}\!A_0$ arises by
extremalization of the functional $E_S$ under the constraint that the wave function is
normalized to unity (i.e. equation (\ref{5.33}) for the present situation), see any
textbook about quantum mechanics, e.g. ref.s\cite{Fl},\cite{Ba} for the Ritz variational
method. But if the present RST eigenvalue systems (\ref{3.47a})-(\ref{3.47d}), or
(\ref{3.50a})-(\ref{3.50d}), resp., do really represent the correct relativistic
generalization of the non-relativistic Schr\"odinger equation, then the general logic
demands that these relativistic mass eigenvalue equations must appear as the variational
equations due to the mass functional $M_Tc^2$ in the same sense as the ordinary
Schr\"odinger equation arises as the variational (i.e.  Euler-Lagrange) equation due to
the non-relativistic Schr\"odinger functional $E_S$! And indeed, if the corresponding
variational procedure for the relativistic functional $M_Tc^2$ (\ref{5.21}) is carried
through with regard of the original relativistic normalization condition (\ref{5.22}), one
arrives just at those RST eigenvalue systems for the ortho- and para-cases, resp. To be
more specific, multiply the normalization condition (\ref{5.22})
\begin{equation}\label{5.37}
  N_D \doteqdot \frac{1}{2} \int d^2\vec r \; \big\{\tilde R_+^{\; 2} + \tilde R_-^{\; 2}
  + \tilde S_+^{\; 2} + \tilde S_-^{\; 2} \big\} - 1 = 0
\end{equation}
by some Lagrangean multiplier $\lambda_D$ and add this to the mass functional $M_Tc^2$
(\ref{5.21}) in order to obtain the modified functional $\tilde M_Tc^2$
\begin{equation}\label{5.38}
\tilde M_T c^2 = M_Tc^2 + \lambda_D\cdot N_D \ .
\end{equation}
Now carry through the variational procedure for the modified functional $\tilde M_Tc^2$
and find the RST eigenvalue systems such as (\ref{3.47a})-(\ref{3.47d}) with merely the
mass eigenvalue $M_*$ being replaced by the Lagrangean multiplier $\lambda_D$, i.e.
\begin{equation}\label{5.39}
M_*c^2 \Rightarrow -\lambda_D \ .
\end{equation}

\indent Summarizing, one can be sure now that one has succeeded in constructing the
correct relativistic generalization of the ordinary, non-relativistic Schr\"odinger
approach; but one further generalizing step has to be done, and this refers to the
interaction potential between the two particles: wheras the ordinary Schr\"odinger
approach is based upon the instantaneous Coulomb potential (see the Schr\"odinger
Hamiltonian $\hat H$ (\ref{2.19''})), the relativistic RST interaction potential
${}^{(p)}\!A_0$ is a dynamical object which has to obey its own field equation (i.e. the
Maxwell equation (\ref{2.35})in the general case)! In this sense, RST is a more
fundamental framework than the ordinary Schr\"odinger quantum mechanics. But more
concretely, this means now that we finally have to generalize the preliminary functional
$\tilde M_Tc^2$ (\ref{5.38}) in such a way that also the Poisson equation (\ref{3.17}) for
the electric potential ${}^{(p)}\!A_0$ arises as an Euler-Lagrange equation due to that
desired more general functional.

\subsection{Principle of Minimal Energy}

Surely, it will not come as a surprise that the functional, which we are after, must have
something to do with the total energy $E_T$ (\ref{5.1}). Therefore it is instructive to
inspect this functional in greater detail, where the magnetic interactions may be
neglected again for the present purposes.

\indent First, observe here that the electrostatic gauge field energy $\hat E_R^{(e)}$
(\ref{5.3}) can be expressed exclusively in terms of the gradient of the electric
potential ${}^{(p)}\!A_0(\vec r)$
\begin{equation}\label{5.40}
\hat E_R^{(e)}=-\frac{\hbar c}{4\pi \alpha_{\rm S}} \int d^3\vec r \; \big\vert\!\big\vert
\vec \nabla \, {}^{(p)}\!A_0(\vec r) \big\vert\!\big\vert^2 \ ,
\end{equation}
because the electric field strength $\vec E_p$ appears as a gradient field whenever the
exchange interactions are missing, i.e. (cf. (\ref{3.14}))
\begin{equation}\label{5.41}
\vec E_p(\vec r) = -\vec\nabla \, {}^{(p)}\!A_0(\vec r)
\end{equation}
(for the generalization of this relationship in the presence of exchange interactions see
ref.\cite{PrMaSo}). Next, consider the contribution of the matter energy $E_D$ (\ref{5.6})
which also contains the electric potential ${}^{(p)}\!A_0(r)$, but in a form different
from the mass equivalent $M_R^{(e)} c^2$ (\ref{5.7}). Furthermore, the electric potential
enters also the mass functional $M_Tc^2$ (\ref{5.21}) again in form of the mass equivalent
$M_R^{(e)}c^2$. Therefore, in order to make the appearance of the electric potential in
the energy functional $E_T$ (\ref{5.1}) explicit, one writes down the latter functional
(under use of the mass functional $M_Tc^2$ (\ref{5.21})) as
\begin{align}\label{5.42}
E_T&=E_D+E_G = 2 (M_T c^2 - M_R^{(e)} c^2) +\hat E_R^{(e)} + \hat E_R^{(m)} \nonumber \\
&=2\, \mathcal Z^2\cdot Mc^2+4\, T_{kin}+ \hat E_R^{(e)} + \hat E_R^{(m)} + 2\, M_R^{(m)} c^2 \ .
\end{align}

\indent But this, indeed, is a rather amazing result because $E_T$ contains now the
electric potential ${}^{(p)}\!A_0(\vec r)$ exclusively in form of the {\it gauge field
  energy} $\hat E_R^{(e)}$ (\ref{5.40}), whereas the mass functional $M_T c^2$
(\ref{5.21}) contains the potential ${}^{(p)}\!A_0(\vec r)$ exclusively in form of the
{\it mass equivalent} $M_R^{(e)}c^2$ (\ref{5.25})! And this implies that neither the mass
functional $M_T$ (\ref{5.21}) nor the energy functional $E_T$ (\ref{5.42}) would lead us
to the electric Poisson equation (\ref{3.17}) as the Euler-Lagrange equation of the
corresponding variational procedure.

\indent Evidently, in order to escape from this dilemma, one has to put in here some new
idea. This is done now by imposing a second constraint, besides the normalization
condition (\ref{5.37}), namely the {\it Poisson identity} (\ref{5.4}). Consequently we
multiply both constraints (\ref{5.4}) and (\ref{5.37}) with the Lagrangean multipliers
$\lambda_G$ and $\lambda_D$, resp., and add this to the original energy functional $E_T$
(\ref{5.42}) in order to thus arrive at the modified energy functional $\tilde E_T$:
\begin{align}\label{5.43}
\tilde E_T &= E_T +2\, \lambda_D\cdot N_D + \lambda_G\cdot N_G \nonumber \\
&=2\, \mathcal Z^2 \cdot Mc^2 + 4\, T_{kin} + \hat E_R^{(e)} + 2\, \lambda_D\cdot N_D + \lambda_G\cdot N_G \ .
\end{align}
Here the renormalization factor $\mathcal Z^2$ is given by (\ref{5.23}), the kinetic
energy $T_{kin}$ ($=T_r+T_{\vartheta}$) by (\ref{5.24a})-(\ref{5.24b}), and finally the
gauge field energy $\hat E_R^{(e)}$ by (\ref{5.40}). Moreover, the magnetic terms are
omitted because we are satisfied for the moment with the electrostatic approximation.
However, carrying through now the variational procedure for the final result $\tilde E_T$
(\ref{5.43}) actually yields just the (fully relativistic) eigenvalue equations
(\ref{3.47a})-(\ref{3.47d}) or (\ref{3.50a})-(\ref{3.50d}), resp., with the first
Lagrangean multiplier $\lambda_D$ being given as before by equation (\ref{5.39}) and the
second one ($\lambda_G$) by
\begin{equation}\label{5.44}
\lambda_G=\frac{\hbar c}{\alpha_{\rm S}} \ .
\end{equation}
And additionally, the variational procedure for $\tilde E_T$ (\ref{5.43}) with respect to
the electric potential ${}^{(p)}\!A_0(\vec r)$ (and magnetic potential
${}^{(p/b)}\!A_\phi(\vec r)$) yields the desired Poisson equations, i.e. the electric one
(\ref{3.17}) and also its magnetic counterpart (to be suppressed here).

\indent Summarizing, it is possible to convert the general Hamiltonian-Lagrange action
principle ($\delta W_{\rm RST} = 0$) to an energy minimalization principle ($\delta \tilde
E_T = 0$) for the bound states. The corresponding Euler-Lagrange equations are just the
relativistic mass eigenvalue equations plus the Poisson equations for the interaction
potentials of electric and magnetic type. Clearly, such a principle of minimal energy for
bound states provides now a powerful approximation technique when looking for the binding
energy. We will readily exploit this pleasant result, but will be also satisfied with the
corresponding non-relativistic approximations.  To this end it is merely necassary to
resort to the non-relativistic approximation of those terms emerging in the relativistic
energy functional $\tilde E_T$ (\ref{5.43}).

\indent First, the non-relativistic forms of the renormalization factor $\mathcal Z^2$ and
the kinetic energy $T$ have already been specified by equations (\ref{5.29}) and
(\ref{5.35}), resp., where the spin-down component $\tilde S_+$ is put to zero so that the
winding energy $E_W$ (\ref{5.30}) does also vanish. Second, the constraint term $N_D$ due
to the normalization condition is omitted completely (also in its non-relativistic form
(\ref{5.33})) because as a possible trial function $\tilde R_(r)$ we resort to the type
(\ref{4.7}) which is a priori normalized to unity. However, in contrast to this, the
Poisson constraint term $N_G$ must be retained in any case. The reason is that if one
takes the value of the wanted functional $\tilde E_S$ upon some trial field
configuration, which is not an exact solution of the corresponding variational equations,
then the Poisson constraint keeps the (naturally inaccurate) trial value of $\tilde E_S$
in the neighborhood of its proper minimum (see the example below). Thus the
non-relativistic version $\tilde E_S$ of the energy functional $\tilde E_T$ (\ref{5.43})
is ultimately found as
\begin{align}\label{5.45}
  \tilde E_S &= 2\, E_{kin} + \hat E_R^{(e)} + 2\, (M_R^{(e)}c^2 - \hat E_R^{(e)} )\\
  &= \frac{\hbar^2}{2M} \int d^2\vec r \; \Big(\frac{d\tilde R(r)}{dr} \Big)^2 +
  \frac{\hbar c}{\alpha_{\rm S}} \int dr\, r^2 \Big(\frac{d {}^{(p)}\!A_0(r)}{dr}\Big)^2 -
  \hbar c\int dr\, r \, {}^{(p)}\!A_0(r) \tilde R(r)^2 \ . \nonumber
\end{align}

\indent Here it is very instructive to consider the difference between the conventional
Schr\"odinger one-particle functional $E_S$ (\ref{5.36}) and the present two-particle RST
result $\tilde E_S$ (\ref{5.45}). It is true, the kinetic two-particle energy ($2\,
E_{kin}$) is twice the conventional one-particle counterpart as displayed in equation
(\ref{5.36}), however the electrostatic interaction energy is described in a radically
different way: whereas the conventional Schr\"odinger approach $E_S$ (\ref{5.36}) relies
here exclusively upon the mass equivalent $M_R^{(e)}c^2$ (\ref{5.25}) in its
non-relativistic form, the RST counterpart $\tilde E_S$ (\ref{5.45}) describes the
interaction energy in terms of {\it both} the field energy $\hat E_R^{(e)}$ {\it and} its
mass equivalent $M_R^{(e)}c^2$! Clearly, for an exact solution of the RST energy
eigenvalue problem both contributions $\hat E_R^{(e)}$ and $M_R^{(e)} c^2$ are identical
(on account of the Poisson identity) so that they cancel for the non-relativistic RST
functional $\tilde E_S$ (\ref{5.45}) which then adopts the conventional Hartree-Fock form.
However for (non-exact) trial configurations, the difference of the field energy $\hat
E_R^{(e)}$ and its mass equivalent $M_R^{(e)}c^2$ becomes important for the degree of
accuracy of the corresponding trial value of $\tilde E_S$. This is the origin of the fact
that RST predictions have the potential to come closer to the experimental numbers than
the conventional Hartree-Fock approach. Clearly, such a pleasant result must be
exemplified now by a brief numerical treatment of the positronium groundstate.

\section{Numerical Demonstration}
\indent Being satisfied for the moment with the non-relativistic approximation, one
inserts the trial function $\tilde R(r)$ (\ref{4.7}) together with the Struve-Neumann
potential ${}^{[1]}\!A_0(r)$ (\ref{4.17})-(\ref{4.18}) into the non-relativistic form
$\tilde E_S$ (\ref{5.43}) of the energy functional which then becomes an ordinary function
of the length paramter $r_*$ ($\tilde E_S(r_*)$, say). The positronium ground-state energy
$E_0$ emerges then as the minimal value of the function $\tilde E_S(r_*)$ which may occur
for $r_* = r_0$:
\begin{subequations}\label{6.1}
\begin{align}
&E_0 = \tilde E_S \Big \vert_{r_* = r_0} \label{6.1a} \\
&\frac{d\tilde E_S(r_*)}{dr_*} \Big \vert_{r_* = r_0} = 0\label{6.1b} \ .
\end{align}
\end{subequations}
This RST groundstate energy $E_0$ must then be compared to the corresponding groundstate
energy $E_{HS}$ predicted by the conventional Hartree-Schr\"odinger approach in order to
test which one of both approaches yields the more accurate numbers in comparison with the
observational data.

\subsection{RST Prediction}
\indent Concerning the non-relativistic RST functional $\tilde E_S$ (\ref{5.45}), it is
highly informative to consider any of its contributions separately. First, the
one-particle kinetic energy $E_{kin}$ is taken from (\ref{5.29}) or (\ref{5.36}) and is
always of the general form
\begin{align}\label{6.2}
  E_{kin} = \frac{\hbar^2}{4M} \int d^2\vec r \; \bigg(\frac{d\tilde R(r)}{dr}&\bigg)^2 =
  2\, \epsilon_{kin} \, \alpha_{\rm S}^2\,Mc^2 \, \frac{1}{y_*} \nonumber \\
  (y_*\doteqdot r&_*/a_{\rm B}) \ .
\end{align}
Here $\epsilon_{kin}$ is a real number which is associated with the exponential trial
function $\tilde R(r)$ (\ref{4.7}), through
\begin{align}\label{6.3}
\epsilon_{kin}=\int\limits_{0}^{\infty} dy\;y\, &e^{-2y}=\frac{1}{4} \nonumber \\
(y\doteqdot r&/r_*) \ ,
\end{align}
and the typical atomic energy ("atomic unit", $a.u.$) is given by
\begin{equation}\label{6.4}
\alpha_{\rm S}^2 \, Mc^2 = \frac{e^2}{a_B} \cong 27,\!21 \,[eV] \ .
\end{equation}
Thus the one-particle energy $E_{kin}$ (\ref{6.2}) appears as
\begin{equation}\label{6.5}
E_{kin} \cong \frac{13,\!61}{y_*^2}\,[eV] \ .
\end{equation}

\indent Next, the electrostatic field energy $\hat E_R^{(e)}$ (\ref{5.40}) is found in a
similar way by use of the Struve-Neumann potential ${}^{[1]}\!A_0(r)$
(\ref{4.17})-(\ref{4.18}) as
\begin{equation}\label{6.6}
  \hat E_R^{(e)} = -\frac{\hbar c}{\vpas} \int d^3\vec r \;
  \bigg(\frac{d\,{}^{[1]}\!A_0(r)}{dr}\bigg)^2 = -2\,\hat \epsilon_R^{(e)}\, \alpha_{\rm S}^2 \, Mc^2 \cdot \frac{1}{y_*} \ .
\end{equation}
The real number $\hat \epsilon_R^{(e)}$ is given in terms of the dimensionless
Struve-Neumann screening factor $f_{SN}(x)$ (\ref{4.18}) as
\begin{equation}\label{6.7}
\hat \epsilon_R^{(e)} = \int\limits_0^\infty dx\; x^2
\bigg[\frac{d}{dx}\Big(\frac{f_{SN}(x)}{x}\Big)\bigg]^2 \simeq 0,\!3822 \ .
\end{equation}
Thus the field energy $\hat E_R^{(e)}$ is found as
\begin{equation}\label{6.8}
\hat E_R^{(e)} \simeq - \frac{20,\!80}{y_*} \, [eV] \ .
\end{equation}

\indent And finally, the mass equivalent $M_R^{(e)} c^2$ (\ref{5.25}) is found to appear
in its non-relativistic form as follows:
\begin{equation}\label{6.9}
  M_R^{(e)}c^2 \Rightarrow -\frac{\hbar c}{2} \int d^2\vec r \; {}^{[1]}\!A_0(r)\tilde
  R(r)^2 = -2 \mu_R^{(e)}\, \alpha_{\rm S}^2\, Mc^2 \cdot \frac{1}{y_*}\ ,
\end{equation}
with the real number $\mu_R^{(e)}$ being given by 
\begin{equation}\label{6.10}
\mu_R^{(e)} = \int\limits_0^\infty dx \; e^{-x}\, f_{SN}(x) \simeq 0,\!4348 \ ,
\end{equation}
which then yields for the desired mass equivalent
\begin{equation}\label{6.11}
M_R^{(e)} c^2 \simeq - \frac{23,\!66}{y_*} \, [eV] \ .
\end{equation}

\indent Observe here that the kinetic energy $E_{kin}$ ($\sim y_*^{-2}$) is positive,
whereas the electric contributions $\hat E_R^{(e)}$ ($\sim y_*^{-1}$) and $M_R^{(e)} c^2$
($\sim y_*^{-1}$) are negative so that the expected equilibrium point (\ref{6.1b}) of the
energy $\tilde E_S(y_*)$ (\ref{5.43})
\begin{align}\label{6.12}
\tilde E_S(y_*) &= \frac{27,\!21}{y_*^{\, 2}} - \frac{20,\!80}{y_*} + 2\cdot
\Big(-\frac{23,\!66}{y_*}+\frac{20,\!80}{y_*} \Big) \ \ [eV] \nonumber \\
&= \frac{27,\!21}{y_*^{\, 2}} -\frac{26,\!52}{y_*} \ \ [eV]
\end{align}
does really exist and is found as
\begin{equation}\label{6.13}
y_0 =\doteqdot \frac{r_0}{a_{\rm B}} = \frac{4\,\mu_R^{(e)} - \epsilon_R^{(e)}}{\epsilon_{kin}} \simeq 1,\!95 \ ,
\end{equation}
see fig.2. Therefore the corresponding groundstate energy $E_0$ (\ref{6.1a}) of positronium is obtained as
\begin{equation}\label{6.14}
E_0= -6,\!48 \, eV \ .
\end{equation}
This differs from the experimental value ($\simeq 6,\!80 \, [eV]$) by $0,\!32 \, [eV]$ and
thus falls within the error estimate ($\simeq 1\, [eV]$) of the Appendix. The important
point here is that the bracket term in the energy functionals (\ref{5.45}) and
(\ref{6.12}) is an essential element of our present RST treatment of the positronium
groundstate because it contributes roughly $3 \, [eV]$ ($\simeq 50\%$) to the binding
energy. Since however this term is missing in the conventional Hartree-Schr\"odinger
approach, one expects that the latter approach will predict the groundstate energy with an
error of (roughly) $50\%$!

\subsection{Hartree-Schr\"odinger Prediction}
\indent According to the general belief, the conventional Schr\"odinger equation
\begin{equation}\label{6.14'}
\hat H \Psi(\vec r_1, \vec r_2) = {}^{(S)}\!E_0 \, \Psi(\vec r_1, \vec r_2)
\end{equation}
would yield the exact groundstate energy ${}^{(S)}\!E_0$ (apart from relativistic and QED
effects) if this eigenvalue problem could be solved exactly. However, these exact
solutions are not known in general (apart from such special cases as (\ref{2.19''})); and
therefore one has to resort to certain approximation techniques, e.g. the Hartree-Fock
method \cite{Fl,Ba}. But since the spin effects are neglected for the present
purposes, that method reduces to the Hartree approach which is formally based upon the
conventional one-particle Schr\"odinger equation ({\it"Hartree-Schr\"odinger approach"}).
The essential point with this approximation method refers to trying a simple product
ansatz for the two-particle wave function $\Psi(\vec r_1, \vec r_2)$ (\ref{6.14'}) in
terms of one-particle wave functions $\psi(\vec r_1)$ and $\psi(\vec r_2)$, i.e. one tries
for the present groundstate problem
\begin{equation}\label{6.15}
\Psi(\vec r_1, \vec r_2) = \psi(\vec r_1) \cdot \psi(\vec r_2) \ .
\end{equation}

\indent This ansatz may be inserted into the conventional action integral
$W_S$~\cite{Fl,Ba}
\begin{equation}\label{6.16}
W_S = \int\!\!\!\!\int d^3\vec r_1\, d^3\vec r_2 \; \Psi^*(\vec r_1, \vec r_2) \, \hat H \, \Psi(\vec r_1,\vec r_2) \ ,
\end{equation}
which adopts its stationary values ($\delta W_S=0$), within the set of normalized wave functions
\begin{equation}\label{6.16'}
  N_S \doteqdot \int\!\!\!\!\int d^3\vec r_1\, d^3\vec r_2 \; \Psi^*(\vec r_1, \vec r_2)
  \cdot \Psi(\vec r_1,\vec r_2) -1 = 0 \ ,
\end{equation}
just for the solutions of the Schr\"odinger eigenvalue problem (\ref{6.14'}). However for
the trial functions (\ref{6.15}), the Schr\"odinger action (\ref{6.16}) becomes converted
to its Hartree-Schr\"odinger form $W_{HS}$ which looks as follows:
\begin{equation}\label{6.17}
W_{HS} = 2\,{}^{(S)}\!E_{kin} + V_{HS} \ ,
\end{equation}
where the one-particle kinetic energy ${}^{(S)}\!E_{kin}$ is given as usual by
\begin{equation}\label{6.18}
{}^{(S)}\!E_{kin} = \frac{\hbar^2}{2M} \int d^3\vec r \; \vec \nabla \psi^*(\vec r) \cdot \vec \nabla \psi(\vec r)
\end{equation}
and the electrostatic interaction energy $V_{HS}$ by
\begin{equation}\label{6.19}
  V_{HS} = -e^2 \int\!\!\!\!\int d^3\vec r_1 \,d^3\vec r_2 \; \frac{\vert\psi(\vec
    r_1)\vert^2 \cdot \vert\psi(\vec r_2)\vert^2}{\vert\!\vert \vec r_1 - \vec r_2 \vert\!\vert} \ .
\end{equation}

\indent Obviously, the Hartree-Schr\"odinger action $W_{HS}$ (\ref{6.17}) represents the
energy of the two-particle system and therefore is the conventional counterpart of our
non-relativistic RST energy functional $\tilde E_T$ (\ref{5.43}). Therefore the
corresponding energy minimalization due to the method of Lagrangean multipliers
($\delta(W_{HS} + \lambda_S\cdot N_S)=0$) yields the following one-particle eigenvalue
equation of the Schr\"odinger type:
\begin{equation}\label{6.20}
-\frac{\hbar^2}{2M} \Delta \psi(\vec r) - \hbar c {}^{(S)}\!A_0(\vec r) \cdot \psi(\vec r)
= - \lambda_S \cdot \psi(\vec r) \ .
\end{equation}
Here the Lagrangean multiplier $\lambda_S$ plays the role of the energy eigenvalue and the
electric interaction potential ${}^{(S)}\!A_0(\vec r)$ is given in terms of the wave
function $\psi(\vec r)$ through
\begin{equation}\label{6.21}
{}^{(S)}\!A_0(\vec r) = \alpha_{\rm S} \int d^3\vec r\,' \; \frac{\vert \psi(\vec
  r\,')\vert^2}{\vert\!\vert\vec r-\vec r\,'\vert\!\vert} \ .
\end{equation}
It is true, both the energy eigenvalue equation (\ref{6.20}) and the interaction potential
(\ref{6.21}) look very similar to their non-relativistic RST counterparts, see equations
(\ref{5.31a})-(\ref{5.31b}) and (\ref{3.52}), resp. But a closer inspection will readily
reveal very important differences.

\indent First, observe that the conventional eigenvalue equation (\ref{6.20}) is based
upon the ordinary Laplacean $\Delta$ in spherical polar coordinates (see (\ref{3.63}) for
its radial part), whereas the corresponding RST eigenvalue equation (\ref{3.62}) relies
upon the spherical polar coordinate $r$ as a cylindrical variable. Correspondingly, the
RST normalization condition (\ref{4.5}) for the wave function $\tilde R(r)$ uses a {\it
  two-dimensional} integral, whereas the conventional formalism refers to a
three-dimensional integration, i.e.
\begin{equation}\label{6.22}
\int d^3\vec r \; \big \vert \psi(\vec r)\big \vert^2 = 1 \ .
\end{equation}
For instance, if we adopt again an exponential trial function with length parameter $r_s$
\begin{equation}\label{6.23}
\psi(\vec r) = \frac{1}{\sqrt{\pi\, r_s^{\,3}}} \, e^{-r/r_s} \ ,
\end{equation}
which obeys the conventional normalization (\ref{6.22}), this trial function must
necessarily differ from its RST counerpart $\tilde R(r)$ (\ref{4.7})! But nevertheless,
the kinetic energies turn out to be identical, i.e. for the conventional case (\ref{6.18})
one finds
\begin{align}\label{6.24}
{}^{(S)}\!E_{kin} = \frac{1}{2} \alpha_{\rm S}^2 \, Mc^2 &\cdot \frac{1}{y_s^2} \simeq
\frac{13,\!61}{y_s^2} \ \, [eV] \nonumber \\
(y_s \doteqdot& r_s/a_B) \ ,
\end{align}
which actually is the same as its RST counterpart (\ref{6.2})-(\ref{6.3}).

\indent However, the crucial difference between both approaches does refer to the
electrostatic interaction potentials ${}^{(S)}\!A_0(\vec r)$ (\ref{6.21}) vs.
${}^{[1]}\!A_0(r)$ (\ref{4.17})-(\ref{4.18}). For the conventional case (\ref{6.21}) one
finds by use of the trial wave function (\ref{6.23})
\begin{equation}\label{6.25}
{}^{[S]}\!A_0(r) = \Big(1-e^{-2r/r_s}\Big)\frac{\alpha_{\rm S}}{r} - \frac{\alpha_{\rm S}}{r_s}\, e^{-2\,r/r_s} \ ,
\end{equation}
and this is of a different type in comparison to the RST case ${}^{[1]}\!A_0(r)$, see
fig.1: the conventional potentials of the form (\ref{6.21}) have vanishing field strength
${}^{(S)}\!E_r$ ($\doteqdot d\,{}^{(S)}\!A_0(r) / dr$) at the origin ($r=0$), whereas the
corresponding RST field strengths adopt a non-vanishing value, cf. (\ref{4.26})! Of
course, this circumstance must have its consequences for the interaction energy $V_{HS}$
(\ref{6.19}) concentrated in the electrostatic field:
\begin{equation}\label{6.26}
V_{HS}=-\hbar c \int d^3\vec r \; {}^{(S)}\!A_0(r)\,\psi(r)^2 =  - \frac{5}{8}\,
\alpha_{\rm S}^2 \,Mc^2 \cdot \frac{1}{y_s} \simeq -\frac{17,\!00}{y_s} \ \,[eV] ,
\end{equation}
and this is now essentially different from the corresponding RST interaction energy (\ref{5.45})
\begin{equation}\label{6.27}
V_{RST} \doteqdot 2\,M_R^{(e)}c^2 - \hat E_R^{(e)} \simeq -\frac{26,\!52}{y_*} \ \, [eV] \ ,
\end{equation}
cf. (\ref{6.12}). Thus the conventional Hartree-Schr\"odinger
energy $W_{HS}$ (\ref{6.17}) is finally found as
\begin{equation}\label{6.28}
W_{HS}(y_s) = \frac{27,\!21}{y_s^2} - \frac{17,\!00}{y_s} \ \, [eV] \ ,
\end{equation}
in contrast to the RST case $\tilde E_S(y_*)$ (\ref{6.12}). Whereas in the latter case the
RST energy adopts its minimal value (\ref{6.14}) for $y_0=1,\!95$, the minimal value of
the conventional energy $W_{HS}$ (\ref{6.28}) occurs for $y_0=3,\!2$ and is found as
$E_0\simeq -2,\!65 \, [eV]$, see fig.2. Clearly this is a very poor result when compared to
the corresponding RST prediction of $6,\!48 \, [eV]$ (\ref{6.14}) and to the experimental
value of $6,\!80\,[eV]$, resp., see fig.2.

\indent Summarizing on the basis of the present groundstate calculations, it appears
natural to assume that in the general situation RST will be superior to the Hartree-Fock
approach, as far as the numerical coincidence of the theoretical predictions and the
experimental values is concerned. Thus as the next task, there remains to be settled the
(non-relativistic) competition between the conventional Schr\"odinger approach and RST.
Here, as the Appendix demonstrates, the anisotropy of the RST field configurations must be
taken into account. Indeed, such a competition seems to be of profound philosophical
relevance, for if RST could predict the same numbers as the standard Schr\"odinger theory,
the latter approach would lose its status of uniqueness!

\appendix*
\section{Spherically Symmetric Approximation}

\indent In order to get some feeling of the extent to which the first anisotropic
correction ${}^{(I\!I)}\!A_0(r,\vartheta)$ (\ref{4.12}) is smaller than the spherically
symmetric contribution ${}^{(I)}\!A_0(r)$ (\ref{4.11}), it is very instructive to consider
a typical example, which admits to exactly calculate all the terms in question.

\indent Here, as the prototype of a well-localized charge distribution $\tilde R(r)^2$,
one may adopt the following step function
\begin{equation}\label{A.1}
  \tilde R^{\;2}(r) \Rightarrow {}^{(c)}\!\tilde R^{\;2}(r) = \begin{cases}
    {\displaystyle\frac{4}{\pi\cdot r_c^2}} \ , &0\leq r \leq r_c \\ 0 \ , & r > r_c \ \ . \end{cases}
\end{equation}
It should be obvious that this (non-relativistic) charge density actually obeys the
normalization condition (\ref{4.5}) for any fixed length parameter $r_c$. Furthermore,
substituting this step density into the spherically symmetric approximation
${}^{(I)}\!A_0(r)$ (\ref{4.11}) yields the following model potential ${}^{[c]}\!A_0(r)$:
\begin{equation}\label{A.2}
{}^{(I)}\!A_0(r) \Rightarrow {}^{[c]}\!A_0(r) = \frac{2\,\alpha_{\rm S}}{r_c^{\;2}} \big\{\sqrt{r_c^2+r^2}-r\big\} \ .
\end{equation}

\indent Of course, this potential ${}^{[c]}\!A_0(r)$ adopts the required Coulomb form
(\ref{3.52}) in the asymptotic region ($r\gg r_c$), see fig.1. If the length parameter
$r_c$ is chosen to be infinitesimally small ($r_c\rightarrow 0$), the Coulomb form does
fill the whole three-space ($0 < r < \infty$). The screening factor ($f_c$, say) is found
for general value of $r_c$ as
\begin{equation}\label{A.3}
f_c(r)=2\,\frac{r}{r_c}\Big\{\sqrt{1+\big(\frac{r}{r_c}\big)^2}-\frac{r}{r_c}\Big\}
\end{equation}
qand thus is seen to tend to unity ($f_c(\infty)=1$) in the asymptotic region ($r\gg
r_c$). On the other hand, the screening factor $f_c$ tends to zero at the origin
($r\rightarrow 0$); and therefore the model potential ${}^{[c]}\!A_0(r)$ (\ref{A.2})
remains finite for $r\rightarrow 0$, i.e.
\begin{equation}
{}^{[c]}\!A_0(0) = \frac{2\,\alpha_{\rm S}}{r_c} \ ,
\end{equation}
cf. the analogous behaviour of the Struve-Neumann potential ${}^{[1]}\!A_0(0)$
(\ref{4.25}). Obviously these are the common features of all those interaction potentials
${}^{(I)}\!A_0(r)$ (\ref{4.11}) due to the exotic states! (see fig.1).

\indent Next one wishes to check for the chosen step density (\ref{A.1}), to what extent
the first anisotropy correction ${}^{(I\!I)}\!A_0(r)$ (\ref{4.14}) really is smaller than
the spherically symmetric potential ${}^{[c]}\!A_0(r)$ (\ref{A.2}). For this purpose, one
substitutes the step density into the radial factor (\ref{4.14}) and thus finds
\begin{equation}\label{A.5} 
{}^{(I\!I)}\!A_0(r) \Rightarrow {}^{[cc]}\!A_0(r) =
  \frac{8\,\alpha_{\rm S}}{3\pi\,r_c^2} \,\Big\{r-\frac{1}{2} r^2\cdot \frac{2r^2+3
    r_c^2}{\sqrt{r^2+r_c^2}^{\,3}}\Big\} \ .
\end{equation}
And indeed, this anisotropic correction vanishes at the origin ($r=0$); and at infinity
($r\rightarrow \infty$) it tends to zero faster ($\sim r^{-3}$) than the Coulomb potential
($\sim r^{-1}$), see fig.1. Therefore the neglection of the first anisotropic correction
seems to be justified.

\indent However, perhaps a better measure of the relative magnitudes consists in the field
energy $\hat E_R^{(e)}$ (\ref{5.40}). Substituting there the decomposition (\ref{4.10}) of
the electrostatic potential ${}^{(p)}\!A_0(\vec r)$ into the radial part
${}^{(I)}\!A_0(r)$ and in its anisotropic corrections, one finds a similar decomposition
of the electrostatic energy $\hat E_R^{(e)}$, i.e
\begin{equation}\label{A.6}
\hat E_R^{(e)}= {}^{(I)}\!\hat E_R^{(e)} + {}^{(I\!I)}\!\hat E_R^{(e)} + \dots
\end{equation}
where the first energy contribution ${}^{(I)}\!\hat E_R^{(e)}$ is due to the
spherically-symmetric potential ${}^{(I)}\!A_0(r)$
\begin{equation}
\label{A.7}
{}^{(I)}\!\hat E_R^{(e)}=-\frac{\hbar c}{\vpas} \int d^3\vec r\;\big\vert\!\big\vert \vec
\nabla {}^{(I)}\!A_0(r) \big\vert\!\big\vert^2 \ ,
\end{equation}
and similarly the second contribution is built up by the spherically-symmetric potential
together with the first anisotropic correction
\begin{equation}\label{A.8}
{}^{(I\!I)}\!\hat E_R^{(e)}=-\frac{\hbar c}{2\pi \alpha_{\rm S}} \int d^3\vec r \; \vec
\nabla {}^{(I)}\!A_0(r) \sdot \vec \nabla {}^{(I\!I)}\!A_0(\vec r) \ .
\end{equation}
But the point here is now that both potentials ${}^{(I)}\!A_0(r)$ and
${}^{(I\!I)}\!A_0(\vec r)$ can be determined exactly for the step function density, see
(\ref{A.2}) and (\ref{A.5}); and therefore the corresponding energy integrals (A.7)
and (A.8) are also known exactly, namely
\begin{subequations}
  \begin{align}
    {}^{(I)}\!\hat E_R^{(e)} &\Rightarrow {}^{[c]}\!\hat E_R^{(e)} = - (2\pi-\frac{16}{3}
    ) \frac{e^2}{r_c} \simeq - 0,\!9499\, \frac{e^2}{r_c} \label{A.9a} \\
    {}^{(I\!I)}\!\hat E_R^{(e)} &\Rightarrow {}^{[cc]}\!\hat E_R^{(e)} = -
    \frac{8}{3}(\frac{4}{3}-\frac{13\pi}{32} ) \frac{e^2}{r_c} \simeq - 0,\!1522
    \,\frac{e^2}{r_c} \label{A.9b} \ .
\end{align}
\end{subequations}

\indent Thus the first anisotropic energy correction (A.9b) amounts to (roughly)
$16\%$ of the spherically symmetric energy (A.9a)! This says that the restriction to
the spherically-symmetric approximation lets us expect an uncertainty of our positronium
groundstate result up to $1\, eV$ near the experimental value (of $6.8 \,eV$).

\pagebreak

\enlargethispage{2\baselineskip}

\begin{figure}
\begin{center}
\epsfig{file=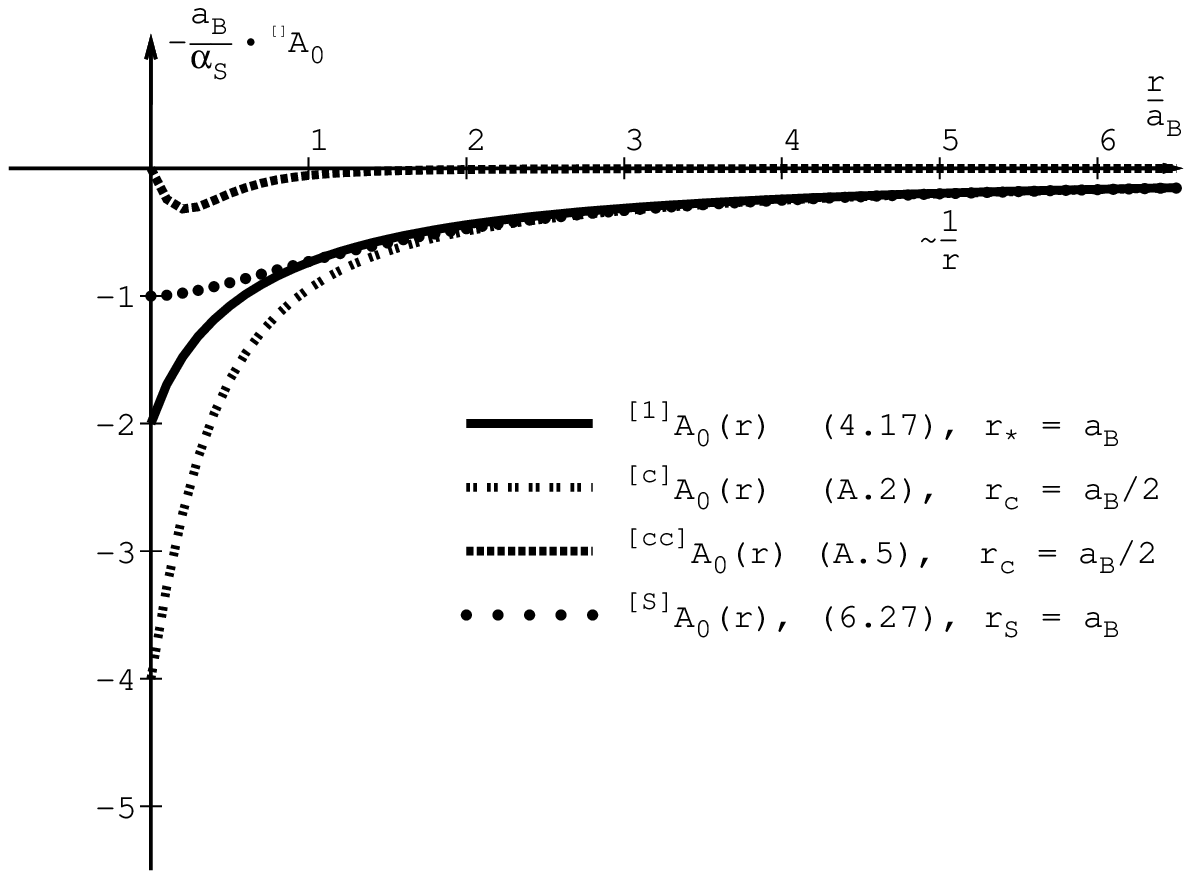,angle=0, width=17cm}
\end{center}
\end{figure}
\noindent
\textbf{Fig. 1:}\hspace{0.5cm} \emph{\large\bf \ Spherically Symmetric Approximation}
\indent
\vspace{0.5cm}

The Struve-Neumann potential~$\Ai$ (4.17) (solid line) has a finite value (4.25)
at the origin~$(r=0)$, as well as the corresponding field strength~$\Er$ (4.26). The
Coulomb form~$\sim 1/r$ (3.58) is adopted at spatial infinity~$(r\leadsto\infty$). These
features are common to all potentials due to a charge distribution of the type (4.2),
e.g. also for the potential model~$\Ac$~(A.2) generated by the cut-off charge~(A.1) 
(broken line). The first anisotropic correction~$\Accc$~(A.5) (intersected line) is
considerably smaller than the spherically symmetric approximation~$\Ac$ (A.2) and is 
therefore neglected for the present rough estimate. The conventional Hartree-Schr\"odinger
potential~$\AS$~(6.27) (dotted line) has vanishing field strength at the origin, in
contrast to the RST potentials~$\Ai$ and~$\Ac$.

\pagebreak
\enlargethispage{2\baselineskip}

\begin{figure}
\begin{center}
\epsfig{file=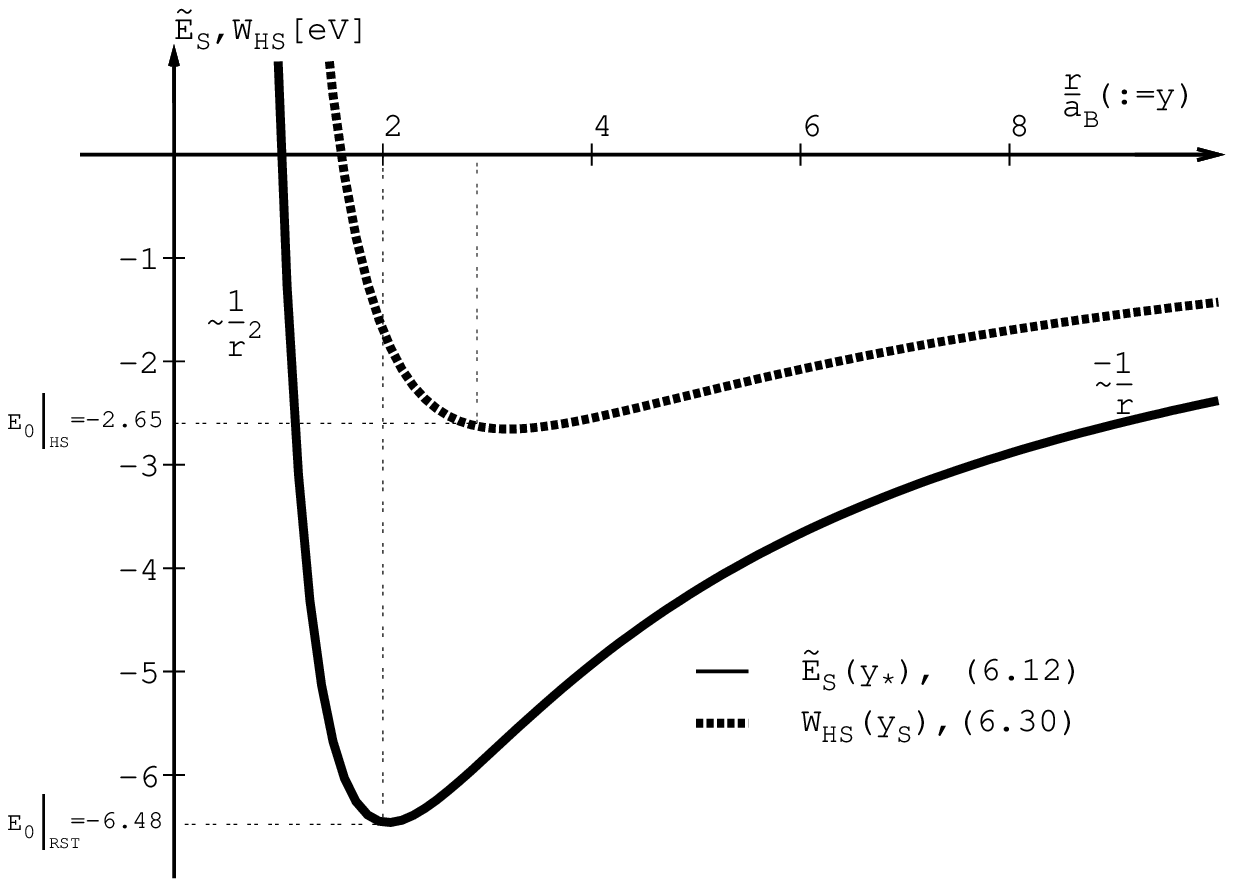,angle=0, width=17cm}
\end{center}
\end{figure}
\noindent
\textbf{Fig.\ 2:}\hspace{0.5cm} \emph{\large\bf \ RST Energy~$\tilde{E}_{\rm S}$ (6.12) and
  Hartree-Schr\"odinger \phantom{Fig.\ 1:\hspace{0.5cm} } Energy~$W_{\rm HS}$ (6.30)}
\indent
\vspace{0.5cm}

The groundstate energy~$E_0$ (6.1a) is obtained from the minimum (6.1b) of the energy curves.
RST and the conventional Hartree-Schr\"odinger approach agree in the kinetic
energies~$E_{\rm kin}$ but differ in the interaction energies~$V_{\rm HS}$ and~$V_{\rm
  RST}$, see equations (6.28)-(6.29). This is the reason why the RST prediction (-6,48 [eV])
for the groundstate energy is closer to the experimental value (-6.80 [eV]) than the
Hartree-Schr\"odinger prediction (-2.65 [eV]).

\pagebreak

\end{document}